\documentclass[longauth]{aa} 

\usepackage[varg]{txfonts}
\usepackage{graphicx}
\usepackage{array}
\usepackage{lscape}                                
\usepackage{txfonts}
\usepackage{natbib}
\usepackage{longtable}
\usepackage{color}
\usepackage{supertabular}
\usepackage[draft]{hyperref}
\usepackage[flushleft]{threeparttable}
\usepackage{multirow}
\usepackage{lscape}
\usepackage{booktabs}

\newcommand{\kms} {km\,s$^{-1}$}

\newcommand{\vsini} {$v$\,sin\,$i$}

\newcommand{\vmacro} {$v_{\rm mac}$}
\newcommand{\vrad} {$v_{\rm rad}$}
\newcommand{\Teff} {$T_{\rm eff}$}
\newcommand{\grav} {log\,{\em $g$}}
\newcommand{\gravt} {log\,{\em $g_{\rm true}$}}

\newcommand{\micro} {$\xi_{\rm t}$}
\newcommand{\helio} {Y$_{\rm He}$}
\newcommand{\fastwind} {{\sc fastwind}}
\newcommand{\ioni}[2]{{#1\,\sc{#2}}}
\newcommand{\OIII}{O\,{\sc iii}\,$\lambda$5592}
\newcommand{\msol}{$M_{\odot}$}
\newcommand{\vinf}{$v_{\infty}$}
\begin{document}
%
\title{The IACOB project}

\subtitle{V. Spectroscopic parameters of the O-type stars in the modern grid of standards for spectral classification
} 

\author{G. Holgado\inst{1,2}, S. Sim\'on-D\'iaz\inst{1,2}, R.H. Barb\'a\inst{3}, J. Puls\inst{4}, A. Herrero\inst{1,2}, N. Castro\inst{5}, M. Garcia\inst{6}, J. Ma\'iz Apell\'aniz\inst{7}, I. Negueruela\inst{8}, C. Sab\'in-Sanjuli\'an\inst{3,9}}

\institute{Instituto de Astrof\'isica de Canarias, E-38200 La Laguna, Tenerife, Spain.
             \and
             Departamento de Astrof\'isica, Universidad de La Laguna, E-38205 La Laguna, Tenerife, Spain.
             \and
             Departamento de F\'isica y Astronom\'ia, Universidad de la Serena, Av. Juan Cisternas 1200 Norte, La Serena, Chile.
             \and
             Universit\"atssternwarte, Scheinerstr. 1, D-81679 M\"unchen, Germany.
             \and
             Department of Astronomy, University of Michigan, 1085 S. University Avenue, Ann Arbor, MI 48109-1107, USA.
             \and
             Centro de Astrobiolog\'ia (INTA-CSIC), Departamento de Astrof\'isica. Ctra. Torrej\'on a Ajalvir km.4, E-28850 Torrej\'on de Ardoz (Madrid), Spain.
             \and
             Centro de Astrobiolog\'ia, CSIC-INTA, campus ESAC, camino bajo del castillo s/n, E-28692 Madrid, Spain.
             \and
             Departamento de F\'isica, Ingenier\'ia de Sistemas y Teor\'ia de la Señal, Escuela Polit\'ecnica Superior, Universidad de Alicante, Carretera de San Vicente del Raspeig s/n, 03690 San Vicente del Raspeig, Alicante, Spain.
             \and
             Instituto de Investigaci\'on Multidisciplinar en Ciencia y Tecnolog\'ia, Universidad de La Serena, Ra\'ul Bitr\'an 1305, La Serena, Chile.
            }
		   
\offprints{gholgado@iac.es}

\date{Date}

\titlerunning{Spectroscopic parameters of the O-type standards for spectral classification}
\authorrunning{Holgado et al.}

%
\abstract
{The IACOB and OWN surveys are two ambitious complementary observational projects which have made available a large multi-epoch spectroscopic database of optical high resolution spectra of Galactic massive O-type stars.}
{As a first step in the study of the full sample of (more than 350) O stars surveyed by the IACOB and OWN projects, we have performed the quantitative spectroscopic analysis of a subsample of 128 stars included in the modern grid of O-type standards for spectral classification. The sample comprises stars with spectral types in the range O3\,--\,O9.7 and covers all luminosity classes.}
{We use the semi-automatized \textsc{iacob-broad} and \textsc{iacob-gbat/fastwind} tools to determine the complete set of spectroscopic parameters that can be obtained from the optical spectrum of O-type stars. A quality flag is assigned to the outcome of the {\sc iacob-gbat/fastwind} analysis for each star, based on a visual evaluation of how the synthetic spectrum of the best fitting \fastwind\ model reproduces the observed spectrum.
We also benefit from the multi-epoch character of the IACOB and OWN surveys to perform a spectroscopic variability study of the complete sample, providing two different flags for each star accounting for spectroscopic binarity as well as variability of the main wind diagnostic lines.}
{We obtain -- for the first time in a homogeneous and complete manner -- the full set of spectroscopic parameters of the "anchors" of the spectral classification system in the O star domain. 
We provide a general overview of the stellar and wind parameters of this reference sample, as well as updated recipes for the SpT\,--\,\Teff\ and SpT\,--\,\grav\ calibrations for Galactic O-type stars. We also propose a distance-independent test for the wind-momentum luminosity relationship. 
We evaluate the reliability of our semi-automatized analysis strategy using a subsample of $\sim$40 stars extensively studied in the literature, and find a fairly good agreement between our derived effective temperatures and gravities and those obtained by means of more traditional ``by-eye'' techniques and different stellar atmosphere codes. The overall agreement between the synthetic spectra associated with the {\sc iacob-gbat/fastwind} best fitting models and the observed spectra is good for most of the analyzed targets, but 46 stars out of the 128 present a particular behavior of the wind diagnostic lines that cannot be reproduced by our grid of spherically symmetric unclumped models. These are potential targets of interest for more detailed investigations of clumpy winds and/or the existence of additional circumstellar emitting components contaminating the wind diagnostic lines (e.g. disks, magnetospheres). Last, our spectroscopic variability study has led to the detection of clear or likely signatures of spectroscopic binarity in 27\% of the stars and small amplitude radial velocity variations in the photospheric lines of another 30\%. Additionally, 31\% of the investigated stars show variability in the wind diagnostic lines.}
{}

%
\maketitle
%
%
\section{Introduction}\label{section1}

We are immersed in the era of large spectroscopic surveys of massive OB-type stars. This enormous observational effort, in combination with the availability of mature stellar atmosphere codes for massive stars \citep[see an extensive list of codes and associated references in][]{Puls15}, is enabling a considerable -- and still on-going -- increase in the amount of available information about stellar parameters, magnetism, and multiplicity in the full O and B star domain (e.g., the series of papers from the VLT-FLAMES Massive Stars and Tarantula Surveys, and other international collaborations such as GOSSS, OWN, MiMeS, IACOB, BOB, or BinaMIcS\footnote{References for these surveys are available through the ADS using the abbreviated survey name.}).

The \textit{Gaia} revolution is also here, and we are very close to the epoch when the \textit{TESS} and \textit{PLATO} satellites are expected to drive a major breakthrough in the field of asteroseismology of massive stars. 
Last, hopefully soon, new space missions observing the ultraviolet and infrared spectral windows (such as \textit{WSO}, \textit{JWST}, \textit{Spica}, and \textit{LUVOIR}) will provide fresh momentum to the investigation of the intricate characteristics of the strong radiatively driven winds dominating the early phases of evolution of these massive stellar objects (Kudritzki \& Puls 2000).

In addition to the interest per se within the field of massive stars, this unprecedented observational enterprise is motivated by the huge impact that our knowledge of the basic physical properties and the evolution of these stars have on many and diverse aspects of the study of the Cosmos \citep[e.g., star formation, chemodynamical evolution of galaxies, re-ionization of the Universe; see ][]{Herbig62, Elmegreen77, Preibisch07, Prantzos08, Tenorio-Tagle06, Bromm09, Robertson10}. They are also the progenitors of the most extreme stellar objects known in the Universe (e.g., hyper-energetic supernovae, Wolf-Rayet stars, luminous blue variables, massive stellar black holes, neutron stars, magnetars, massive X- and $\gamma$-ray binaries), and the origin of new studied phenomena such as long duration $\gamma$-ray bursters \citep{Woosley06} or the recently detected gravitational waves produced by a merger of two massive black holes \cite[][incl. LIGO and Virgo Collaborations]{Abbott16}.

In spite of the remarkable advances in the modeling and spectroscopic analysis techniques of these stars in the last two decades, our knowledge of these important (but complex) astrophysical objects has been limited until very recently to conclusions extracted from the analysis of single-epoch medium resolution spectroscopic observations of relatively small samples \citep[see, however, recent efforts by the VLT-FLAMES Tarantula Survey, BOB and MiMeS collaborations,][]{Evans11, Morel14, Wade16}. The IACOB project \citep[][]{Simon11c,Simon14,Simon15} is an ambitious long-term observational project driven by the compilation and scientific exploitation of a large database of high-resolution multi-epoch spectra of Galactic OB stars. One of the immediate objectives of the project is to perform a thorough empirical characterization of the whole sample (including spectroscopic-physical parameters and abundances). Eventually, this wealth of information will be conveniently used to investigate the impact that parameters in addition to mass, rotation and stellar winds -- such as binarity/multiplicity, magnetic fields, or stellar oscillations -- have on the physical and wind properties of massive stars, as well as on the evolution of these important astrophysical objects. In this endeavor, the IACOB project has established a strong collaboration with the complementary southern high-resolution survey OWN \citep{Barba10, Barba17} and will be actively participating in the WEAVE \citep{Dalton16} spectroscopic survey of OB stars in the Galactic plane.

An important part of the work developed in the first stages of the IACOB project is the quantitative spectroscopic analysis of the whole sample of stars available in the IACOB and OWN surveys. At present, both databases together comprise more than 700 Galactic O- and B-type stars and $\sim$10\,000 spectra. We are making progress in the analysis of the complete IACOB+OWN sample of O-type stars \citep{Holgado17}. In this paper, we first concentrate on those targets among them historically selected as standards for spectral classification. We start with this reference sample in order to present our methodology of analysis as well as to assess the level of agreement found between the results obtained by means of our semi-automatized analysis strategy and those provided by more traditional ``by-eye'' techniques. We obtain -- for the first time in a homogeneous and complete manner -- the full set of spectroscopic parameters\footnote{That can be obtained from the optical spectrum of O-type stars.} of the "anchors" of the spectral classification system using a single-snapshot observation per target. In addition, we benefit from the multi-epoch character of the IACOB and OWN surveys to perform a spectroscopic variability study of the complete sample.

The structure of this paper is as follows. The sample and the observations are described in Sect.~\ref{section2}. Section~\ref{section3} presents the spectroscopic analysis tools we used to extract line-broadening, spectroscopic parameters and multi-epoch information. Sections~\ref{section4} and \ref{section5} constitute the core of the paper. There, we present and discuss the results and the global properties of the sample. Concluding remarks and future prospects are found in Sect.~\ref{section6}.

\section{Sample definition and observations}\label{section2}

\begin{table*}
\caption{Summary of the spectroscopic observations used in this work} 
\label{tablesurveys} 
\centering 
\begin{tabular}{l l l c c c c}
	\hline\hline
	\noalign{\smallskip}
	Survey       & Telescope    & Instrument & Resolving power & Range [\AA] & \# Spectra & \# Stars\\ 
	\hline
	\noalign{\smallskip}
	IACOB        & NOT2.56m      & FIES       & 46\,000 \& 25\,000 & 3750-7250  & 342  &  57  \\
	IACOB(sweG)  & MERCATOR1.2m  & HERMES     & 85\,000            & 3770-9000  & 498  &  28  \\
	OWN          & ESO2.2m       & FEROS      & 46\,000            & 3530-9210  & 279  &  78  \\
	CAF\'E-BEANS & CAHA2.2m      & CAF\'E     & 65\,000            & 3930-9220  & 97   &  31  \\ 
	\hline
	\noalign{\smallskip}
	TOTAL        &              &            &                    &            & 1216 & 195* \\ 
	\hline
	\noalign{\smallskip}
	\multicolumn{7}{l}{* 128 stars after removing stars in common}
	\end{tabular}
\end{table*}

\cite{Walborn90} presented the first comprehensive digital atlas of optical spectra for spectral classification of OB stars. It comprised a total of 75 standard objects with spectral types O3\,--\,B3 (--\,B8 at Ia) that were organized following the MK system (and some developments not considered before). Several updates and additions have occurred since then  \cite[see][and references therein]{Walborn2002}, the last one being developed in the framework of the Galactic O-Star Spectroscopic Survey \citep[GOSSS,][]{Maiz11, Sota11, Sota14, Maiz16}.

This work is based on the O-type stars included in the GOSSS OB2500 v2.0 grid of standards\footnote{We note that the GOSSS project has made available an updated version of the grid of O-type standards during the development of this work \citep{Maiz16}, including a few more stars and changes in the spectral classification for three luminosity class III stars that were shifted to luminosity class IV.} defined in \cite{Maiz15}. The grid comprises 131 Galactic stars with spectral types in the range O2\,--\,O9.7 (all luminosity classes) from both -- Northern and Southern -- hemispheres.

We performed an exhaustive search for spectra in the three modern high resolution spectroscopic databases IACOB\footnote{Also including the IACOB-sweG subsample, obtained with the HERMES spectrograph with the initial objective to investigate the spectroscopic behavior of MK standards earlier than B9 in
the \textit{Gaia}-RVS spectral range \citep{Negueruela15b}.} \citep{Simon11b, Simon11c, Simon15}, OWN \citep{Barba10,Barba17}, and CAF\'E-BEANS \citep{Negueruela15} and found a total of 1216 spectra for 128 of the 131 standard stars for spectral classification. These databases include spectra collected with four different instruments: FIES \citep{Telting14}, HERMES \citep{Raskin04}, FEROS \citep{Kaufer97}, and CAF\'E \citep{Aceituno13}. A global overview of the number of stars and spectra available from each survey is presented in Table \ref{tablesurveys}, where the corresponding resolving power ($R$) and wavelength range of the spectra are also indicated. As expected from the general philosophy of these surveys, most of the stars have more than 2 spectra, obtained at different epochs. Whenever available, all the multi-epoch spectra per target were used to roughly investigate spectroscopic variability (Sect.~\ref{subsectionVar}), but only the spectrum with best quality -- in terms of signal-to-noise ratio, S/N -- was considered for the quantitative spectroscopic analysis\footnote{This subset of spectra can be accessed online via the IACOB webpage: \href{http://www.iac.es/proyecto/iacob/}{http://www.iac.es/proyecto/iacob/}} leading to the stellar parameters (Sect.~\ref{section31}). 

A more detailed overview of the final sample of stars included in this work is presented in Table~\ref{tableStandards}. Apart from the name(s) of each star, an abridged spectral classification\footnote{http://gosc.iaa.es for detailed classifications.} -- following Table 1 in \cite{Maiz15} --, and the number of available spectra per instrument, we indicate some specific details about the spectra (one per star) used for the spectroscopic analysis. For completeness, Table~\ref{tableStandards} also includes the 3 stars for which we did not find spectra in any of the surveys. These are mainly southern stars with V magnitude fainter than $\sim$9.5 (two of them have V\,$\approx$\,10\,--\,11), hence making it more difficult to obtain a high-resolution spectrum of similar quality with a small or medium size telescope. Finally, we refer the reader to Tables~\ref{ParamStandards} for more detailed spectral classifications -- directly extracted from \cite{Sota11, Sota14} -- also including information about the qualifiers.

The FIES, HERMES, and FEROS spectra were reduced with the corresponding available pipelines (FIEStool\footnote{\href{http://www.not.iac.es/instruments/fies/fiestool/FIEStool.html}{http://www.not.iac.es/instruments/fies/fiestool/FIEStool.html}}, HermesDRS\footnote{\href{http://www.mercator.iac.es/instruments/hermes/hermesdrs.php}{http://www.mercator.iac.es/instruments/hermes/hermesdrs.php}}, and FEROSDRS\footnote{\href{http://www.eso.org/sci/facilities/instruments/feros/tools/DRS.html}{http://www.eso.org/sci/facilities/lasilla/instruments/feros/tools/DRS.html}}, respectively). The CAF\'E spectra were reduced using a pipeline developed by one of us (JMA). All of them were then homogeneously normalized and corrected from heliocentric velocity using our own IDL routines. The similarity between all the instruments considered -- in terms of resolving power --, the quality of the spectra in terms of S/N, and the homogeneous reduction methods allows us to minimize observational effects on the outcome of our spectroscopic study. In particular, we remark that, in spite of the different resolving power provided by the various instruments, for R$>$25\,000 this parameter becomes non-critical for the spectroscopic determination of stellar parameters and for the kind of spectroscopic variability study performed in this paper.  

Concerning the quality of the whole sample of spectra, most of the spectra (95\%) have a S/N(@4500~\AA) higher than 100, the median of the distribution is $\approx$150, and the maximum S/N achieved is $\approx$250. More detailed information about the S/N of the spectra used to derive the stellar parameters can be found in Table~\ref{tableStandards}. Finally, we note that many of the spectra with S/N(@4500~\AA) below 100 correspond to CAF\'E-BEANS. This survey, which aims at the detection and follow up of multiple systems among a selected sample of $\approx$100 Northern O-type stars \citep{Negueruela15} has a lower requirement in terms of S/N than the other two surveys. In addition, the CAF\'E spectrograph has a lower efficiency in the blue range than HERMES, FIES or FEROS. Therefore, in this paper, the spectra from CAF\'E-BEANS exclusively serve to increase the number of available epochs for the detection of spectroscopic variability in the stars studied.

\section{Methods}\label{section3}

\subsection{Quantitative spectroscopic analysis}\label{section31}

The stellar parameters of our sample of O-type standards for spectral classification were determined using the best S/N spectrum for each star. In this paper, we concentrate on those parameters that can be obtained from the quantitative spectroscopic analysis of the commonly designated {\em optical part} of the spectrum (i.e. 4\,000\,--\,7\,000~\AA). These refer to the projected rotational velocity (\vsini) and the amount of non-rotational broadening (aka. macroturbulence, \vmacro) affecting the line-profiles of each star (i.e. the {\em line-broadening parameters}), plus the {\em spectroscopic parameters}, namely the effective temperature (\Teff), gravity (\grav), wind-strength parameter ($Q$), helium abundance ($Y_{\rm He}$), microturbulence (\micro), and the exponent of the wind velocity-law ($\beta$). Other {\em fundamental parameters}, such as the radius ($R$), the luminosity ($L$), or the mass ($M$) require additional information about distances and extinction. Since we still lack accurate values of distances for most of the O stars in our sample, we postpone the determination of this second set of parameters until we have access to the improved information about parallaxes that will be released by the ESA-\textit{Gaia} mission \citep{Perryman01}. The abundance analysis will also be presented in a separated paper.

This section describes the main analysis strategy we followed. Briefly, we applied standard techniques incorporated to semi-automatized tools designed to perform quantitative spectroscopic analyses of large samples of O-type stars in an homogeneous, objective and relatively fast way.  This information is complemented (in Sect.~\ref{subsectionVar}) with some notes about the investigation of signatures of spectroscopic variability of those targets for which we already had available multi-epoch observations\footnote{By the time we were working on this paper, 67\% of the sample stars have at least 3 IACOB, OWN and/or CAF\'E-BEANS spectra. Another 23\% had available 1 or 2 spectra and we have more than 10 spectra for 20 stars. Since the three surveys are still active, these numbers have certainly increased since then.}.

\subsubsection{Line-broadening parameters}\label{subsectionBroad}

We first used the \textsc{iacob-broad} tool \citep{Simon14b} to determine \vsini\ and \vmacro\ for each star. For the sake of homogeneity, we mainly based the analysis on the \ion{O}{iii}\,$\lambda$5592 line, but had to rely on the \ion{Si}{iii}\,$\lambda$4552 or the \ion{N}{v}\,$\lambda$4603/20 lines in a few cases in which the \ion{O}{iii} line was weak \citep{Simon14,Markova14}. Being aware that the \ion{N}{v}\,$\lambda$4603/20 lines can be affected by the wind, they were only used if the absence of a strong wind was confirmed. 

\begin{figure}
\includegraphics[width=0.49\textwidth]{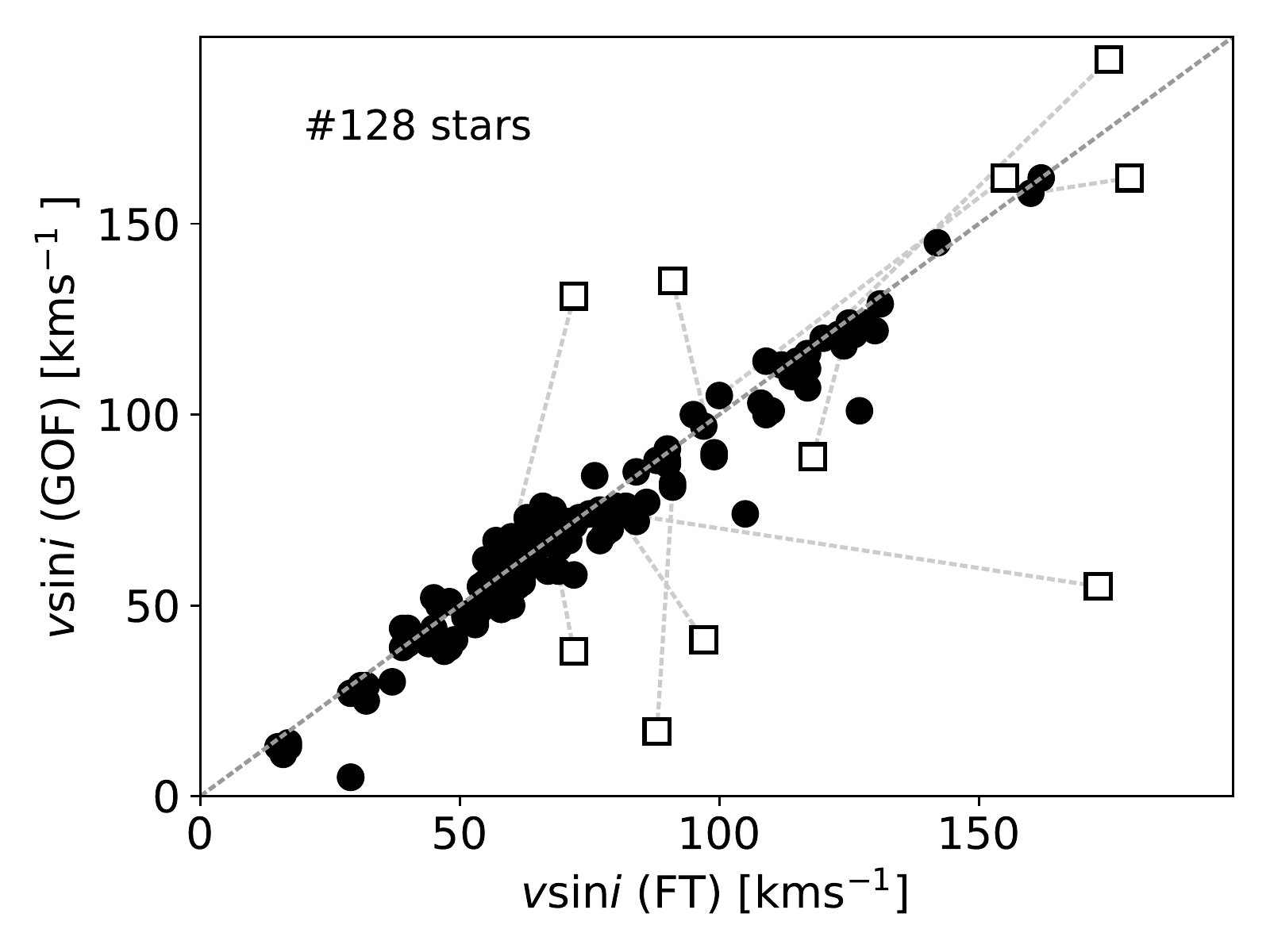}
\caption{Comparison of projected rotational velocities resulting from the FT and GOF analysis strategies incorporated to the {\sc iacob-broad} tool. Open squares indicate the resulting values for those cases in which the line-broadening analysis of the \ion{O}{iii}\,$\lambda$5592 line was considered not reliable. Dotted lines connect these results with those obtained from the analysis of the \ion{Si}{iii}\,$\lambda$4552 or \ion{N}{v}\,$\lambda$4603/20 lines.}
\label{vsin_FTGOF_comp}
\end{figure}

As described in \cite{Simon14b}, the combined Fourier transform (FT) + Goodness of fit (GOF) analysis strategy implemented in \textsc{iacob-broad} provides several solutions for the pair of line-broadening parameters. We used the comparison between \vsini(FT) and \vsini(GOF) as an assessment of the reliability of the results from line-broadening analysis. In addition, this comparison allowed us to identify problematic cases that required further consideration (i.e., those with discrepancies larger than 20 \kms).
This was, for example, the case for thirteen of the stars shown in Fig~\ref{vsin_FTGOF_comp}.
For ten of these targets (those marked with open squares) we found that the discrepancy between \vsini(FT) and \vsini(GOF) was caused by the combined effect of having a weak \ion{O}{iii} line and a low S/N spectrum (all these stars have either an early or a late spectral type). The situation improved in all of them when we considered one of the other available diagnostic lines indicated above. For illustrative purposes we connect with a dotted line the determinations based on the \ion{O}{iii} line and the alternative diagnostic line\footnote{The line used for each star is included in Table~\ref{tableStandards}.}. 
Regarding the remaining three targets with a difference between the two derived values of \vsini\ larger than 20~\kms\ (see Fig.~\ref{vsin_FTGOF_comp}) a closer inspection of the \textsc{iacob-broad} graphical output allowed us to identify clear (in one of them, HD~57236) and likely (in the other two, CPD~-59~2600 and HDE~298429) signatures of a broad-line secondary component affecting the \ion{O}{iii} line. In these cases, while the FT is not very much affected, the GOF solution tries to mimic the composite line-profile by increasing \vmacro, hence resulting in a lower \vsini.

Once the problematic cases were corrected or discarded, we ended up with a set of \vsini(FT) and \vsini(GOF) values that are in fairly good agreement, with $\sigma$($\Delta$\vsini)\,=\,5.3~\kms. The final values considered for the quantitative spectroscopic analysis with {\sc iacob-gbat} (see Sect.~\ref{subsectionGbat}) are quoted in Table~\ref{ParamStandards}. Basically, we kept the pair \vsini(GOF)\,--\,\vmacro(GOF) for all but the three problematic stars indicated above. For the latter, we used \vsini(FT)\ and the associated \vmacro(FT+GOF)\footnote{\vmacro\ obtained from the fit of the line assuming a fixed \vsini\ value derived from the FT.} values while also keeping in mind that we are likely (or clearly in one of the stars) dealing with a composite spectrum.

\subsubsection{Radial velocities}\label{subsectionvrad}

\begin{figure}
\includegraphics[width=0.49\textwidth]{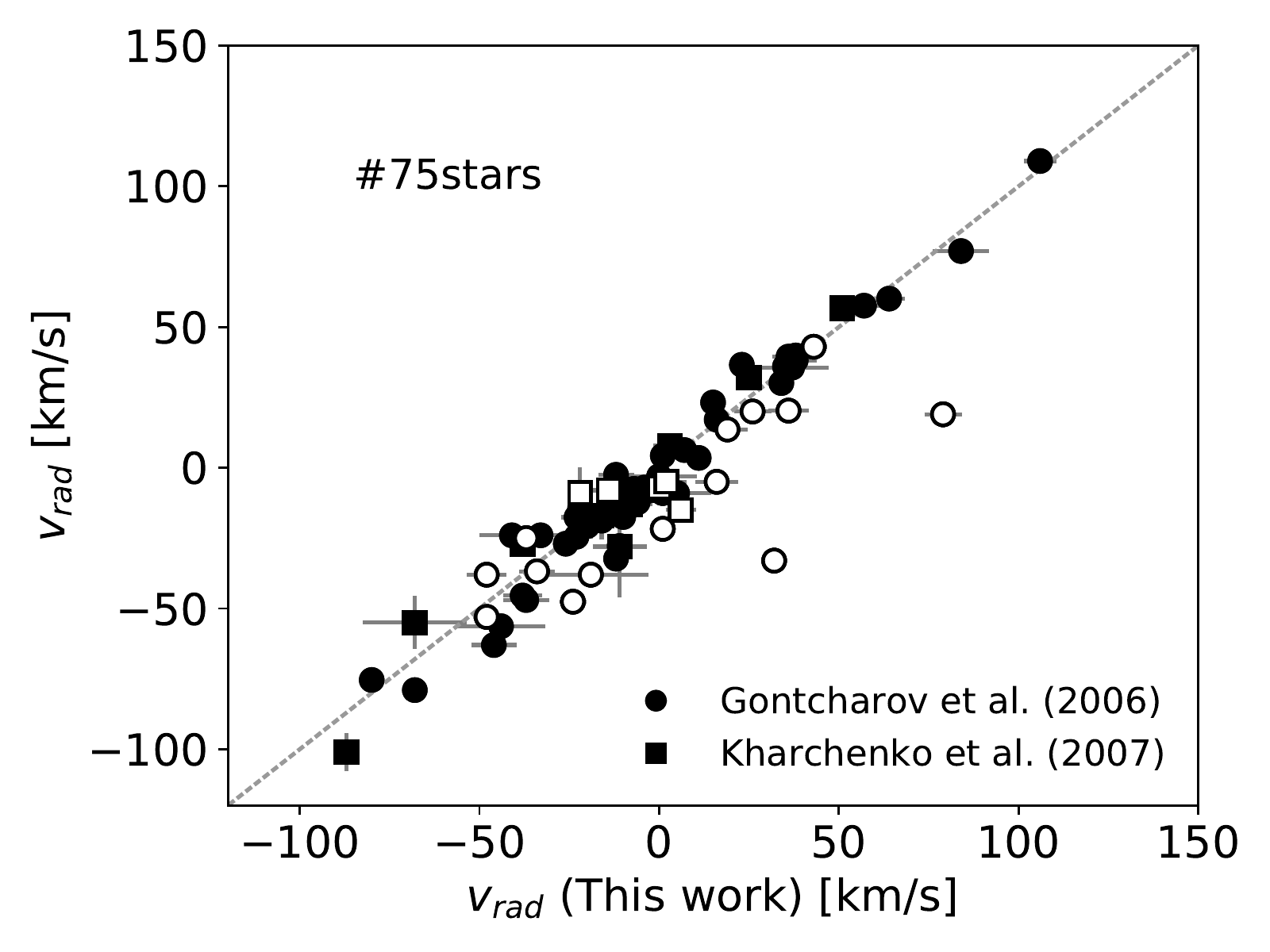}
\caption{Comparison of the \vrad\ values determined in this work with those available in the literature. Open points indicate stars for which we have detected clear or likely signatures of spectroscopic binarity (see Sect. \ref{subsectionVar}). We note that these studies do not provide values of \vrad\ for 53 of the stars in our sample. Lines show the uncertainty of the individual \vrad\ measurements.}
\label{vrad_SIMBAD}
\end{figure}

Rough initial estimations of the radial velocities (\vrad) associated with the spectra selected to perform the quantitative spectroscopic analysis were obtained from direct comparison of the observed location of the core of the \ion{He}{i}\,$\lambda$5875 line (identified ``by-eye'') and its corresponding laboratory wavelength. These values were used to correct the spectra from radial velocity before launching \textsc{iacob-gbat} (see Sect.~\ref{subsectionGbat}). Then, in a second iteration, the \fastwind\ synthetic spectrum of the best fitting model for each star (convolved with the corresponding \vsini, \vmacro, and $R$) was used to improve the radial velocity determination by means of a simple (iterative) cross-correlation technique. To this end, we considered the following initial set of lines:  \ion{He}{i}\,$\lambda$4026, \ion{He}{i}\,$\lambda$4387, \ion{He}{i}\,$\lambda$4471, \ion{He}{i}\,$\lambda$4713, \ion{He}{i}\,$\lambda$4922, \ion{He}{i}\,$\lambda$5015, \ion{He}{i}\,$\lambda$5047, \ion{He}{i}\,$\lambda$5875, \ion{He}{ii}\,$\lambda$4200, \ion{He}{ii}\,$\lambda$4541, \ion{He}{ii}\,$\lambda$4686, and \ion{He}{ii}\,$\lambda$5411. Each line was considered separately, and the associated spectral window selected automatically using the corresponding synthetic line. Those lines that were in emission or absent in the observed spectrum were eliminated from the very beginning on a star-by star base. Then we followed an iterative process by which those lines providing \vrad\ measurements 2 $\sigma$ larger than the mean of all lines were removed. Lastly, final values and associated uncertainties (compiled in Table~\ref{tableStandards}, already corrected from heliocentric velocity) were obtained from the mean and standard deviation of individual \vrad\ measurements acquired from the lines that survived.

We found that the total number of \ion{He}{i-ii} lines considered is sufficient to compensate for situations like, e.g., the absence of \ioni{He}{i} lines in early type stars or the exclusion of the \ion{He}{i}\,$\lambda$5875 and \ion{He}{ii}\,$\lambda$4686 lines when they appear in emission. Indeed, we found that, even in the most complex cases, at least 7 lines are available for the final computation of the radial velocity. We note that, following this strategy, we do not intend to provide the most accurate values of \vrad\ that can be extracted from these high-resolution spectra, but to end up with a radial velocity correction good enough for the purpose of the spectroscopic parameter determination. In particular, we consider that the typical uncertainty in \vrad\ (the standard deviation from using all lines) obtained for most of the stars is $\sim$5\,\kms, with a few critical cases reaching somewhat larger values ($\sim$10\,--\,25 \kms).

Figure~\ref{vrad_SIMBAD} compares our derived values with those included in the catalogs by \cite{Gontcharov06} and \cite{Kharchenko07} for 75 stars in common (see also last column in Table~\ref{tableStandards}). There is a general good agreement (the standard deviation of the differences is $\sim$\,10~\kms), with the highest discrepancies found in stars for which we have detected clear or likely signatures of spectroscopic binarity (see Sect. \ref{subsectionVar}).

\subsubsection{Spectroscopic parameters}\label{subsectionGbat}

\begin{table}[!t]
	\caption{Parameter space covered by the grid of \fastwind\ models at solar metallicity.} 
	\label{gridFAST}
        \centering
        \begin{threeparttable}
            \begin{tabular}{lll}
		\hline \hline
        \noalign{\smallskip}
		Parameter & Range  & Step size \\
		\hline
        \noalign{\smallskip}
		\Teff\                  [K]    &     22\,000\,--\,55\,000  &         1\,000  \\
		\grav\                  [dex]  &     2.6\,--\,4.4      &         0.1   \\
		\micro\                 [\kms] &     5\,--\,20         &         5     \\
		$Y_{\rm He}$ \tnote{a}         &     0.06, 0.10, 0.15, 0.20, 0.25, 0.30 & Irregular \\
		log~$Q$  \tnote{b}             &     -11.7, -11.9, -12.1, -12.3, -12.5  & Irregular \\
		                               &     -12.7, -13.0, -13.5, -14.0, -15.0 & \\
		$\beta$                        &     0.8\,--\,1.2  & 0.2 \\
		\hline
	\end{tabular}
		\begin{tablenotes}
		 \item[a] $Y_{\rm He}$ = $N$(He)/$N$(H) 
		 \item[b] $Q$\,=\,$\dot{M}$/($v_{\infty}R$)$^{1.5}$; $\dot{M}$ in $M_{\odot}~yr^{-1}$, $v_{\infty}$ in \kms, and $R$ in $R_{\odot}$ 
		\end{tablenotes}		
     \end{threeparttable}
\end{table}

The continuously increasing amount of high-quality spectroscopic observations of massive stars provided by different surveys during the last decade (see references in Sect.~\ref{section1}) made clear from the very beginning the necessity to develop semi-automatized techniques which allow for the extraction of information about stellar parameters and abundances from large spectroscopic datasets of OB stars in a reasonable computational time. Some notes on various of the techniques proposed can be found in \cite{Mokiem06, Lefever07, Urbaneja08, Simon11a, Castro12, Irrgang14}. 

\textsc{iacob-gbat} \citep{Simon11a} is a grid-based automatic tool for the quantitative spectroscopic analysis of O-stars developed in the framework of the IACOB project. This tool, that is basically an automation of traditional ``by-eye'' analysis techniques \citep[see, e.g.,][]{Herrero92,Herrero02, Repolust04}, includes an extensive grid of \fastwind\ \citep{Santolaya97,Puls05,Rivero12} models and a variety of IDL programs to handle the observations, perform the analysis by means of a versatile implementation of a $\chi^2$ algorithm, and visualize/evaluate the results. It has been developed under the philosophy of being user-friendly, portable, and fast.

While it has already been used for the analysis of $\approx$100 O dwarfs in the context of the VLT-FLAMES Tarantula Survey \citep{SabinSan14, SabinSan17}, as well as in other individual studies of specific targets \citep{GarciaRojas14,Castro15,Negueruela15b}, this is the first publication in which the tool is extensively applied to a large sample of high-resolution spectra of Galactic O-type stars.

We refer the reader to \cite{Simon11a} and \cite{SabinSan14} for a thorough description of the tool and some of the basic ideas about the analysis strategy and the interpretation of results. In this section, we extend further those explanations, including some details concerning the application of \textsc{iacob-gbat} to the analysis of IACOB and OWN data. Finally, we include in Appendices~\ref{ErrorAppend} and \ref{vinfAppend} a few technicalities and updates of the tool not described in previous publications, including the effect of the assumed terminal velocities ($v_{\infty}$) on the outcome of the {\sc iacob-gbat} analysis.

We used the grid of \fastwind\,(v10.1) models for solar metallicity described in \cite{Simon11a}, updated with an extension to somewhat cooler temperatures. During the analysis of the IACOB+OWN sample of standard stars, we found that the lower limit in effective temperature previously considered for the grid (25\,000~K) was not enough to properly constrain this parameter in late-O~supergiants; we hence extended the grid down to 22\,000~K. The final ranges and step size considered for each of the six free parameters of the grid (\Teff, \grav, $Y_{\rm He}$, \micro, $\beta$, log~$Q$) are summarized in Table~\ref{gridFAST}. 

\begin{table}[!t]
	\caption{Diagnostic lines used in the \textsc{iacob-gbat} spectroscopic analysis of our sample of IACOB+OWN spectra} 
	\label{LinesDiag}
	\centering
	\begin{tabular}{cccc}
		\hline \hline
        \noalign{\smallskip}
		\ioni{H}{} & \ioni{He}{i}  & \ioni{He}{ii}  & \ioni{He}{i} + \ioni{He}{ii}\\
		\hline
        \noalign{\smallskip}
		H$\alpha$ &  $\lambda$4387 &  $\lambda$4200 & $\lambda$4026 \\
		H$\beta$  &  $\lambda$4471 &  $\lambda$4541 & $\lambda$6678 + $\lambda$6683\\
		H$\gamma$ &  $\lambda$4713 &  $\lambda$4686 & \\
		H$\delta$ &  $\lambda$4922 &  $\lambda$5411 & \\
		          &  $\lambda$5875 &   & \\
		\hline
	\end{tabular}
\end{table}

As described in \cite{Simon11a}, this grid was created to cover the full range of stellar parameters considered in the case of normal O-type stars with (evolutionary) masses up to $\approx$100~\msol. However, in order to optimize the computational time, the user has the possibility to select a smaller range in each of the free parameters that will be inspected by the tool. 

In the analysis presented in this paper, we initially restricted the range of values considered for \Teff\ and \grav\ using the spectral type and luminosity class of each star and the calibrations by \cite{Martins05} (hereafter MSH05). We basically used the values proposed in Martins' observational calibration and extended the corresponding ranges in these two parameters by $\pm$5\,000~K and $\pm$0.4~dex, respectively. In those cases in which the ranges considered in \Teff\ and/or \grav\ did not allow us to fully sample the lower envelope of the reduced $\chi^2$ distribution up to $\chi^2$+4 \citep[see][]{SabinSan14}, we launched again the third and fourth steps of \textsc{iacob-gbat} extending a bit further the corresponding ranges. 

Concerning the other parameters (\micro, $Y_{\rm He}$, log~$Q$, and $\beta$) we used the full range available. In contrast to some of the previous studies, we did not initially fix the microturbulence \citep[as, e.g., in][]{Repolust04,Markova14} or the $\beta$ parameter \citep[as, e.g., in][]{SabinSan14}.

Table~\ref{LinesDiag} summarizes the complete list of \ion{H}{i} and \ion{He}{i-ii} diagnostic lines considered for the spectroscopic analysis presented in this paper. The lines selected are those traditionally used in the spectroscopic analysis of O-type stars \citep[see, e.g.,][]{Herrero92,Repolust04} plus two more due to our wider wavelength coverage, \ion{He}{i}\,$\lambda$5875 and \ion{He}{ii}\,$\lambda$5411. For the sake of homogeneity, we always used the same set of lines and we gave the same initial weight to all of them. Note, however, that \textsc{iacob-gbat} follows an automatic iterative strategy to provide weights for the diagnostic lines in the computation of the global reduced $\chi^2$ distribution (step 4). This means that not necessarily all diagnostic lines finally have the same weight in the selection of the best fitting model and the final estimation of central values and uncertainties. This iterative process, that is explained in detail in Appendix \ref{ErrorAppend}, accounts for the quality (in terms of S/N) of the various diagnostic lines, and detects those cases where the given lines are not properly fitted by the final global solution, all this in an automatic and objective manner.

Once the range of parameters have been determined and the diagnostic lines selected, we launch \textsc{iacob-gbat} for each spectrum. We use the previously derived values of \vsini\ , \vmacro\ and \vrad\ as input parameters. 

As a first step \textsc{iacob-gbat} will prompt the user to select the spectral window around each of the lines to be considered in the $\chi^2$ computation. During this pre-processing of the spectra the diagnostic lines were locally renormalized and nebular lines, blends and cosmic rays eliminated whenever necessary. Then, \textsc{iacob-gbat} proceeds to the computation of the global reduced $\chi^2$ distributions and the final results (see Appendix \ref{ErrorAppend}). 

As described in \cite{SabinSan14}, \textsc{iacob-gbat} provides as final output the best fitting model within the considered grid, along with information on central values (or upper/lower limits) for each of the 6 free parameters indicated in Table~\ref{gridFAST} and their associated uncertainties. In addition, the tool generates several summary figures allowing us to perform a visual assessment of the results and the quality of the best fitting model solution (see Sect.~\ref{section41}).

\subsection{Spectroscopic variability}\label{subsectionVar}

The common strategy followed in the literature when performing the spectroscopic analysis of Galactic O-type stars is to consider single-epoch observations. However, observational evidence of the existence of spectroscopic features associated with multiple components (due to binary stars or even multiple systems) and other variability phenomena in an important percentage of massive stars has continuously increased in the last decades (e.g, stellar oscillations, wind-strength variability, magnetic fields with variable geometries, etc).

To give some numbers, the most recent surveys specifically devoted to detect high mass binaries \citep{Mason09, Sana11,Sana12,Sana13, Sota14, Barba14,Barba17} indicate that the percentage of stars found in spectroscopic binaries (or multiple systems) among O-type stars is 35-60\% and that the percentage of binary/multiple systems may increase up to 90\% or more if one includes statistical corrections for completeness and visual companions (which may yield composite spectra and/or induce variations in the observed spectra when obtaining multiple epochs under variable seeing conditions\footnote{Also important in this regard is the serendipitous detection outside these surveys of additional companions in targets already identified as double line spectroscopic binaries \citep[e.g.,][]{Simon14, Lorenzo10}.}). 
In addition, the last comprehensive study of the incidence of spectroscopic variability among O-type stars \citep{Fullerton96} detected line-profile variations for 77\% of the considered sample. 

All this has warned us about the necessity to incorporate {\em -- as an important ingredient --} the information provided by multi-epoch observations to the stellar parameter determination process in order to establish the reliability of the outcome resulting from the quantitative spectroscopic analysis of a single snap-shot observation.  Some work in this direction can be found in \cite{Markova05}

In this paper, we used all the spectra available in the IACOB, OWN and CAF\'E-BEANS surveys to detect possible signatures of spectroscopic variability in our sample of analyzed stars. We remark that we did not aim to perform an in-depth characterization of spectral variability in the sample studied. This is definitely outside the scope of this paper; in particular, because the number of multi-epoch spectra available for most of the sample is not enough for these purpose. Instead, our main motivation was to identify variable features in the spectrum of a given star that could affect the reliability of the stellar parameters provided in Table~\ref{ParamStandards}, which resulted from the \textsc{iacob-gbat} spectroscopic analysis of the best S/N spectrum per star (see Sect.~\ref{subsectionGbat}). As an aside, this study also provides complementary information to that available in the literature about binarity/multiplicity and/or other spectroscopic variability phenomena detected in this sample of 128 Galactic O-type stars.

\begin{figure}
\includegraphics[width=0.49\textwidth]{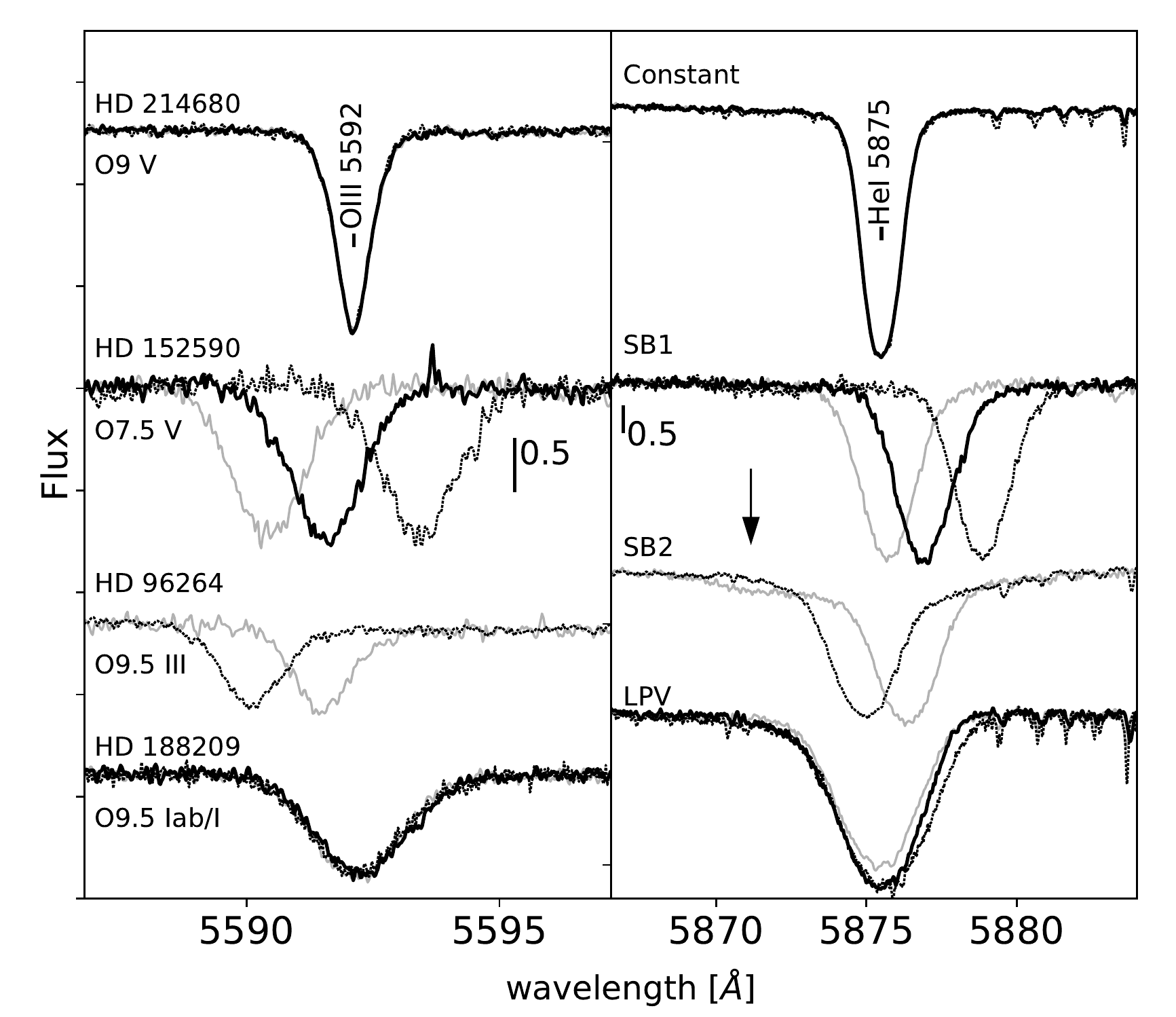}
\caption{A few examples of stars flagged as {\em C}, {\em SB1}, {\em SB2} and {\em LPV} (see Sect.~\ref{subsectionVar} for explanations). The spectra (plotted in different tones of grey) are corrected for heliocentric velocity, but not for telluric lines. The second component of the SB2 (marked with an arrow) is very dim and hence likely not detected at low-resolution.}
\label{BinTypes}
\end{figure}

\begin{figure*}
\centering
\includegraphics[width=0.95\textwidth]{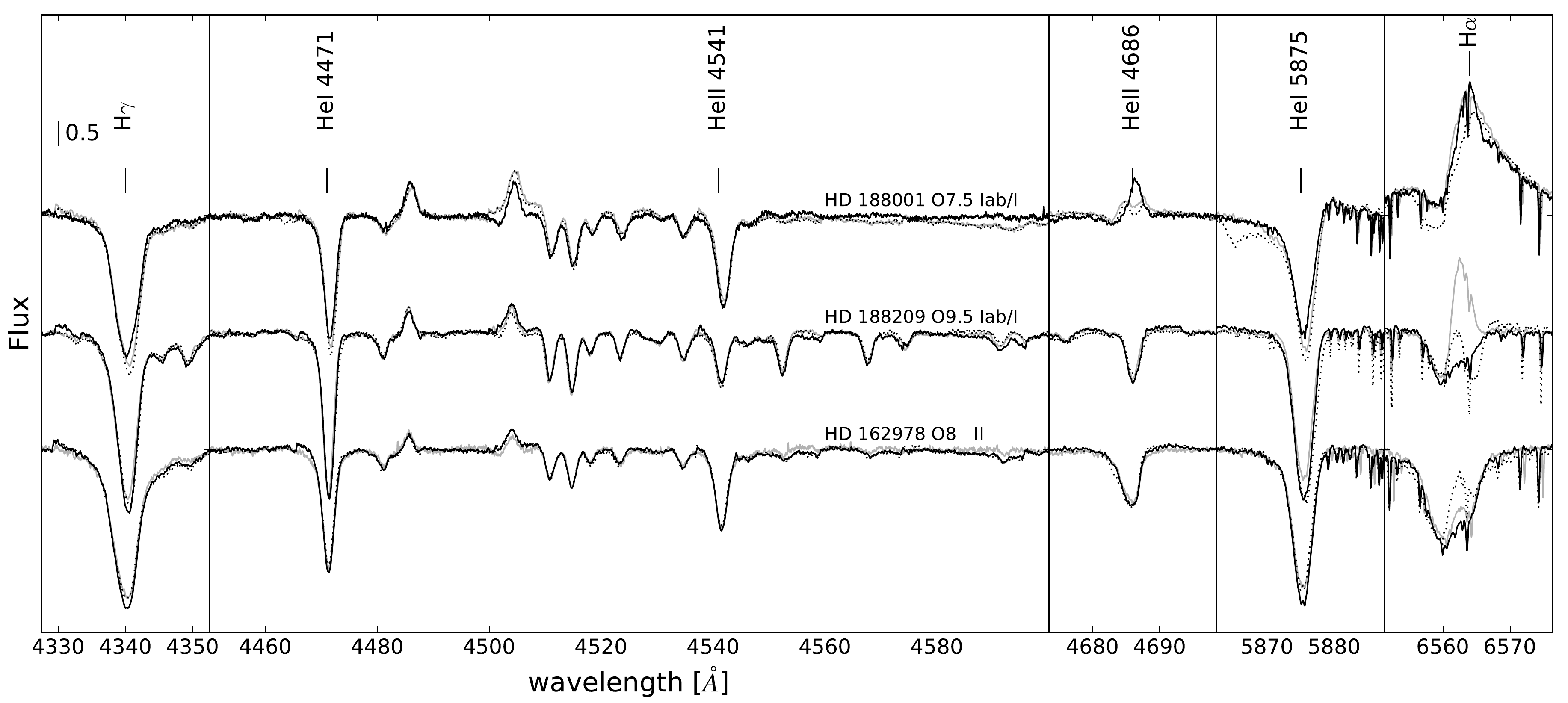}
\caption{Three illustrative examples of stars showing variability in the wind diagnostic lines. The spectra (plotted in different tones of grey) are corrected for heliocentric velocity, but not for telluric lines.
}
\label{VarTypes}
\end{figure*}

In a first iteration, we considered 5 diagnostic lines (\OIII, \ion{He}{i}\,$\lambda$4387, H$\alpha$, \ion{He}{ii}\,$\lambda$4686, and \ion{He}{i}\,$\lambda$5875) and used the interstellar \ion{Na}{i}\,$\lambda\lambda$5890, 5895 lines as a sanity check for instrumental effects and the accuracy of the heliocentric velocity correction\footnote{Internal consistency checks performed by the IACOB+OWN+CAF\'E-BEANS collaboration have shown that the dispersion in velocity resulting from the analysis of the interstellar \ion{Na}{i} lines when combining FIES, FEROS, HERMES, and CAF\'E spectra is below 0.5 \kms.}. By overplotting all the available spectra per target -- zooming into the spectral regions where the diagnostic lines are located -- we could qualitatively assign one or several of the following flags to each star: 
\begin{itemize}
	\item presumably constant ({\em C}) star, if we have more than 2 spectra, and we do not visually detect a clear shift in radial velocity (or  a variation in the shape of the line profile) in any of the diagnostic lines;
	\item spectroscopic binary, including cases with one ({\em SB1}) and two ({\em SB2}) components detected. We note that in all detected SB2 stars in our sample the secondary component is very dim. This is expected as the standards were defined avoiding composite spectra, but using observations with a lower resolution than our study;
	\item stars showing clear variability in any of the two main wind-diagnostic lines -- H$\alpha$ and/or \ion{He}{ii}\,$\lambda$4686 --, separating the cases when the lines are in emission ({\em WVe}) or in absorption ({\em WVa});
	\item line-profile variability ({\em LPV}), when small variations of the line-profiles -- which cannot be clearly assigned to wind variability or binarity effects -- are detectable by eye from the inspection of a sufficiently large number of spectra;
\end{itemize}

Some illustrative examples of each of these cases are presented in Figs.~\ref{BinTypes} and \ref{VarTypes}. In addition, we labeled as {\em MD (More Data needed)} those stars for which we were not able to clearly assign any of the above~mentioned flags. The latter situation was mainly found in stars for which we have available only 1 or 2 spectra.

In a second iteration, once we had obtained the best fitting model resulting from the {\sc iacob-gbat} analysis, we refined the classification of those stars flagged as {\em C}, {\em LPV}, and {\em SB1} using a more objective (quantitative) criterion. To this end, we determined -- following the method described in Sect.~\ref{subsectionvrad} -- the radial velocity of all available spectra for each star and computed the associated dispersion -- $\sigma$(\vrad) -- and the peak-to-peak amplitude -- $\Delta$\vrad\ -- of all measurements. Based on this information, and taking into account the outcome of the qualitative assessment of variability previously performed, we decide to flag those stars with $\sigma$(\vrad) in the range 2.5\,--\,10~\kms, and $\Delta$\vrad\ in the range  5\,--\,20~\kms\ as {\em LPV}, while those cases with lower (higher) dispersion were marked as {\em C} ({\em SB1}), respectively. 

The outcome from this investigation of line-profile variability has been added to Table~\ref{ParamStandards}. A global summary of the number of stars labeled with the various variability flags considered is presented in Table~\ref{Table_GP}.

We also had access to the information about variability in Galactic O-type stars compiled by the OWN project in the last years (see last column in Table~\ref{ParamStandards}). This independent study (Barb\'a et al., in prep.) benefits from a much larger number of epochs for most of the (Southern) stars, including spectra gathered with high-resolution spectrographs at La Silla Observatory, Las Campanas Observatory, Cerro Tololo Inter-American Observatory (Chile), and the Complejo Astron\'omico El Leoncito (Argentina) \citep[see][]{Gamen07}. In this case, the following flags were used: {\em SB1}, {\em SB2}, {\em VAR}, {\em VAR?}, and {\em C}, defined in their work under similar criteria \citep{Barba10}.

\begin{threeparttable}[!t]
	\caption{Summary of the results in the \textsc{iacob-gbat} spectroscopic analysis of our sample of IACOB+OWN spectra} 
	\label{Table_GP}
	\centering
	\begin{tabular}{ccccccc}
		\hline \hline
		\noalign{\smallskip}
		Variability flag& C & LPV & SB1 & SB2 & WV  & MD  \\ 
		\hline
		\noalign{\smallskip}
		\# stars & 29 & 38 & 28 & 7 & 40 & 29 \\  	
        \noalign{\smallskip}
		\hline \hline
		\noalign{\smallskip}
		Quality flag\tnote{a} &  Q1 & Q2  & Q3 & Q4 & & \\
		\hline
		\noalign{\smallskip}
		\# stars & 75 & 16  & 30 & 7 & &  \\		 
		\hline
		\noalign{\smallskip}
		\multicolumn{4}{l}{128 stars in total}
	\end{tabular}
		\begin{tablenotes}
		 \item[a] As explained in Sect.~\ref{section41}, related to the fit-quality.
		\end{tablenotes}		
     \end{threeparttable}    

\section{Results}\label{section4}

The results of the {\sc iacob-broad} and {\sc iacob-gbat} analysis of the 128 O-type stars are summarized in Table~\ref{ParamStandards}. The complete list of stars is divided in four different groups. The first three are stars for which we consider that the outcome of {\sc iacob-gbat} is reliable (in general terms, see, however, next section). In each of these three groups (which correspond to luminosity classes V+IV, III, and II+I, respectively), stars are sorted by spectral type. The fourth group includes the stars for which a reliable determination of the stellar parameters could not be achieved, due to the absence of \ion{He}{i} lines or presence of SB2 signs, and the three stars with no available high resolution spectrum. In addition to the line-broadening and spectroscopic parameters and their associated uncertainties (whenever available), we quote the spectral classification of the stars \citep[following][]{Maiz16}, the flag indicating the quality of the final fit resulting from the quantitative spectroscopic analysis (see Sect.~\ref{section41}), and the type of variability detected from inspection of the multi-epoch spectra available in the IACOB, OWN, and CAF\'E-BEANS surveys (see Sect.~\ref{subsectionVar}). 

We note that all the information included in Table~\ref{ParamStandards} is directly obtained from the analysis or inspection of the available spectra except for \gravt, the gravity corrected from centrifugal acceleration. In this case we used the calibrations presented in MSH05, combined with the spectral classification associated with each star to obtain information about the stellar radii, needed to compute the centrifugal correction \citep{Herrero92, Repolust04}.

Table~\ref{ParamStandards} is also complemented with Figs. \ref{All0} to \ref{All115}, where we show -- for each of the spectra analyzed-- the overall agreement between the best fitting \fastwind\ model and the observed spectrum for all diagnostic lines considered in the spectroscopic analysis.

\subsection{General notes on the outcome of the \textsc{iacob-gbat} analysis}\label{section41}

{\sc iacob-gbat} is an automatized tool that has the advantage to provide a more complete and objective exploration of the parameter space (compared to more traditional {\em ``by-eye''} techniques). However, the outcome of the {\sc iacob-gbat} analysis must not be taken blindly as a valid result without final supervision by the user, as should be for any automatized method. In our case, this includes (1) the inspection of the $\chi^2$ distributions for each of the parameters considered and, more importantly, (2) a final visual assessment of the overall agreement between the proposed best fitting model and the observed spectrum. On the one hand, the former allows us the detection, e.g., cases in which only upper or lower limits can be obtained, or situations in which a certain parameter cannot be constrained \cite[see examples in][]{SabinSan14}; on the other hand, the visual inspection is a mandatory step to identify cases in which the resulting values for certain (or the whole set of) parameters are not reliable due to, e.g., limitations of the grid of \fastwind\ models used (for example, the fact that we are using 1D unclumpled models) or the misidentification of a composite spectrum as if it was associated with a single star. 

After a careful inspection of all the analyzed spectra, we decided to summarize the outcome of the \textsc{iacob-gbat} analysis using the following quality flags: (Q1) acceptable fit with no major remarks, (Q2) \ion{He}{ii}\,$\lambda$4686 appearing in the observed spectrum as an inverse P-Cygni profile or showing a double peak, (Q3) H$\alpha$ and \ion{He}{ii}\,$\lambda$4686 not fitting at the same time, (Q4) no \ion{He}{i} available to properly constrain the effective temperature of the star. Although the Q2\,--\,Q4 flags are not mutually exclusive, we remark that only one of them was assigned to each individual star, giving priority to Q4 over the other flags and to Q2 over Q3. A summary of the number of stars cataloged within any of these categories is presented in Table~\ref{Table_GP}.  

\begin{figure}
\includegraphics[width=0.49\textwidth]{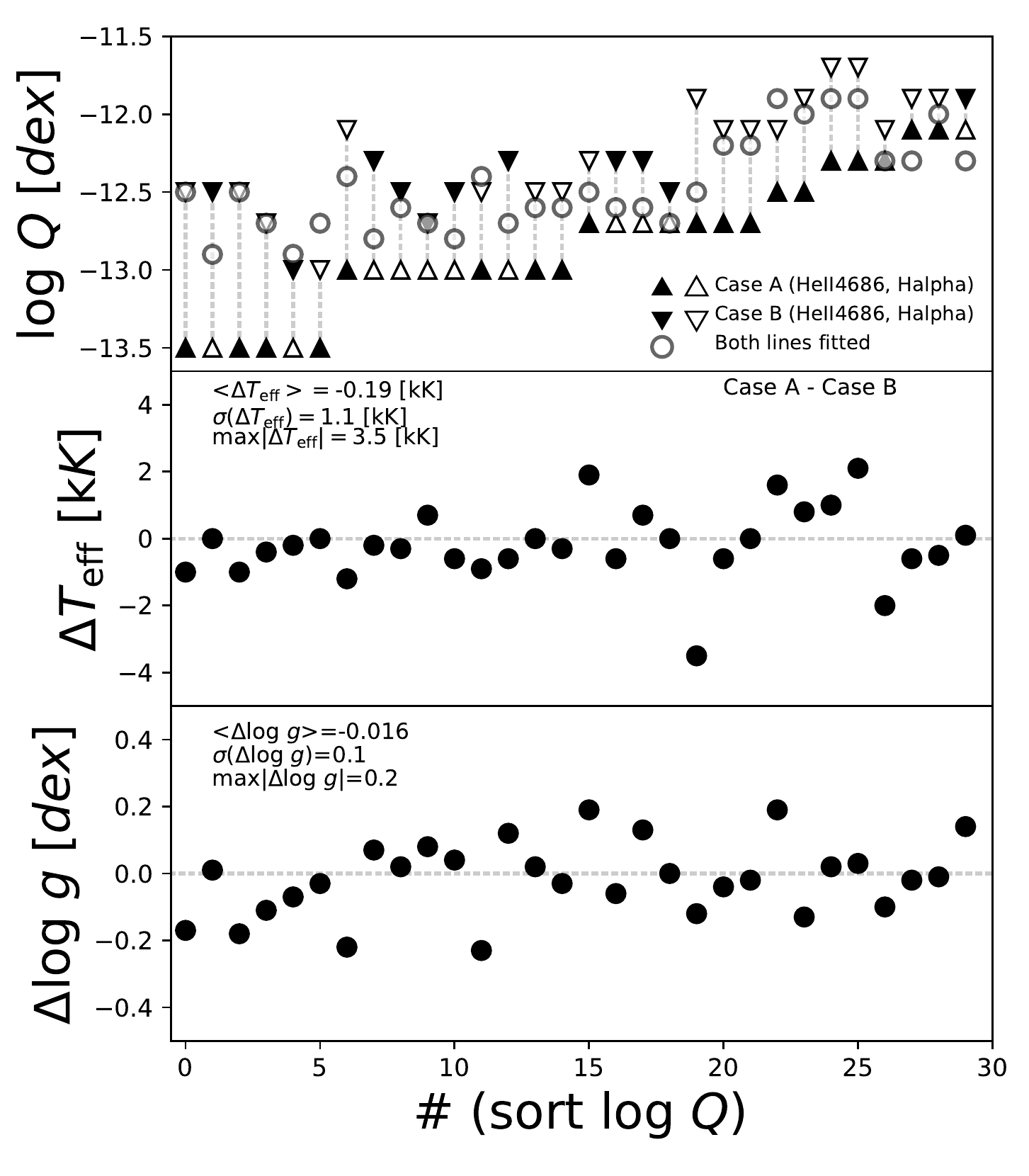}
\caption{Difference in the resulting \Teff\ (middle panel) and \grav\ (bottom panel) when fixing log~$Q$ to those values needed to fit the \ion{He}{ii}\,$\lambda$4686 or H$\alpha$ lines, in the sample of stars labeled as Q3. In the top panel the weaker (stronger) wind is represented by a triangle with the peak up (down). Filled triangles are log~$Q$ obtained from \ion{He}{ii}\,$\lambda$4686, and open triangles with H$\alpha$. Open circles are the values obtained when fitting both lines at the same time automatically. Stars are sorted by increasing  minimum log~$Q$ in the horizontal axis. Standard deviation, maximum and mean of the differences also included in the middle and bottom panels.	
}
\label{CompForcQ}
\end{figure}

Results presented in Table~\ref{ParamStandards} must be taken with care for stars not flagged as Q1 (good fit). In particular, we note that this table does not include the resulting parameters of the \textsc{iacob-gbat} analysis for those stars labeled as Q4. These stars are 7 early O-type stars and, as already mentioned elsewhere \cite[e.g.,][]{Rivero12}, these objects require the use of nitrogen lines to obtain a more reliable effective temperature (and hence the complete set of parameters). Regarding the reliability of the parameters provided in Table~\ref{ParamStandards} for stars flagged as Q2 or Q3, we refer the reader to notes below, and results presented in Table~\ref{ParamStandards3} and Fig.~\ref{CompForcQ}.

\begin{figure}
\includegraphics[width=0.49\textwidth]{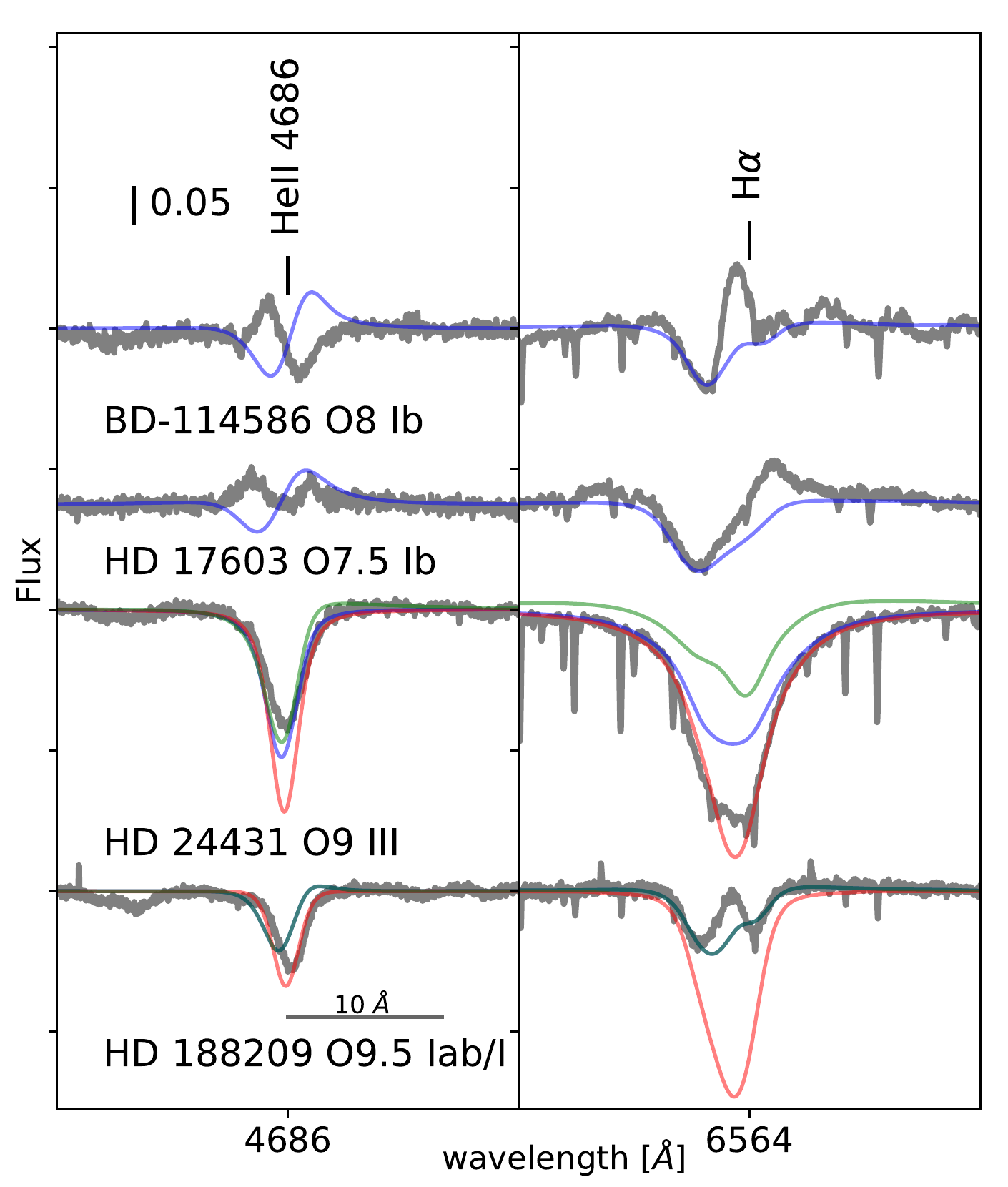}
\caption{Four illustrative examples of stars in which a good fit could not be achieved. Solid blue lines correspond to the synthetic spectra of the best fitting \fastwind\ model resulting from the {\sc iacob-gbat} analysis. For HD\,24431 and HD\,188209 (stars labeled with the Q3 quality flag) we also include the synthetic spectra of two models with different values of the $Q$ parameter (see text for details) where green(red) corresponds to higher (lower) $Q$, respectively. For HD\,188209 green and blue lines are overlapped.}
\label{TipQual}
\end{figure}

Some examples of stars cataloged as Q2 and Q3 are included in Fig.~\ref{TipQual} (see also Figs.~\ref{ExamplesQ2} to \ref{ExamplesQ3_4}). For each star, we show the region of the observed spectrum where the \ion{He}{ii}\,$\lambda$4686 and H$\alpha$ lines are located along with some \fastwind\ synthetic lines used for reference. BD~$-$11\,4586 and HD\,17603 are two illustrative cases in which \ion{He}{ii}\,$\lambda$4686 appears as an inverse P-Cygni profile or a double peak, respectively (i.e. they have been flagged as Q2). In this case, we overplot the synthetic profiles corresponding to the best fitting model using both profiles as valid diagnostic lines. These are two clear situations in which \ion{He}{ii}\,$\lambda$4686 should not be used as diagnostic line for defining the luminosity class and/or determining the wind-strength Q-parameter (i.e. the mass loss rate) of the star. Interestingly, and despite having been selected as standards for spectral classification, we have identified 16 stars (most of them having luminosity class I or II) where one of these two spectroscopic features affecting the \ion{He}{ii}\,$\lambda$4686 are present. We note, however, that this is likely due to the fact that spectral classification is normally performed using low resolution ($R\sim$2500) spectra, where these subtle effects affecting the line profiles are not easily detected (see, e.g., Fig.~\ref{HeII4686+R2500}). 


The most likely explanation of the observed behavior of the \ion{He}{ii}\,$\lambda$4686 in the stars flagged as Q2 (see complete list in Fig.~\ref{ExamplesQ2}) is the presence of a stellar disk or other type of emitting material surrounding the star. High rotation, the coupling of the magnetic field and the stellar wind \citep[e.g.,][]{Breysacher00,ud-Doula02,Walborn2002,Martins15,Castro17}, and binary interaction \citep[e.g.][]{Teodoro16} can cause this type of large scale circumstellar structures.

HD~24431 and HD~188209 illustrate the situation found in those (30) stars we have flagged as Q3. There is no model in our grid of \fastwind\ models producing a simultaneous fit to H${\alpha}$ and \ion{He}{ii}\,$\lambda$4686. In these cases {\sc iacob-gbat} tries to find a compromise between all the diagnostic lines that provide information about the stellar wind (mainly H${\alpha}$ and \ion{He}{ii}\,$\lambda$4686, but also H$_{\beta}$ and \ion{He}{i}\,$\lambda$5875), and ends up (normally) with an intermediate value of the wind-strength Q-parameter (blue lines in Fig.~\ref{TipQual}). Those are the values included in Table~\ref{ParamStandards}. 

There are several effects that could explain this result, all of them related to limitations in the modeling strategy adopted in this work (which is based on spherically symmetric unclumped wind models). 
In Sect.~\ref{logQsect}, we further discuss this point but before we are able to benefit from the versatility of {\sc iacob-gbat} to perform an investigation of the effect of fixing log~$Q$ to the two extreme values required to fit H${\alpha}$ and \ion{He}{ii}\,$\lambda$4686 in the rest of spectroscopic parameters. The main results are presented in Table~\ref{ParamStandards3} and, graphically, in Fig.~\ref{CompForcQ}. We mainly concentrate on the effect on the derived effective temperature and gravity.
The main results of this formal exercise can be summarized as follows:
\begin{itemize}
 \item  At first glance, there is not a clear systematic trend in which of the two wind diagnostic lines requires a higher value of log~$Q$ to be fitted individually (see, however, Sect.~\ref{clumping}). 
 \item The difference in the extreme log~$Q$ values can be up to 1 dex in some cases 
 \item The standard deviation of the difference in the derived \Teff\ and \grav\ when fixing log~$Q$ to one of the two extreme values is on the order of the uncertainties associated with these quantities; however, more critical situations, with differences up to 3500 K and 0.2 dex in \Teff\ and \grav\, respectively, are also found as specific cases.
\end{itemize}


Globally, even though a more detailed modeling of individual cases -- with a simultaneous fit of H${\alpha}$ and \ion{He}{ii}\,$\lambda$4686 -- will help to better constrain the mass loss rates associated with these stars, we can conclude that we do not expect an important effect on the derived effective temperatures and gravities in most cases.

As a final remark, we warn that labeling a given result from {\sc iacob-gbat} as Q1 does not necessarily imply that the star is single. We have found 7 cases where our multi-epoch study has allowed us to identify the star as SB2 (see Table~\ref{ParamStandards}), but the analyzed single snapshot spectrum\footnote{We recall that our only criterion to select the spectrum to be analyzed with {\sc iacob-gbat} is having the largest S/N among the available spectra for a given target.} does not show any obvious signature of dealing with a composite spectrum, and the final outcome from {\sc iacob-gbat} (in terms of quality of the global fit) is almost perfect. This reinforces our suggestion to incorporate -- {\em as an important ingredient} -- the information provided by multi-epoch observations to the stellar parameter determination process (see Sect.~\ref{subsectionVar}).

\subsection{General notes on the outcome of the study of spectroscopic variability}

Among the list of 128 O-type standards for spectral classification investigated in this work we identified at least 35 stars with clear or highly likely signatures of being part of binary system using the multi-epoch spectroscopic observations described in Sect.~\ref{section2}. As summarized in Table~\ref{tablesurveys}, we found 7 double lined spectroscopic binaries\footnote{Two of these SB2 stars were not cataloged as such by the GOSC: HD\,57236 and HDE\,229196.}, and 28 stars showing a variability in radial velocity that is unlikely produced by stellar oscillations\footnote{As obtained for the quantitative criteria applied in Sect.~\ref{subsectionVar}.}, and hence were labeled as SB1. All these stars have been identified with open symbols in the various figures presented along this paper. The remaining 65 objects for which we had more than 2 spectra were separated in two main groups ($C$ and $LPV$, comprising 29 and 38 stars, respectively), depending on the degree of variability of the line profiles (see Sect.~\ref{subsectionVar}). While we initially consider these 67 stars as likely single, investigations including a larger number of multi-epoch observations may modify this classification, highlighting the binary status of some of them.

We also found 40 stars with some degree of variability in at least one of the main wind diagnostic lines\footnote{We remind again that we were not able to investigate variability in 30 of the stars in our sample.}. Most of the targets with detected variability in H$\alpha$ and/or \ion{He}{ii}\,$\lambda$4686 correspond to luminosity classes I and II (all spectral types) with a few of them found among early type stars with luminosity classes III, IV, and V. There are two main effects that may impact the investigation performed in this work. The first one refers to the possible impact that variability of the \ion{He}{ii}\,$\lambda$4686 line could have on the identification of the luminosity class of the star, which in O-type stars mainly depends on the strength of this line in emission/absorption \citep[see, e.g., Table 5 in][]{Sota11}. The second one is the impact that the variability of H$\alpha$ and \ion{He}{ii}\,$\lambda$4686 may have on the stellar and wind parameters associated with a given star. 

Concerning the impact on the assigned luminosity class, we have found 7 targets with clearly detected variability of the \ion{He}{ii}\,$\lambda$4686 line (Fig.~\ref{HeII4686+R2500}). This variability is almost negligible in four of them when the spectra are degraded to R=2500. For the rest, we have found that the detected variability is not expected to modify the assigned luminosity class when following the spectral classification criteria summarized in \cite{Sota11}.

As an indication of the potential effect of the variability of H$\alpha$ and \ion{He}{ii}\,$\lambda$4686 on the derived parameters, we refer the reader to the results presented in \cite{Markova05} and Fig.~\ref{CompForcQ} in this paper. In particular, Markova found that the observed variations of H${\alpha}$ typically seen in O supergiants imply changes of $\pm$4\% with respect to the mean value of $\dot{M}$ for stars with stronger winds, and of $\pm$16\% for stars with weaker winds. On the other hand, Fig.~\ref{CompForcQ} shows that the effect on \Teff\ and \grav\ is not expected to be much larger than the typical uncertainties associated with the determination of these parameters. In any case, one should take into account potential variability in the determined stellar and wind parameters from different single snapshot spectra when comparing results from different studies in the literature (see, e.g., Sect.~\ref{subsectionComp} below).

\subsection{Comparison with previous results}\label{subsectionComp}

Many of the stars in our sample have been already investigated elsewhere. We present in this section a comparison of our results with those found in the literature with two main purposes. On the one hand, to validate the reliability of our automation of an analysis strategy that has been traditionally performed {\em ``by-eye''}. On the other hand, to identify possible systematic effects resulting from the use of different stellar atmosphere codes, analysis techniques, or specific {\em single snap-shot} observations.

We concentrate on four parameters (\vsini, \vmacro, \Teff, and \grav) and four of the latest papers performing quantitative spectroscopic analysis of a relatively large sample of Galactic O-type stars.  Specifically, we have found 12, 17, 46, and 36 stars in common with \cite{Repolust04}, \cite{Markova14}, \cite{Simon14b}, and \cite{Martins15}, respectively.


\begin{figure}[!t]
\includegraphics[width=0.49\textwidth]{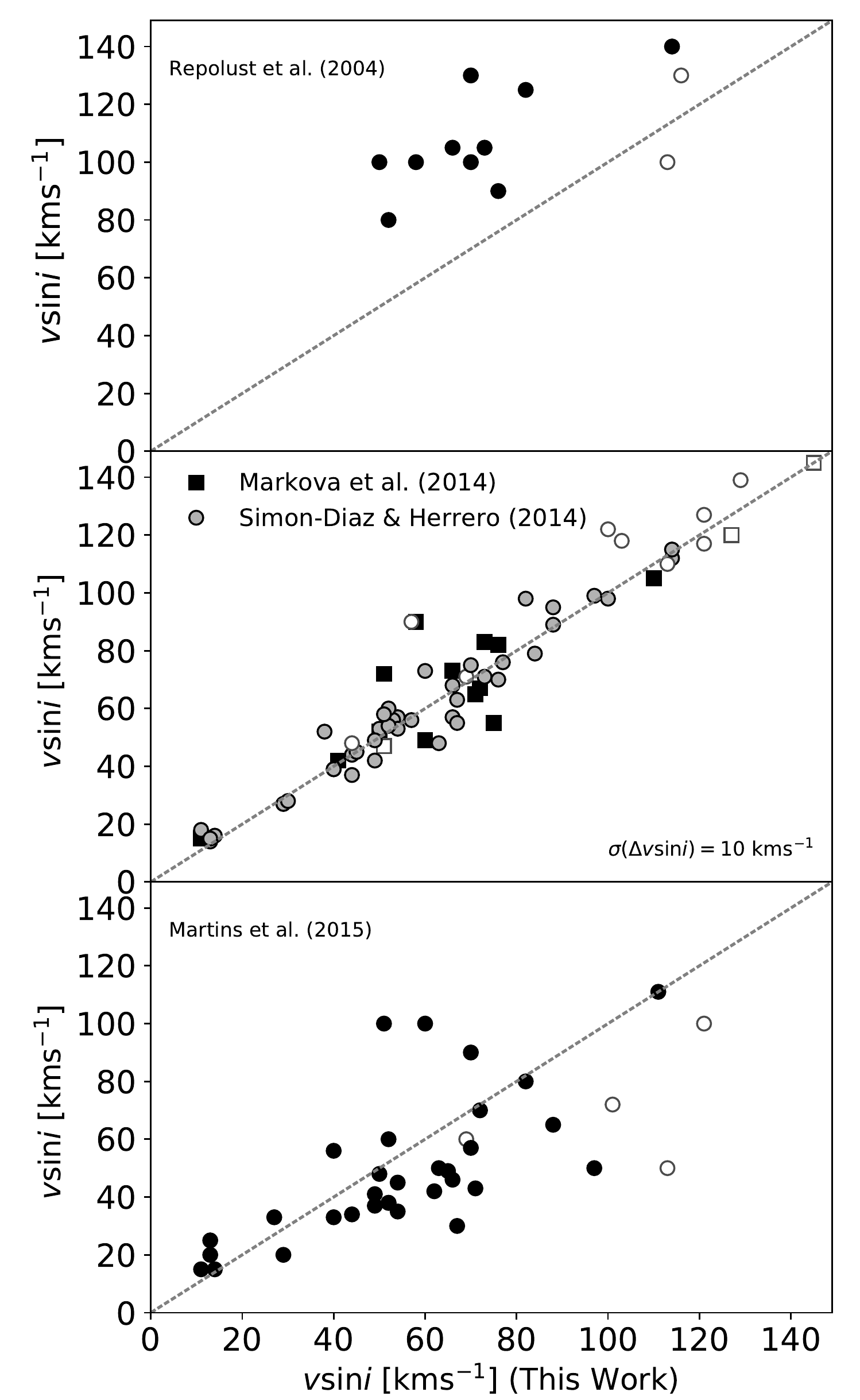}
\caption{Comparison of \vsini\ determinations for a sample of stars in common with four reference papers in the literature. Open symbols indicate stars for which we have detected clear or likely signatures of spectroscopic binarity.}
\label{broad_comparison_vsini}
\end{figure}
	
	
	
	
\begin{figure}[!t]
\includegraphics[width=0.49\textwidth]{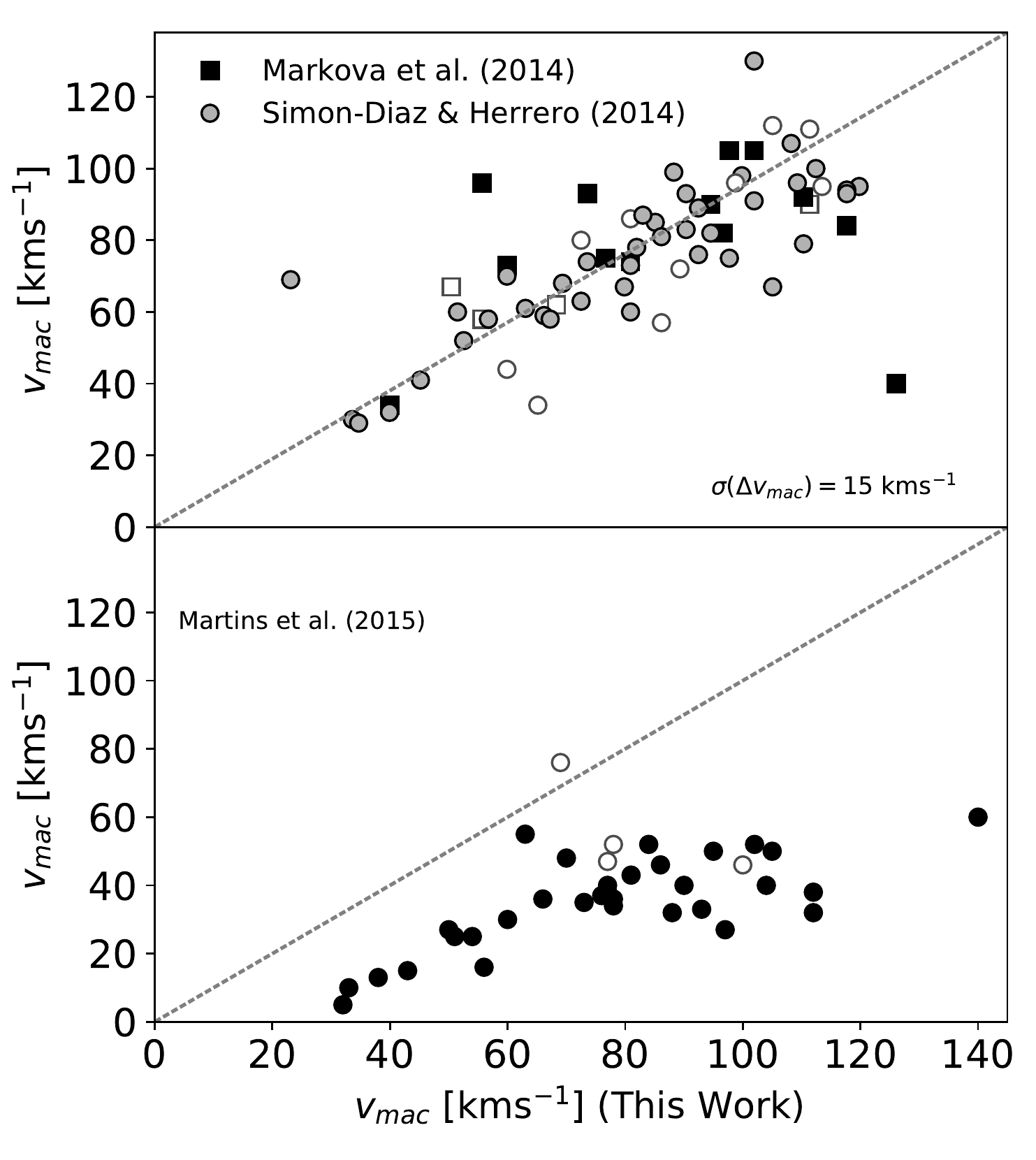}
\caption{Comparison of \vmacro\ determinations for a sample of stars in common with three reference papers in the literature. Open symbols indicate stars for which we have detected clear or likely signatures of spectroscopic binarity. See Sect.~\ref{subsectionCompBroad} for details and discussion.
}
\label{broad_comparison_vmac}
\end{figure}
	

\subsubsection{Line-broadening parameters (\vsini\ and \vmacro)}\label{subsectionCompBroad}

We start with the comparison of projected rotational velocities in Fig.~\ref{broad_comparison_vsini}. Before proceeding to the discussion of results, we indicate that the values considered by \cite{Repolust04} -- top panel -- were obtained from the analysis of medium resolution spectra and assuming that rotation is the only source of line-broadening. The other three studies use analysis techniques and spectra of similar quality as in this paper, with the only difference that \cite{Markova14} and \cite{Simon14b} -- middle panel -- used \textsc{iacob-broad}, while \cite{Martins15} -- bottom panel -- used their own tool to perform the line-broadening analysis. As expected, there is a fairly good agreement between those studies correcting for the effect of macroturbulent broadening, while important discrepancies (up to 50 \kms\ in some critical cases) are found with the values provided in \cite{Repolust04}. As stated in, e.g., \cite{Simon07, Simon14b} and \cite{Markova14}, the derived values of \vsini\ are importantly overestimated in the whole O-type star domain whenever the effect of other broadenings agents on the line-profiles are neglected. 


\begin{figure*}
\centering
\includegraphics[width=0.9\textwidth]{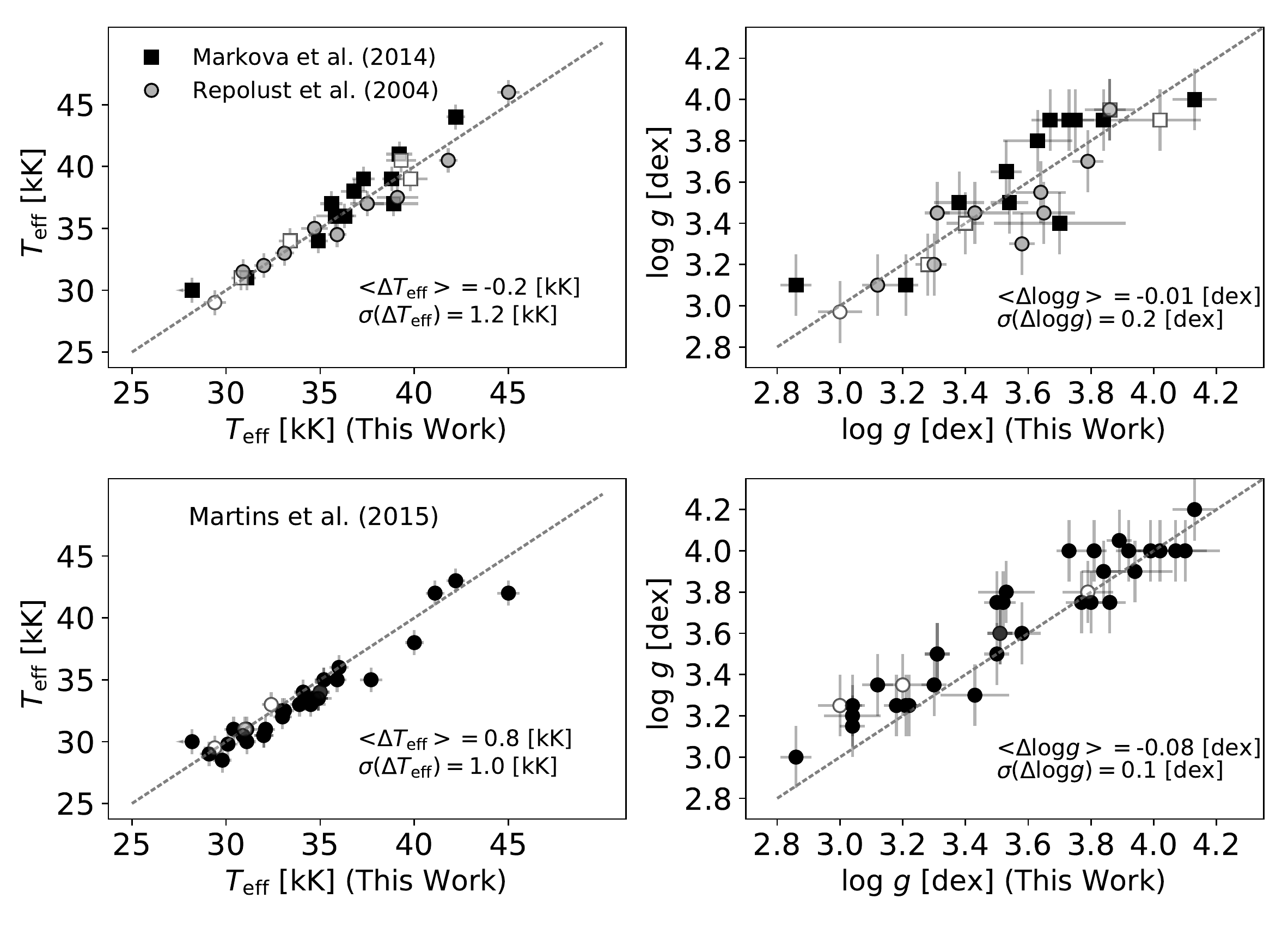}
\caption{Comparison of parameters for stars in common with \cite{Repolust04} and \cite{Markova14}, \fastwind\ studies (Top) and \cite{Martins15} -- {\sc cmfgen} (Bottom). Open symbols indicate stars for which we have detected clear or likely signatures of spectroscopic binarity. The dashed line is the 1:1 relation and each panel shows the mean and standard deviation of differences.}
\label{CompValues}
\end{figure*}


Nine stars present discrepancies in the \vsini\ determination larger than 20~\kms\ when comparing our results with those from \cite{Martins15}. Three of them have been identified as spectroscopic binaries in our study. The use of different spectra could explain the discrepancy. For the other six, after checking again the outcome from {\sc iacob-broad}, we have not found a clear explanation for the disagreement or a reason to modify our results. Indeed, we note that for these stars we find similar values as those previously obtained in \cite{Simon14b} and \cite{Markova14} using different spectra. 

Figure~\ref{broad_comparison_vmac} presents a similar comparison but for the case of the macroturbulent velocity. In this case, \cite{Repolust04} is excluded since they do not consider this parameter in the line-broadening analysis. Again a fairly good agreement is found with \cite{Markova14} and \cite{Simon14b} -- top panel --, indicating the robustness of the results provided by \textsc{iacob-broad}, even when different spectra are analyzed by different users. The clear disagreement with the values computed by \cite{Martins15b} can be easily explained if one takes into account the different definition of the macroturbulent profile considered by \citeauthor{Martins15}, an isotropic definition (corresponding with a Gaussian profile), and the other three works (a radial-tangential profile). As illustrated in \cite{Simon14b}, for a fixed value of \vsini, a Gaussian-type macroturbulent profile implies a value of \vmacro\ $\sim$65\% smaller than when a radial-tangential profile is considered. We refer the reader to \cite{Simon14b} for a detailed explanation of the reason why we prefer to use a radial-tangential definition of the macroturbulent broadening profile.

From the comparison of results presented in the middle panel of Fig.~\ref{broad_comparison_vsini} and the top panel of Fig.~\ref{broad_comparison_vmac}, we conclude that the mean uncertainty in \vsini\ and \vmacro\ associated with the analysis of different single-epoch spectra of the same star is of the order of $\sim$15~\kms\ in both cases. This difference is relevant considering the high resolution  spectra (R$>$ 25 000) used in most of these studies (excluding \cite{Repolust04} with R$\sim$5000), corresponding to limit rotational velocities lower than 12 \kms.

\subsubsection{Effective temperature (\Teff) and gravity (\grav)}

Figure~\ref{CompValues} compares our derived values for \Teff\ and \grav\ with those determined by \cite{Repolust04, Markova14} and \cite{Martins15}. Similarly to our work, the first two studies are based on the stellar atmosphere code \fastwind. In contrast, \citeauthor{Martins15} used {\sc cmfgen} \citep{Hillier98}. 

We find a general good agreement in all cases. These results not only validate our automatized analysis strategy, but also highlight the robustness reached by the quantitative spectroscopic analyses of O-type stars in the last years -- at least regarding the two considered parameters, \Teff\ and \grav.

The comparison presented in the top panels indicates that we are basically reproducing the results obtained by \cite{Repolust04} and \cite{Markova14}, who used the same stellar atmosphere code (\fastwind) and analysis strategy as this work, but a more subjective ``by-eye'' fitting of the main \ion{H}{} and \ion{He}{} diagnostic lines. The mean values of the differences in \Teff\ and \grav\ are $-$200~K and $-$0.01 dex, respectively, well below the formal uncertainties associated with these parameters (see Table~\ref{ParamStandards}). On the other hand, the standard deviations of the differences are 1200~K and 0.2~dex, respectively. These values, that are somewhat larger than the uncertainties resulting from the {\sc iacob-gbat} analysis (especially for the case of log~$g$, that are of the order of $\pm$0.05\,--\,0.10 dex) indicate that the latter are sometimes too optimistic. Other sources of uncertainty such as those associated with stellar variability, or assumed values for those parameters fixed in the analysis (e.g, \vsini, \vmacro, \micro,  $v_{\infty}$) may have a similar, or even more important effect on the final errors associated with these parameters for a given star. Finally, a correct treatment of the normalization of the spectra is an essential point for the derived gravities.

Interestingly, in spite of the different assumptions considered in the set of line-broadening parameters used in the quantitative spectroscopic analysis, we do not find any remarkable difference between the comparison of our results and those by \cite{Repolust04} and \cite{Martins15}. This can be interpreted as indirect evidence indicating that the exact parametrization of the global line-broadening (e.g. by means of one --\vsini\ -- or two parameters -- \vsini\ and \vmacro, or a different definition of the macroturbulent profile) is not going to critically affect the derived \Teff\ and \grav.

When comparing with \cite{Martins15} we need to take into account that we are using a similar set of diagnostic lines, but they are using the {\sc cmfgen} stellar atmosphere code. A closer inspection of the two bottom panels in Fig.~\ref{CompValues} indicates that, although there is a fair agreement -- within the associated derived uncertainties -- between results obtained by these two independent works, there seems to be some hints pointing out to the existence of a systematic difference in the derived effective temperatures and gravities, with {\sc cmfgen} resulting in lower effective temperatures and higher gravities (with mean values of the differences $\sim$800 K and 0.09 dex, respectively). Regarding the gravities, this discrepancy might be attributed to the approximate treatment of the background line opacities in \fastwind. This could lead to an underestimated radiative acceleration in the upper photosphere and, in consequence, to underestimated gravities \citep[see also][]{Massey13}. Regarding the discrepancy in the derived \Teff, such discrepancy has already been found previously \citep[see, e.g.,][]{SimonDiaz2008}, who provide some examples. The origin of this discrepancy is rooted in different predictions for the strength of \ion{He}{ii}\,$\lambda$4200/4541, where {\sc cmfgen} provides deeper profiles in the temperature range between 30 to 35 kK, though the corresponding \ion{He}{i} profiles match perfectly. Thus far, we have no real explanation for this difference. 

Finally, we discuss some of the most discrepant results individually. From the comparison with \cite{Markova14} and \cite{Repolust04}, we highlight HD\,151804 and HD\,193514, two mid O-type supergiants for which differences in \grav\,$\sim$\,0.3 dex are found. HD\,151804 is a star with a very strong and variable wind, a fact that could explain the discrepancy in gravities obtained from different snap-shot spectra. HD\,193514 has also been labeled as WV; although we do not see clear variability affecting the wings of H$\beta$ and H$\gamma$ in our spectra, we cannot discard a similar explanation for this star. From the comparison with \cite{Martins15}, we highlight HD~30614 ($\alpha$ Cam, $\Delta$\grav\,$\sim$\,0.3 dex). This is an O9\,Ia star that we identified as SB1 and, in addition, for which we measured a \vsini\,$\sim$\,110~\kms\ while \cite{Martins15} considered a much lower value of this quantity (50~\kms) in the spectroscopic analysis. A difference of 60~\kms\ in the assumed \vsini\ can well explain a difference of 0.3 dex in the derived gravity (with the lower \vsini\ implying a higher \grav).   

\section{Discussion}\label{section5}

In this section we use the results of our homogeneous and automatized {\sc iacob-gbat} spectroscopic analysis to describe the global properties of the sample, review the most recent and commonly used SpT\,--\,\Teff\ and SpT\,--\,\grav\ calibrations for Galactic O-type stars, and propose a distance-independent test for the wind-momentum luminosity relationship. We remind that, as explained in Sect.~\ref{section41}, the 7 stars labeled as Q4 (no \ion{He}{i} lines available) have been excluded from the figures and discussion presented in this section.

\begin{figure*}[!t]
	\includegraphics[width=\textwidth]{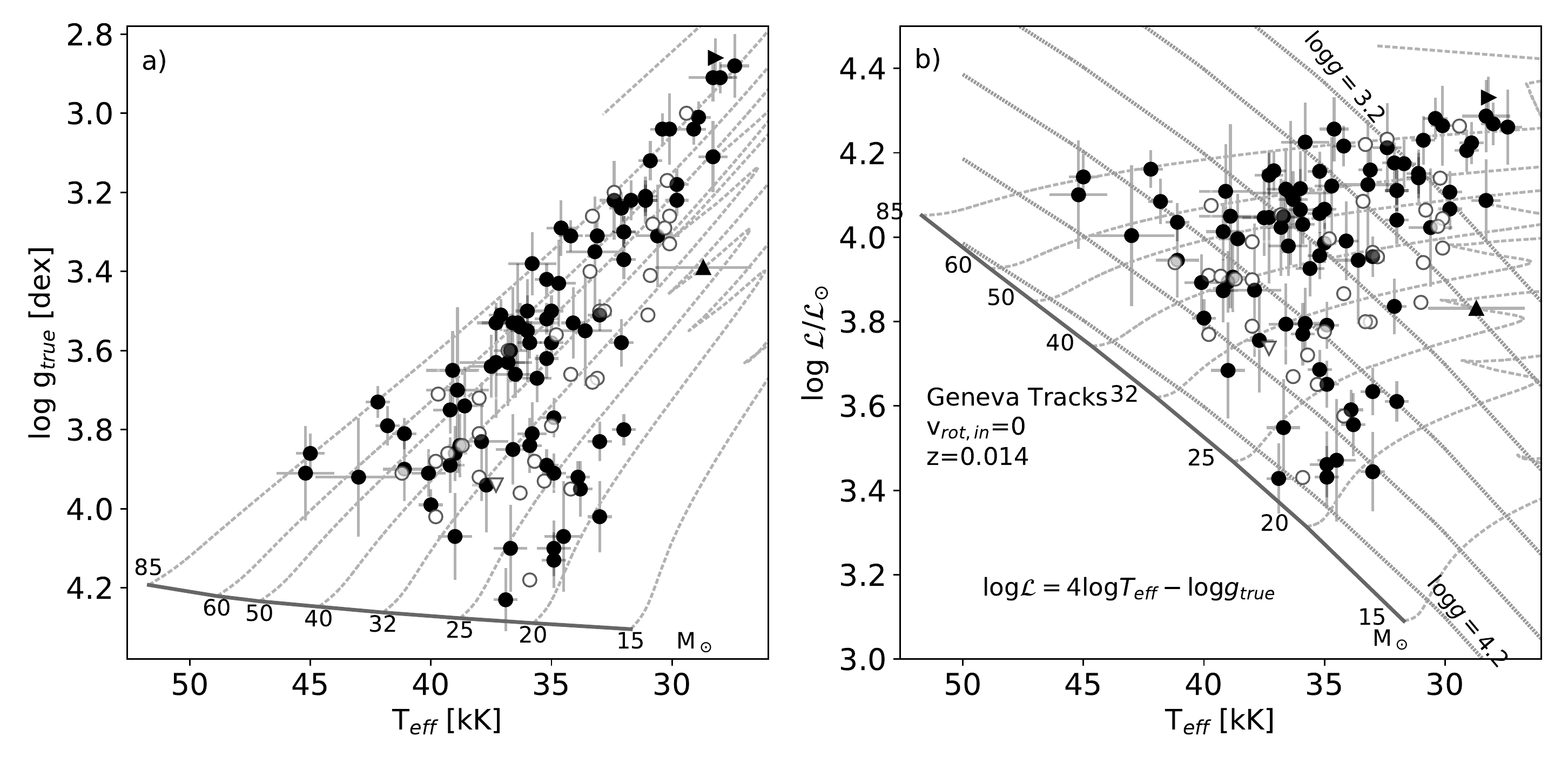}
	\caption{Location of 121 of the stars in our sample in the Kiel (left) and spectroscopic HR (right) diagrams. Open symbols indicate stars for which we have detected clear or likely signatures of spectroscopic binarity. Triangles are stars for which only upper or lower limits in any of the two parameters used to construct these diagrams (\Teff\ and \grav) could be obtained. Individual uncertainties are included as error bars. Evolutionary tracks and position of the ZAMS from the non-rotating, solar metallicity models by \cite{Ekstrom12} and \cite{Georgy13} are included for references. The dotted diagonal lines in the sHR diagram are the isocontours of constant gravity. We note that those 7 early O-type stars flagged as Q4 (see Sect.~\ref{section41}) are not included in the figures}.
	\label{GAP}
\end{figure*}

\begin{figure*}[!t]
	\includegraphics[width=\textwidth]{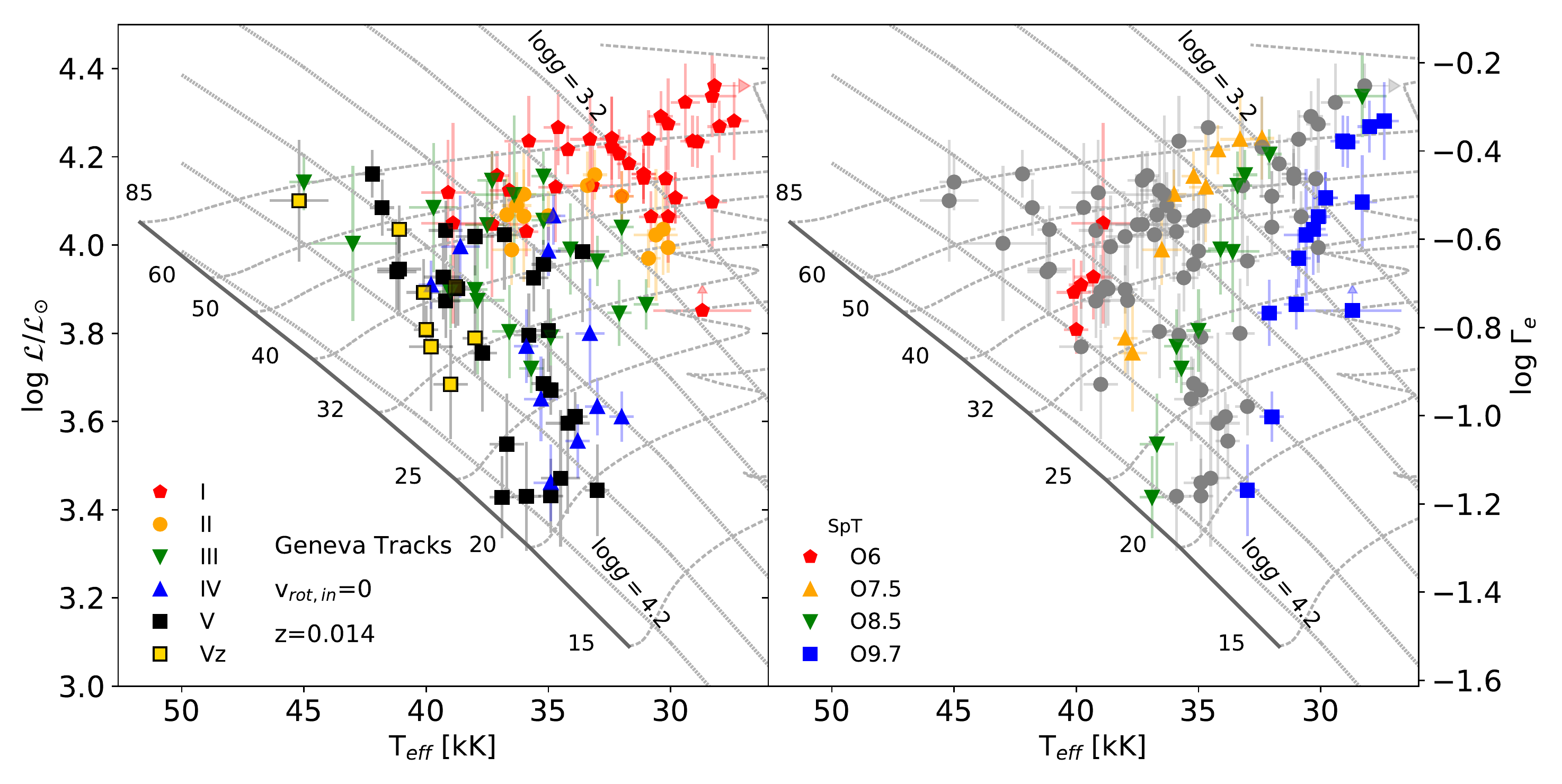}
	\caption{Same as right panel of Fig.~\ref{GAP}, but using various colors and symbols to identify different luminosity classes (left) and specific spectral types (right).The right-hand ordinate axis displays the electron scattering Eddington factor ($\Gamma_{e}$). In this figure we do not differentiate between likely single stars and those targets for which we have detected clear or likely signatures of spectroscopic binarity.
	}
	\label{SPC}
\end{figure*}

\subsection{Kiel and spectroscopic HR diagrams}\label{KsHRD}

Figure \ref{GAP} presents the location of 121 of the stars from our sample in the \grav\,--\,\Teff\ (Kiel) diagram and the spectroscopic HR (sHR) diagram \citep{Langer14}. We also include in both diagrams the Geneva non-rotating evolutionary tracks for solar metallicity \citep{Ekstrom12, Georgy13} for reference purposes.

As already shown elsewhere \citep[see, e.g.,][]{Repolust04, Markova14, Martins15c}, the O-type stars mostly concentrate between the 20 and 85~M$_{\odot}$ evolutionary tracks and are located within the main sequence (if we use non-rotating models as reference!). In the particular case of the sample under study, the expected good coverage of the complete O star domain is challenged by the lack of standard stars close to the zero age main sequence (ZAMS) above $\sim$25~M$_{\odot}$. Furthermore, the existence of this empirical gap, already highlighted in several papers dealing with samples of relatively small size \citep[e.g.,][]{Herrero92, Repolust04, Martins05, Herrero07, Simon14}, is becoming more and more evident with the increase in the number of O-type stars analyzed spectroscopically \cite[see, e.g.,][]{Castro14, SabinSan17}. Therefore, the absence of standard stars in this region of the Kiel and sHR diagram seems to be a natural consequence of a more general observational characteristic of massive stars. Indeed, preliminary analyses of the complete IACOB+OWN sample of $\sim$270 O-type stars labeled as likely single or SB1 using a similar strategy as in this work \citep{Holgado17} confirm the existence of this gap. A more detailed discussion of this result will be presented in a forthcoming paper of the IACOB series (Holgado et al., in prep.), where several hypotheses that have been proposed to explain the lack of observed stars close to the commonly accepted ``theoretical'' location of the ZAMS are discussed. Among them, we already highlight [1] observational limitations due to the fast evolution of massive stars during their early evolution and/or the fact that very young massive stars could be still embedded in their birth cocoon, hampering its detection and hence study in the optical range \citep[e.g.,][]{Garmany82,Herrero07, Castro14} and [2] theoretical considerations about how pre main-sequence objects connect with their main-sequence evolution \citep[e.g.,][]{Bernasconi96,Behrend01,Haemmerle16}.

We separate in the two diagrams presented in Fig.~\ref{GAP} the stars for which we have detected clear or likely signatures of spectroscopic binarity (open symbols) from those identified as likely single (filled symbols). The former group is not located in a specific region of the diagrams; instead, they can be found everywhere in the O star domain. Nevertheless, a firm conclusion must wait until the analysis of the whole IACOB+OWN sample of Galactic O-type stars is completed.

Additionally, in Fig.~\ref{SPC} we show how the standard stars defining the various luminosity classes and spectral types are distributed in the sHR diagram using different colors and symbols. From inspection of the left panel it becomes clear that, in the O star domain, the separation between luminosity classes in the sHR diagram is not defined by strict boundaries but, instead, there are many overlapping regions in which stars with different luminosity classes have similar effective temperatures and gravities. In addition, the combination of information provided in the two panels of Fig.~\ref{SPC} serves us to also illustrate the well-known trends of effective temperature with spectral types and luminosity class, how the various luminosity classes are more clearly separated in terms of stellar parameters in the late O-type star domain, and the intrinsic scatter in \Teff\ and \grav\ expected for any given combination of spectral type and luminosity class.

Based on these figures we can also conclude that the stars that should be filling the gap close to the ZAMS are expected to be found among those classified as luminosity class V and having spectral types earlier than $\sim$O8. Following \cite{Walborn73}, the O~Vz stars should be the most promising candidates. This author suggested that the Vz phenomenon (a spectroscopic peculiarity that is defined by a stronger \ion{He}{ii}\,4686 absorption, relative to other \ion{He}{} lines, compared to that found in typical class V spectra) should be a clear indication of youth, and hence proximity to the ZAMS. 

%
\begin{figure*}[!t]
\centering
\includegraphics[width=0.95\textwidth]{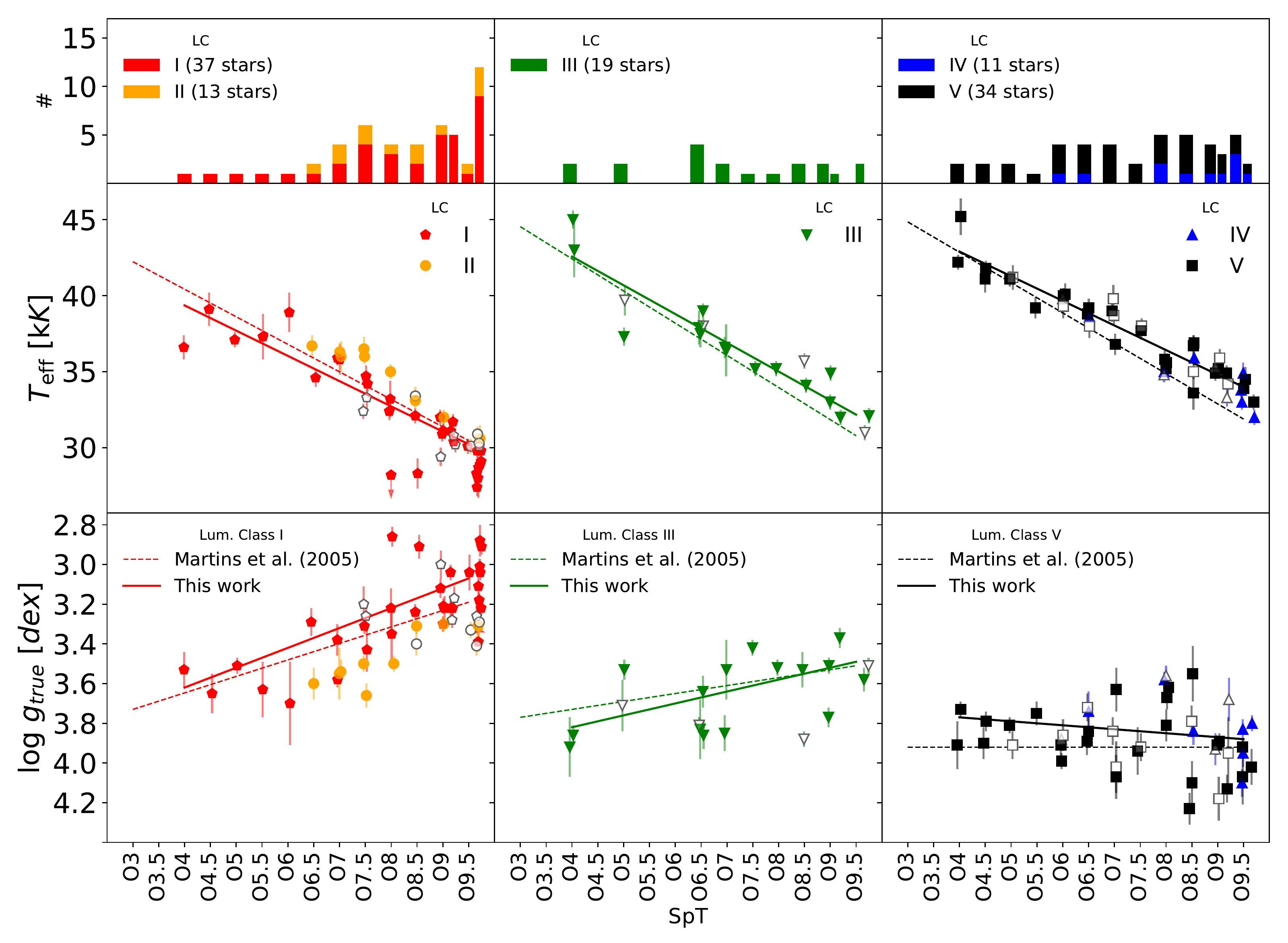}
\caption{SpT--\Teff\ (middle panels) and SpT--\grav\ (bottom panels) calibrations. The global sample is divided by luminosity class groups in columns. Top panels show the number of stars per spectral type bin. The ``observational scales'' proposed by MSH05 are indicated by dashed lines. Calibrations obtained in this work are solid lines. As in previous figures, open symbols indicate stars for which we have detected clear or likely signatures of spectroscopic binarity. Stars with luminosity class Ia are not considered for the calibrations presented in Table~\ref{Cab1}, and as solid lines in these figures. }
\label{Calibs}
\end{figure*}

We evaluate this possibility by considering those stars included in the grid of standards for spectral classification that have been recently confirmed as O~Vz stars following the criterion described in \cite{Arias16}. This subsample of stars (labeled with filled yellow squares in the left panel of Fig.~\ref{SPC}) traces the lower boundary of the distribution of O dwarfs in the sHR diagram; however, they are still 0.2 dex away (in \grav) from the theoretical ZAMS defined by the \cite{Ekstrom12} models. Exploring the dearth of stars close to the ZAMS with a larger sample of O~Vz is hence an interesting line of future work (although see also the clarifying notes about the O~Vz phenomenon in \citeauthor{SabinSan14}, \citeyear{SabinSan14}, and \citeauthor{Arias16}, \citeyear{Arias16}).

Last, we note that the presence of the gap has been mainly outlined up to now by studies making use of \fastwind\ models. If we account for the systematic differences in \Teff\ and \grav\ found when comparing \fastwind\ and {\sc cmfgen} models (see Fig.~\ref{CompValues}), log~$\mathcal{L}$ might rough decrease by -0.12 dex. Would this correction in log~$\mathcal{L}$ be really needed, the gap could be reduced. However, the lower values of \Teff\ required by the {\sc cmfgen} analysis would move again the stars away from the predicted ZAMS.

\subsection{SpT--\Teff\ and SpT--\grav\ calibrations}\label{subsectionCalib}

The calibration of stellar parameters versus spectral type for Galactic O-type stars presented by MSH05 has become a standard reference for the massive star community and also for other related fields such as, e.g., the study of \ion{H}{ii} regions. This calibration replaced the previous ones based on spectroscopic analyses performed with stellar atmosphere codes that did not take into account non-LTE, wind, and line-blanketing effects \citep[e.g.][]{Vacca96}. MSH05 provided several recipes connecting the spectral type and luminosity class of a given star with its effective temperature and spectroscopic gravity (along with other parameters). In particular, they used a compilation of the results -- available in the literature -- from the  optical spectroscopic analysis of a sample of Galactic O stars using non-LTE spherically expanding models including line-blanketing to build what they called the ``observational scale''. Despite its novelty and importance, this calibration is based on a heterogeneous set of spectroscopic observations (in terms of quality and resolution), stellar parameter determinations and spectral classifications. In particular, the parameters gathered by MSH05 combine results from \cite{Herrero02} and \cite{Repolust04} obtained from \fastwind, and from \cite{Martins05b}, who used {\sc cmfgen}. In addition, it relies on a relatively small sample of targets (45 stars distributed among dwarfs, giants, and supergiants). 

In this section we revisit the SpT--\Teff\ and SpT--\grav\ calibrations proposed by MSH05 using a much larger sample of stars (114, see below) that (1) have been observed spectroscopically in an homogeneous way, (2) have been analyzed homogeneously with state-of-the-art models and techniques, (3) constitute part of the modern grid of O-type standards for spectral classification, and whose spectral classification has been recently reviewed in an homogeneous way by the GOSSS team. 

Results are summarized in Fig.~\ref{Calibs}, where we separate the sample in three main blocks by luminosity class: supergiants and bright giants (left, 50 stars), giants (middle, 19 stars), and subgiants and dwarfs (right, 45 stars). The top panel shows the number of stars per spectral type bin in each luminosity class category while the middle and bottom panels present the SpT\--\Teff\ and SpT\--\grav\ calibrations, respectively, together with the ``observational scales'' proposed by MSH05. 
Once more, we identify those stars labeled as SB1 with open symbols (SB2 are excluded from these figures). In addition, and similarly to previous figures, we do not include those 7 targets labeled as Q4 (see Sect.~\ref{section41}). As a consequence of the latter, we end up with no targets with spectral types earlier than O3.5 and the number of stars in the spectral type bins O4 and O4.5 are slightly reduced.

We find an overall good agreement with the calibrations presented by MSH05; however, a few points of interest deserve further attention. The most critical one is the systematic offset towards higher \Teff\ (up to $\sim$3\,000~K in some cases) that we obtain in the late O-type giants, subgiants and dwarfs with respect to Martins' observational scale. The offset tends to become smaller and even disappears towards the early-type stars. A similar result was already obtained in \cite{Simon14} using a smaller sample of Galactic O-type dwarfs observed in the framework of the IACOB project. The fact that many stars in the O9 spectral type bin have been more recently reclassified as O9.2, O9.5, and O9.7, plus the combined effect of the intrinsic scatter in \Teff\ present in each SpT bin and the low number statistics used to define the observational scale may also have an important impact on the defined calibrations \citep[see a more detailed discussion in][]{Simon14}. Other explanations related to, e.g., discrepancies in the stellar parameters derived with different stellar atmosphere codes or by different people can be ruled out as the main reason of the systematic offset in the late O-type regime in view of the results presented in Fig.~\ref{CompValues}: the agreement between the various studies using \fastwind\ is almost perfect, and the mean difference in \Teff\ between our results and those obtained by \cite{Martins15} using {\sc cmfgen} is $\sim$900 K. In addition, we note that the ``observational scales'' presented in MSH05 were mainly obtained using results from analyses performed by that time with {\sc fastwind}.

In these regards, it is also interesting to see that the scatter in the SpT--\Teff\ and SpT--\grav\ calibrations discussed in \cite{Simon14} still remains although we are now limiting our sample to the standard stars for spectral classification. We thus confirm that this scatter is a natural consequence of the way the spectral classification process is defined (see also Fig.~\ref{SPC} and notes in Sect.~\ref{KsHRD}) and not necessarily due to caveats related to the classification of O-type stars with intermediate and extreme \vsini\ values.

A last point of attention concerns those four late O-type supergiants for which we have obtained values of \grav$_{\rm true}\,\lesssim\,$2.9 dex and effective temperatures lower than 28\,500 K (see leftmost panels in Fig.~\ref{Calibs}). These stars\footnote{HD~151804: O8~Iaf, HDE~303492: O8.5~Iaf, HD~105056: ON9.7~Iae, HD~195592: O9.7~Ia.} can be considered as clear outliers from the calibrations. Interestingly, all of them have been classified as Ia, have been flagged as Q2 or Q3 (plus also WV), and are among those targets for which we have obtained the largest values of log~$Q$ (i.e. they have the strongest winds among the analyzed stars). In addition, the derived values of \Teff\ and \grav\ imply values of the Eddington factor ($\Gamma_{\rm e}$\,=\,$L$/$L_{\rm Edd}$) close to 0.5. While there is the possibility that late-O stars with luminosity class Ia could form a separated group in the calibrations, given the points above we cannot discount that this result is just a consequence of the limitations of our analysis strategy for stars approaching the Eddington limit (see further notes in Appendix~\ref{vinfAppend}).

We provide new linear calibrations based on the sample of stars analyzed in this paper in Table~\ref{Cab1}. To obtain them we have discarded all stars for which we have detected clear or likely signatures of spectroscopic binarity.  In addition, for the calibrations of supergiant stars, we exclude those stars identified as luminosity class Ia (see notes above).

We find that our SpT\,--\,\grav\ calibration for dwarfs always results in gravities lower than the value proposed by MSH05 (\grav\,=\,3.9~dex). However, we remark that assuming a unique value per SpT of this parameter in the O dwarfs is an oversimplified recipe since it actually ranges between 4.2 and 3.5 dex \cite[see also notes in][]{Simon14}.This specific calibration must be handled carefully to avoid misinterpretations. The spectral classification was defined arbitrarily discreet and we are using a linear scale.


\begin{table}[t!]
\caption{Linear fits of the observed data presented in Fig.~\ref{Calibs}. \Teff\ (top) and \grav\ (bottom) as a function of spectral types for three luminosity classes, with SPT the spectral type number of O-type stars.}
\label{Cab1}
	\begin{tabular}{r@{\hskip 0.01in}l}
		\Teff\ [kK] $=$	& $\left\{\begin{tabular}{@{\ }ll} -1.66  $\times$ SPT + 46.00 & (I, 25 stars) \\-1.89  $\times$ SPT + 50.13  & (III, 15 stars) \\ -1.62  $\times$ SPT + 49.38 & (V, 25 stars) 
		\end{tabular}\right.$ \\ 
\\
	\grav\ [dex] $=$	& $\left\{\begin{tabular}{@{\ }ll} -0.10  $\times$ SPT + 4.02 & (I, 25 stars) \\-0.06  $\times$ SPT + 4.06  & (III, 15 stars) \\ { }0.02  $\times$ SPT + 3.69 & (V, 25 stars) 
	\end{tabular}\right.$ \\ 
	\end{tabular}
\end{table}


\begin{figure*}
	\includegraphics[width=\textwidth]{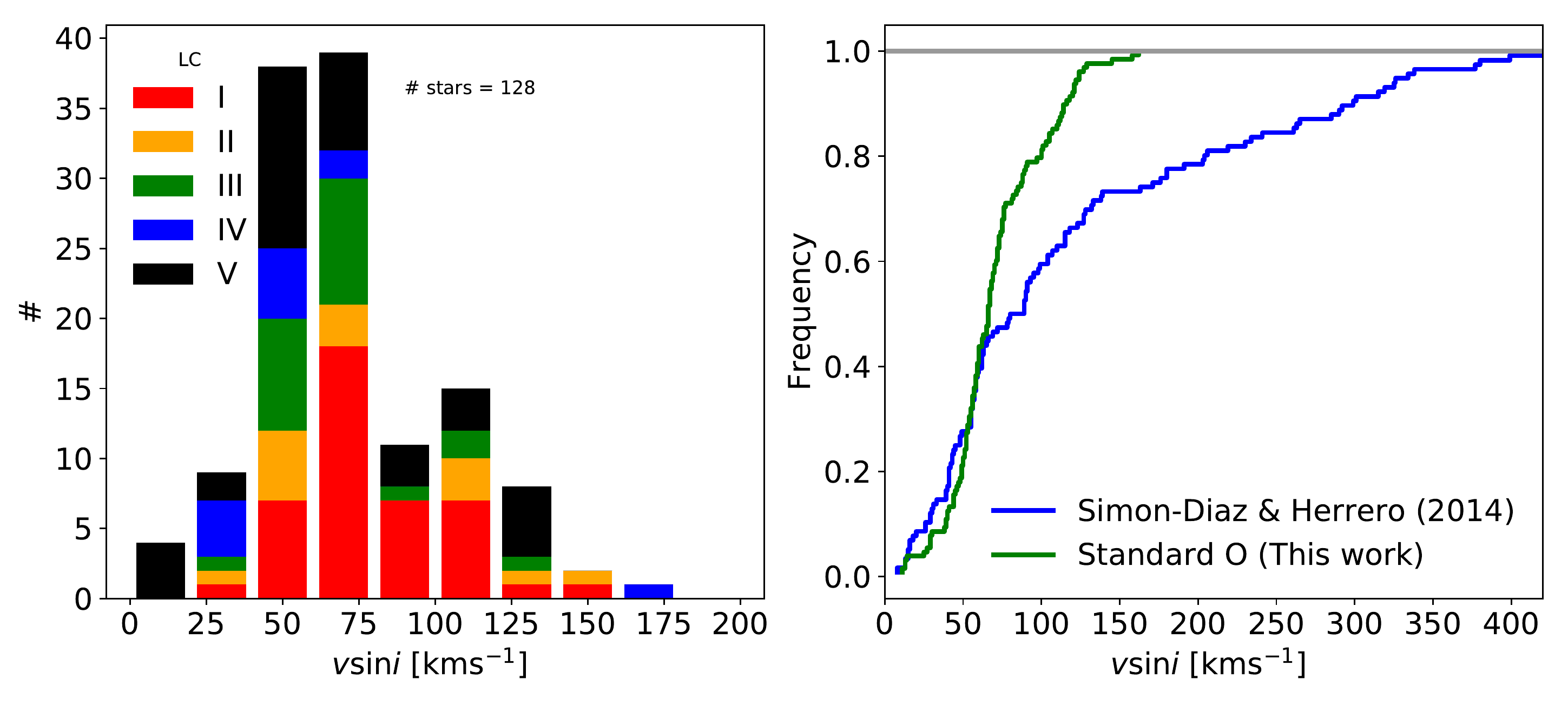}
	\caption{[Left] Distribution of \vsini\ in the sample. Different colors have been assigned to different luminosity classes. [Right] Cumulative histogram of two samples. The green one is our sample, the standard stars. The blue one is a more general sample of O stars, studied in \cite{Simon14b}.
	}
	\label{vsini_Num}
\end{figure*}

\subsection{\vsini\ distribution}

Figure~\ref{vsini_Num} presents the distribution of projected rotational velocities of the complete sample of O-type stars considered in this work in the form of histogram (left panel) and cumulative distribution (right panel). For comparison purposes, we also show the cumulative distribution resulting from the line-broadening analysis of the more general sample of Galactic O-type stars presented in \cite{Simon14b}.

The distribution peaks around 40\,--\,80~\kms\ and covers a range in \vsini\ between $\sim$\,10 and 250~\kms. More specifically, 95\% of the sample stars have \vsini$<$120\kms. 

When we compare this distribution with that resulting from the study of \cite{Simon14b} it becomes clear that the sample of O-type standards for spectral classification is affected by an important selection effect and is not representative of the global spin distribution of Galactic O-type stars. This is, however, not surprising since -- almost by definition -- these stars are selected among the ones with narrower lines, and hence lower \vsini. Therefore, one has to be very careful when extracting conclusions from the study of this sample, especially in the context of stellar evolution and regarding relative percentages of, e.g., fast/slow rotators, detected spectroscopic binaries, or magnetic stars.   

Figure~\ref{vsini_Num} also shows that while almost 75\% of the stars in the sample considered by \cite{Simon14b} have a projected rotational velocity below 120~\kms, there are 27 stars among the list of standards above this limit. This means that it will not be very strange to find situations in which a spectroscopic template for a given standard has broader lines than the star to be classified (even at a resolving power of 2500, in the GOSSS project). It would hence be ideal to try to find new standards with a lower \vsini\ to replace these 27 stars. 

\subsection{\vmacro\ distribution}

Our results regarding the macroturbulent broadening basically mimic those presented and discussed in \cite{Simon17}. We hence directly refer the reader to that paper, plus \cite{Markova14} and \citet{Godart17}, for a thorough discussion about this parameter in the full O and B star domain.

\subsection{Helium abundance}

\begin{figure}[!t]
\centering
\includegraphics[width=0.45\textwidth]{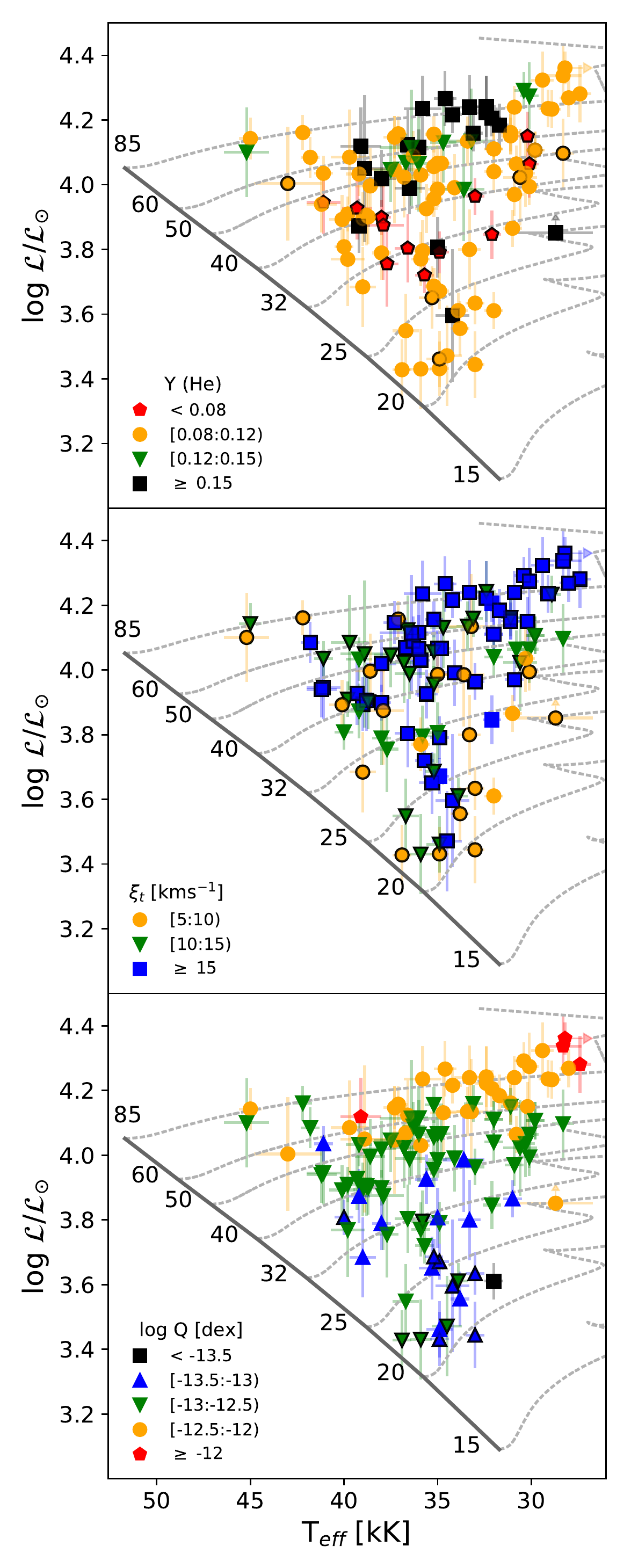}
\caption{Distribution of the helium abundance (top), microturbulence (middle) and wind-strength Q-parameter (bottom) in the sHR diagram. Stars for which it was only possible to obtain upper or lower limits for any of these quantities are indicated by symbols with the border in black. The range in the legend corresponds to the central value in case of a well-determined value, or to the upper or lower limit otherwise.}
\label{He_micro_Q}
\end{figure}

The top panel in Fig.~\ref{He_micro_Q} shows the distribution of helium abundances of our sample of O-type stars in the sHR diagram. We separate the sample in 4 abundance bins using different symbols and colors. Basically, we consider as stars with normal He abundance those having \helio\,=\,0.10$\pm$0.02 (yellow circles) and then distinguish between slightly (\helio\,=\,0.12\,--\,0.15, green triangles) and highly (\helio\,$>$\,0.15, black squares) enriched stars in helium. Finally, the fourth bin comprises stars with helium abundances below the commonly considered baseline for massive stars (\helio\,$<$\,0.08, red pentagons).   

\begin{figure}[t!]
		\centering
		\includegraphics[width=\hsize]{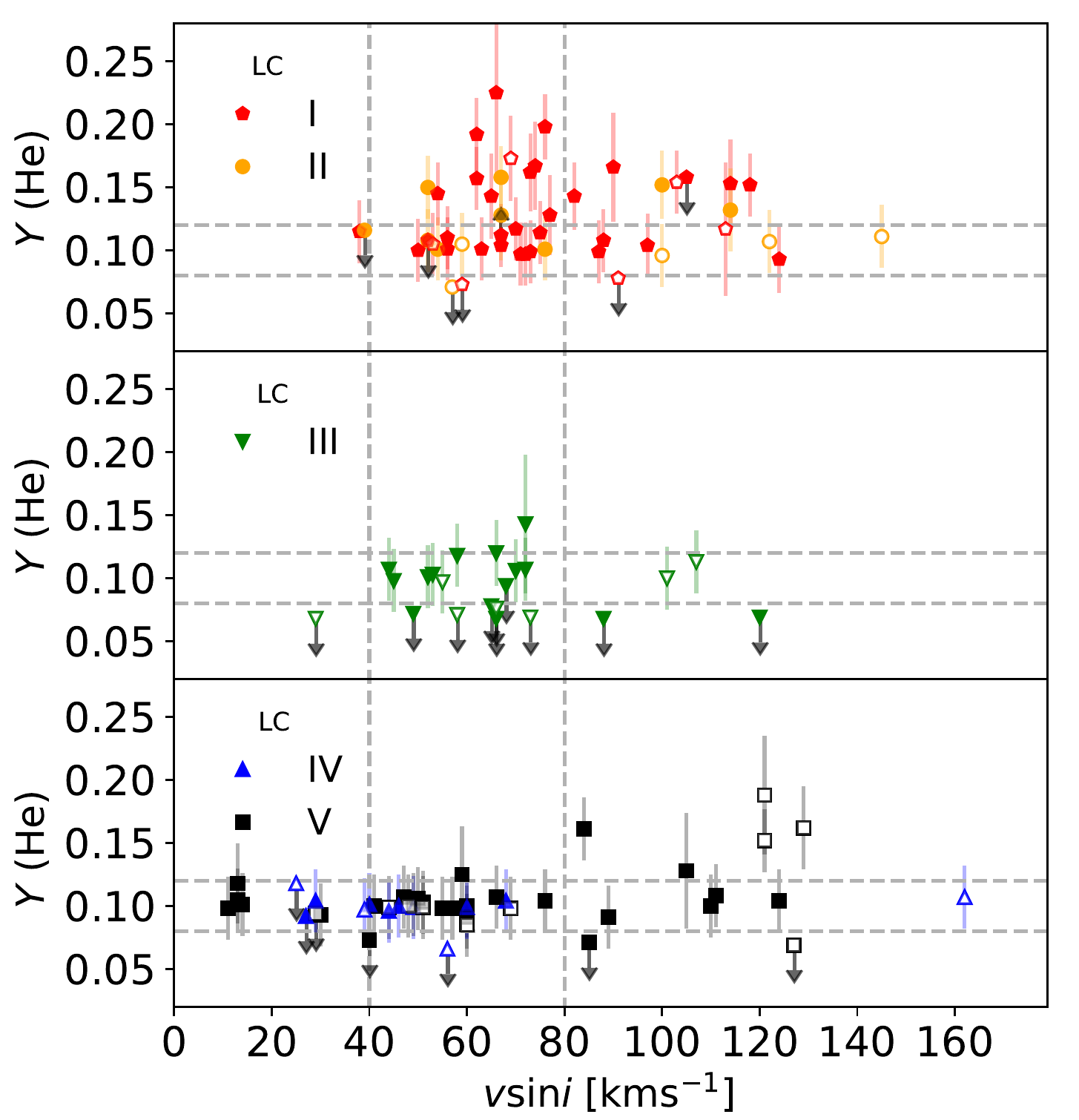}
		\caption{Helium abundance as a function of projected rotational velocity for the O Standard stars sample. Upper and lower limits are represented by arrows. Different colors and shapes represent luminosity classes. Open symbols indicate stars for which we have detected clear or likely signatures of spectroscopic binarity. Dashed gray lines delimit the range of values in \vsini\ and $Y_{\rm He}$ in which most of the stars are concentrated: 50\,--\,100 \kms\ and 0.08\,--\,0.12, respectively.
		}
		\label{vsini_He}
	\end{figure}

Figure~\ref{vsini_He} presents a different view to complement the same information, this time connecting the derived helium abundances with \vsini\ and separating the sample in the various luminosity class groups.  

Being aware of the important observational biases affecting the sample of stars under study, we just highlight the following main points regarding helium abundances in the sample of O-type standards for spectral classification:

\begin{itemize}
 \item More than half of the targets ($\sim$65\%) have what we have called normal helium abundances. This subgroup of stars is distributed all around the O star domain and covers the whole range of projected rotational velocities. This mean that 35\% of the so-called standards stars have anomalous abundances.
 \item Among the stars with clear He enrichment ($\sim$23\%) we mainly find O supergiants and bright giants with masses above $\sim$40~M$_{\odot}$, naturally. We note, however, that a similar number of O stars with luminosity classes I and II are found to have normal helium abundances. 
 \item Only 6 dwarfs and giants have \helio\,$>$\,0.12, most of them having \vsini\ in the range 70\,--140\,~\kms. Three of them are O dwarfs labeled as SB1. The presence of these enriched stars close to the ZAMS is not easily explicable with the current paradigm of rotational mixing of single stars.
 \item Regarding the 15 stars (13\% of the full sample) for which the {\sc iacob-gbat} analysis has resulted in He abundances below 0.08, most of them are either giants or relatively evolved dwarfs. The only three O supergiants and bright giants with \helio\,$<$\,0.08 have been found to be spectroscopic binaries, but no clear correlation between binarity signs and low He abundances can be extracted from our study for the stars within the other luminosity classes (see also Sect.~\ref{logQsect}). 
\end{itemize}

\subsection{Microturbulence}      

	
State-of-the-art studies link the origin of microtubulence to subsurface convection caused by the iron opacity peak \citep{Cantiello09}. However, the observational assessment of this hypothesis has been exclusive concentrated in the B-star domain. It still needs to be further evaluated in stars with higher masses.

The efforts devoted to investigate and empirically characterize microturbulence in the O-star domain have been really scarce since the works by \cite{McErlean98, Smith98} and \cite{Villamariz00}. The main reason is probably related to the fact that, in contrast to the case of early B-type stars, the number of metal lines available to determine accurate values of this parameter (e.g., using the curve of growth method) is much more limited in O stars. Some information can be obtained from the study of the \ion{He}{} lines; however, as illustrated by the outcome of our {\sc iacob-gbat} spectroscopic analysis, this is a hard task, since one can only obtain lower or upper limits for this parameter in most cases (see Table~\ref{ParamStandards}).

As a consequence, most of the quantitative spectroscopic analyses of O-type stars consider microturbulence as a fixed parameter \citep[see, e.g.,][]{Herrero02, Repolust04, Markova14}.
Only a few works have recently started to leave microturbulence as a free parameter \cite[e.g.,][]{Mokiem05,SabinSan14,Ramirez17}; however, the quality and wavelength coverage of the spectra considered in those studies are more limited than in our case.

The middle panel of Fig.~\ref{He_micro_Q} shows the distribution in the sHR diagram of the values of microturbulence resulting from the {\sc iacob-gbat} analysis of the sample of O-type stars investigated here. As indicated above, we could only determine upper or lower limits for the majority of the targets. This highlights the necessity to extend the range of microturbulence values in our grids (towards higher values) and, in addition, the inherent limitations of the use of \ion{He}{} lines for the determination of microturbulence. However, the diagram already provides a first rough general overview of the behavior of microturbulence in a region of the HR diagram that has been vaguely explored until now. 

As previously outlined by \cite{Massey13}, O supergiants systematically imply quite a high value of microturbulence. Interestingly, we also find that there is a non negligible number of stars in the whole O star domain (even dwarfs) with relatively high values of microturbulence. This may have important consequences for the reliability of the projected rotational velocities determined using FT techniques, as discussed by \cite{Simon14}, and might explain why we are still empirically detecting certain thresholds in the measured values of \vsini\ for O-type star.

A more detailed investigation of microturbulence in O stars, exploring the availability of other diagnostic lines better suited to obtain more accurate estimations of this parameter is one important line of future work. As indicated by Markova et al. (subm.), a proper treatment of turbulence in the modeling of O-type stars might be of ultimate importance for a proper characterization of the density structure of the stellar photosphere.

\subsection{Wind-strength Q-parameter} \label{logQsect}
	
\begin{figure}
	\centering
	\includegraphics[width=\hsize]{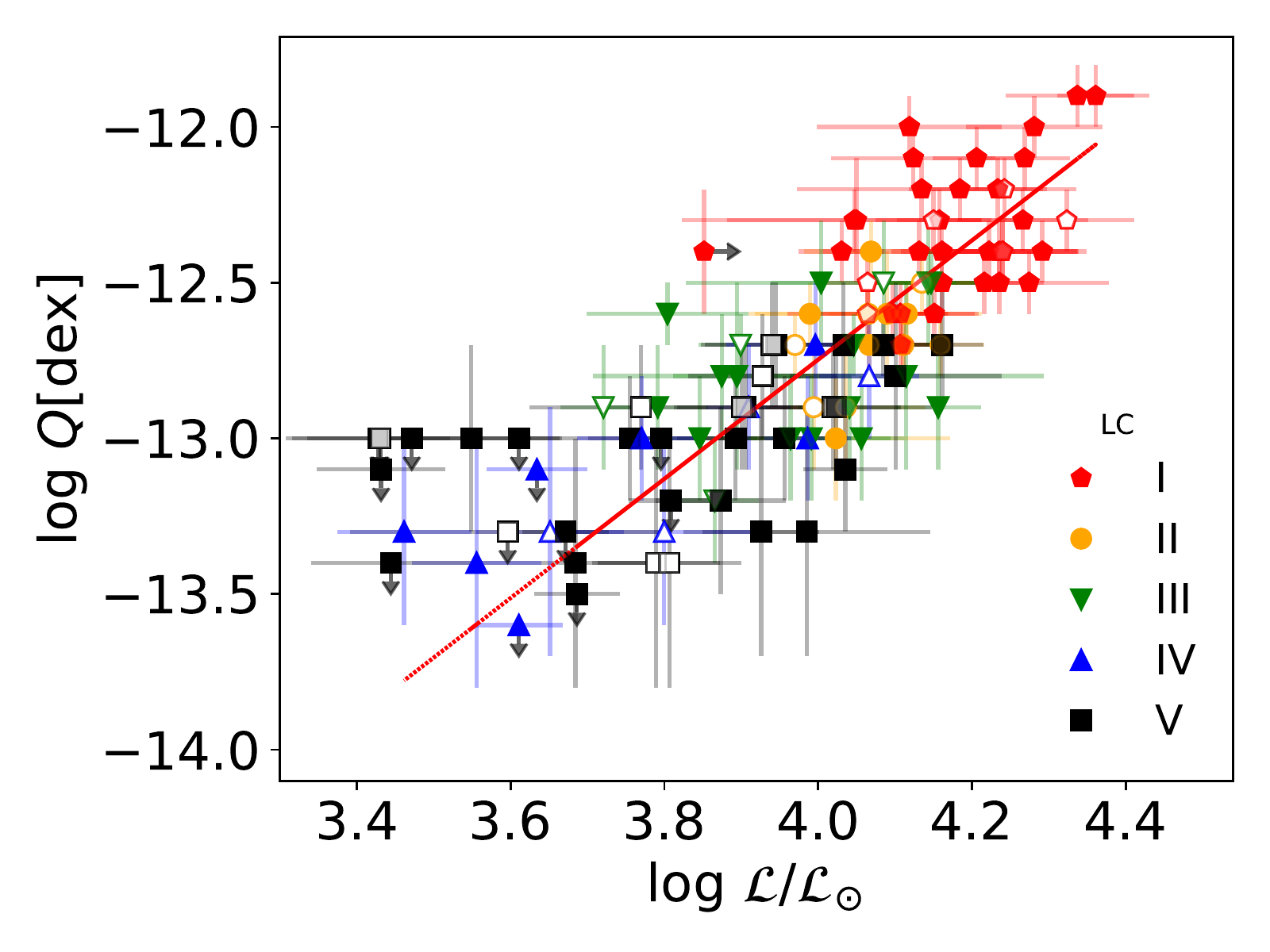}
	\caption{log~$Q$ with respect of parameter $\log\mathcal{L}$, spectroscopic representative of the luminosity, for the O standard stars.  Upper and lower limits are represented by arrows. Different colors and shapes represent luminosity classes. Open symbols indicate stars for which we have detected clear or likely signatures of spectroscopic binarity. The solid line represents the linear regression of the values, excluding stars marked with an open symbol and/or an arrow.
	}
	\label{logQ_Lum}
\end{figure}

The bottom panel in Fig.~\ref{He_micro_Q} depicts the distribution of the wind-strength $Q$-parameter in the sHR diagram. Inspection of this diagram indicates a positive correlation between log$Q$ and log$\mathcal{L}$ which is further confirmed in Fig.~\ref{logQ_Lum}. We note the use of a different color coding in each of the two related figures. While in the bottom panel of Fig.~\ref{He_micro_Q} symbols are colored following various ranges of increasing log~$Q$, the different colors are used in Fig.~\ref{logQ_Lum} to separate the various luminosity classes. As in previous figures, open symbols correspond to stars for which we have identified clear or likely signatures of spectroscopic binarity. In addition, downward arrows indicate stars for which {\sc iacob-gbat} only provides upper limits for the value of log~$Q$. 
As expected, these mainly refer to stars with log$\mathcal{L/L_{\odot}}\lesssim$3.8 (i.e., the late O dwarfs). Lastly, we note that the error bars associated with log~$Q$ shown in Fig.~\ref{logQ_Lum} only reflect the uncertainties in this parameter obtained during the {\sc iacob-gbat} fitting process. Other sources of uncertainties resulting from the discrepancies between the terminal velocities assumed in the grid of {\fastwind} models coupled with {\sc iacob-gbat} and the actual values of this parameter for each analyzed target (see notes in Appendix~\ref{vinfAppend}) are not considered.

In the following, we show how the diagram presented in Fig.~\ref{logQ_Lum} can be used as a distance-independent test of the wind-momentum luminosity relationship \citep[WLR,][and references therein]{Kudritzki00} in the O star domain. The theory of radiatively-driven winds predicts that the modified stellar wind momentum ($D_{\rm mom}=\dot{M}v_{\infty}R^{0.5}$) depends directly on luminosity through the WLR, defined as 
\begin{equation}\label{WLR}
 \log{D_{\rm mom}} = x \log{L/L_{\odot}} + \log{D_{\rm 0}} \\
\end{equation}
where $x$ is defined as 1/$\alpha'$ being $\alpha'=\alpha-\delta$\footnote{$\alpha'$ corresponds to the slope of the line-strength distribution
function ($\alpha$), corrected for ionization effects ($\delta$) \citep{Puls96}.}, and $D_{0}$ is a constant if the considered objects have the same metallicity and a similar effective number of driving lines, which, e.g., is roughly valid for O-stars as discussed here \citep[][and references therein]{Puls96}\footnote{Indeed, Eq.\ref{WLR} would contain an additional mass dependence if \textit{x} was quite different from 3/2.}.

Now, by considering Eq. \ref{WLR} and the definition of $D_{\rm mom}$, $Q$, and $\mathcal{L}$, we can obtain the following expression (see Appendix~\ref{appendixLinear})

\begin{equation}
\log{Q}=x\log{\mathcal{L}}+\frac{3}{4}\log{\frac{M}{R}} + f(x,M) 
\label{EqlogQ}
\end{equation}
where $f(x,M)$ depends on $x$, $M$, and several constants, and its mass
dependence vanishes if $x \approx 2$

Since $M/R$ is not changing too much over the considered region of the HR diagram -- log($M/R$) varies between 0.2 and 0.8 in the Geneva non-rotating evolutionary tracks for solar metallicity --, a roughly linear correlation between log~$Q$ and $\log\mathcal{L}$ is expected, with $x$ being the slope. We remark, however, that this is not as exact as the WLR, since the term $M/R$ is indeed varying.

We then applied a method of orthogonal least squares to perform a linear regression of the data presented in Fig.~\ref{logQ_Lum}. We took into account errors in both axes individually but not their correlation, as that treatment is not simple and we do not expect a major effect. We exclude from the regression stars with clear or likely signatures of spectroscopic binarity, and also those stars for which only upper or lower limits in log\,$Q$ could be obtained. We found a slope value of 1.91$\pm$0.25, which is in fairly good agreement with the range of values of $x$ (=\,1.51\,--\,2.18) provided by \citet{Herrero02}, and in particular concordance with \citet{Mokiem07} ($x$ =~1.86$\pm$0.20 without considering clumping correction). Taking into account uncertainties, our value of $x$ is in agreement with studies providing higher values of $x$, hence lower values of $\alpha'=1/x$. Regarding the current paradigm, the acceleration arising from optically thick and from all lines is very similar to the theoretically predicted value ($x=1.826$) in \cite{Vink00}.

\subsection{The exponent of the wind-velocity law ($\beta$)}

As indicated by \cite{Puls96}, the exponent of the wind-velocity law is a crucial parameter for the determination of mass-loss rates. We hence decided to explore the possibility of obtaining information about this parameter, together with the rest of free parameters, by allowing $\beta$ to vary during the fitting process.

Even accounting for the optimal quality of the compiled spectroscopic dataset (in terms of resolution and S/N), we have found that the {\sc iacob-gbat} analysis is not able to provide reliable constraints on this parameter (at least within the considered range of values, i.e., 0.8\,--1.2). The resulting reduced $\chi^2$ distributions for this parameter are degenerate in $\sim$40\% of the stars in the sample, and only rough upper/lower limits could be found in another $\sim$50\%. In addition, despite only a lower limit for $\beta$ could be obtained, we see some hints of supergiants been dominated by cases in which the best fitting model has an associated value of $\beta$\,=\,1.2.

	
\begin{figure}
\includegraphics[width=0.49\textwidth]{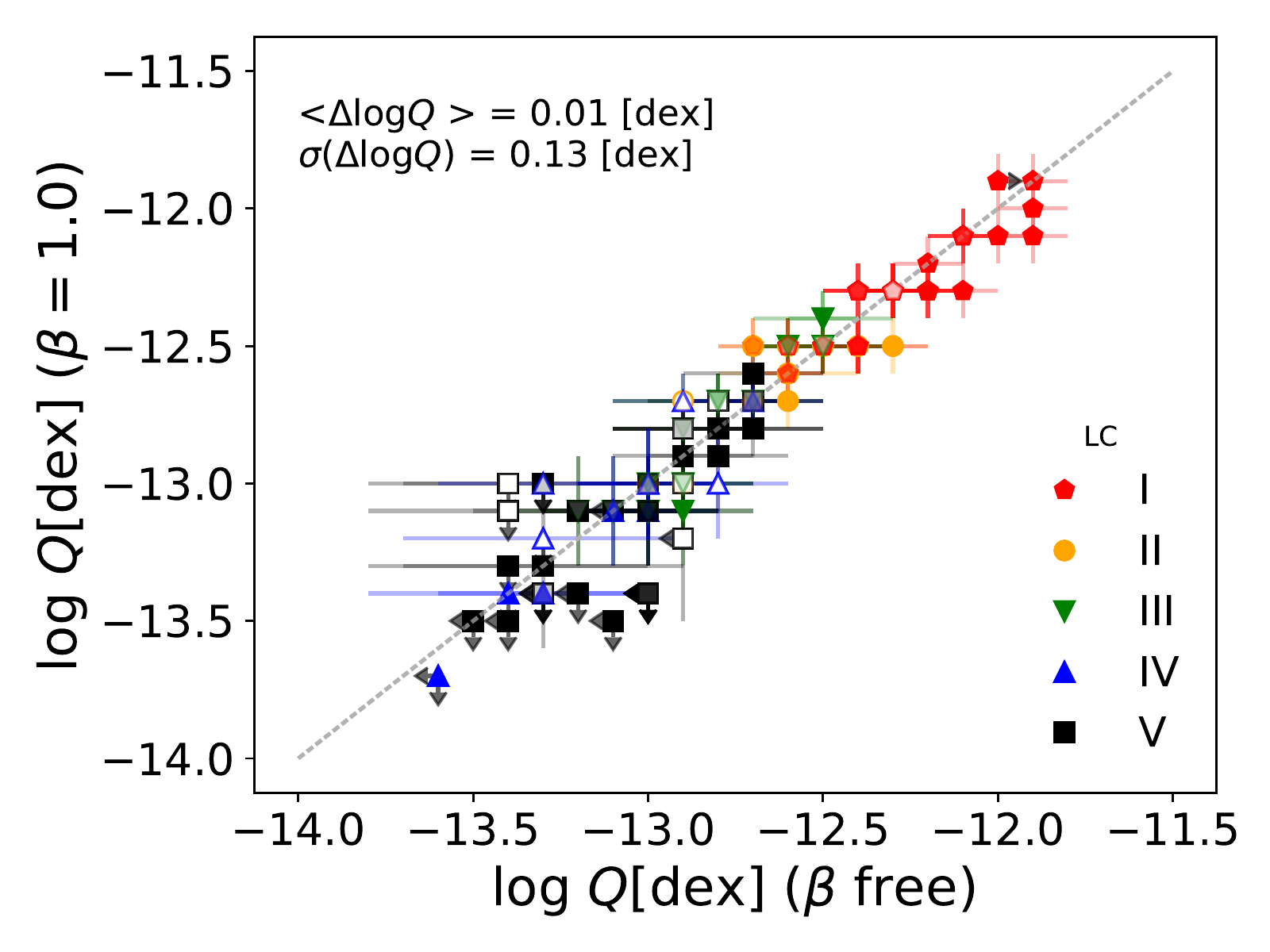}
\caption{Comparison of log$Q$ values  derived from the {\sc iacob-gbat} analysis leaving $\beta$ as a free parameter (x-axis), and fixing this parameter to 1 (y-axis). The figure includes results for all stars in the sample. Different colors and symbols indicate different luminosity classes. Open symbols indicate stars for which we have detected clear or likely signatures of spectroscopic binarity. The dashed line is the 1:1 relation, and we show the mean and standard deviation of differences.}
\label{logQ_comparison}
\end{figure}
	

In view of the difficulty of constraining this parameter in more than 90\% of the sample, we decided to additionally investigate what is the effect of fixing $\beta$ to a given value. We focus on the effect on the central values and associated uncertainties obtained for the Q-parameter in that case. To this end, we launched the {\em Step 4} of {\sc iacob-gbat} (see Appendix~\ref{ErrorAppend}) for all stars, but fixing $\beta$ to 1, and compared the results with those obtained when leaving $\beta$ as a free parameter. Results are summarized in Fig.~\ref{logQ_comparison}, where we separate stars by luminosity class using different symbols and colors. On the one hand, as expected, the uncertainties provided by {\sc iacob-gbat} for the Q-parameter are smaller when $\beta$ is fixed. The main reason is because in the latter case, the degeneracy between $\beta$ and log~$Q$ is broken. On the other hand, while a certain dispersion ($\sigma$\,=\,0.13 dex) in the difference of values associated with the best fitting model is found -- in agreement with previous findings by \cite{Puls96} for the case of stars with thin winds --, there is not a clear correlation between positive and negative differences in the derived log$Q$ values and luminosity class or wind strength.

Therefore, fixing $\beta$ to 1 seems a reasonable assumption when analyzing large samples of optical spectra of O-type stars including targets covering all luminosity classes. In this case, however, an additional uncertainty of $\sim$\,0.15\,--\,0.20 dex must be added to the formal uncertainties resulting from the fitting process.

\subsection{A final note on the stars labeled as Q3. Could wind clumping solve the problem?}\label{clumping}

In Sect.~\ref{section41}, we introduced stars labeled as Q3 as those in which a simultaneous fit to H$\alpha$ and the \ion{He}{ii}\,$\lambda$4686 could not be achieved with any synthetic spectrum generated from our grid of unclumped {\sc fastwind} models. Since these two lines are the main spectroscopic features in the optical spectra of O-type stars providing information about the wind-strength Q-parameter, we stressed that the values of log~$Q$ indicated in Table~\ref{ParamStandards} should be treated with caution (differences up to 1 dex in log~$Q$ can be found in some extreme cases depending on whether we rely on H$\alpha$ or \ion{He}{ii}\,$\lambda$4686). In this section, we further evaluate the possibility that the situation found in those stars labeled as Q3 was actually linked to limitations of our spectroscopic analysis using unclumped models.

 The main effect of introducing (optically thin) clumping in the modeling of the stellar wind is a global reduction of the derived values of the mass loss rate ($\dot{M}$) by a factor $\sqrt[]{f_{cl}}$ \citep[see, e.g.,][]{Repolust04,Puls06,Puls08,Sundqvist14}. Concerning the outcome of a quantitative spectroscopic analysis based on the H and He lines in the optical regime, in most cases, the inclusion of clumping in the models only affects the derived value of log~$Q$, but the overall fit-quality barely changes (at least if one assumes a clumping factor that is spatially constant in the wind-line forming region). However, there are also some situations in which H$\alpha$ and \ion{He}{ii}\,$\lambda$4686 react differently to clumping. This may help to solve the problem indicated above.

In the O star domain this was illustrated, for example, in the study of central stars of planetary nebulae with effective temperatures similar to the main sequence O-type stars by \cite{Kudritzki06} and \cite{Urbaneja08}.
They found that, as expected from theory, below \Teff~$\approx$~37000 K the effect of clumping on \ion{He}{ii}\,$\lambda$4686 and H$\alpha$ lines start to become different, due to the different dependence of opacity on density: below the indicated threshold in \Teff\, \ion{He}{iii} begins to recombine, and \ion{He}{ii}\,$\lambda$4686 changes from a $\rho^2$-dependent line to a $\rho$-dependent one, while H$\alpha$ preserves its $\rho^2$-dependent character. As a result, in stars presenting a clumpy wind in this domain of stellar parameters, the H$\alpha$ line will require a larger value of log~$Q$ than the \ion{He}{ii}\,$\lambda$4686 line when performing the spectroscopic analysis using unclumped models. 

With this in mind, we now evaluate the hypothesis that the 30 stars in our sample labeled as Q3 are actually showing observational evidence of having a clumpy wind. 

\begin{figure}
	\centering
	\includegraphics[width=\hsize]{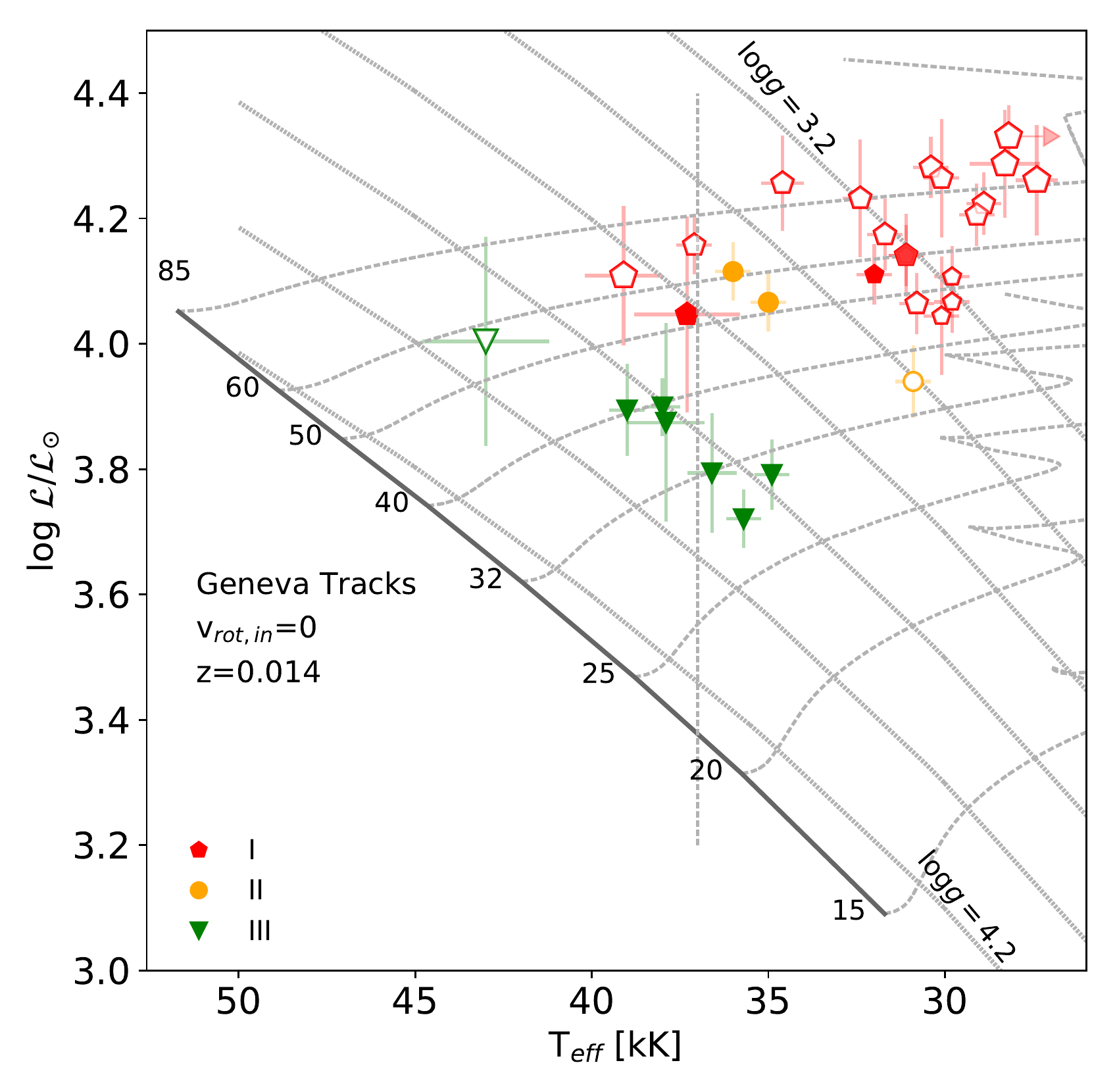}
	\caption{Same as left panel in Fig.~\ref{SPC}, but only for stars with Q3 quality flag. Size depends on the log~$Q$ obtained in the automatic analysis. Open/filled points are stars for which the value of log~$Q$ when fitting only the H$\alpha$ line is larger/smaller than when fitting \ion{He}{ii}\,$\lambda$4686.  The horizontal line at \Teff\,=\,37 kK indicates the boundary of the region where clumping in the wind is expected to affect the two wind diagnostic lines differently (see text).
	}
	\label{TheQ3}
\end{figure}

Figure~\ref{TheQ3} depicts the location of all the stars in this subsample in the sHR diagram. As already shown in Figure~\ref{CompForcQ}, not all of them are consistent with the requirement that H$\alpha$ needs a larger value of log~$Q$ than \ion{He}{ii}\,$\lambda$4686. Indeed, there are 11 stars clearly showing the opposite. We hence separate both subgroups in Fig.~\ref{TheQ3}, using open symbols for those cases in which log~$Q$(H$\alpha$)\,$>$\, log~$Q$(\ion{He}{ii}\,$\lambda$4686), and filled symbols for the opposite situation.

Most of the open symbols are luminosity class I, and they are concentrated in the region (\Teff\,$\lesssim$\,37000~K) where clumping in the wind is expected to affect the two main lines differently. The overall fit for all these stars will hence likely improve when including the effect of wind clumping in the analysis. Interestingly, there are 5 additional stars with luminosity class I and II (filled red and orange symbols) for which {\sc iacob-gbat} provides a larger value of log~$Q$ when fitting only the \ion{He}{ii}\,$\lambda$4686 line. We do not find an obvious explanation at this point, but we notice that all of them have been classified as WVa (variability of H$_{\alpha}$ in absorption).

Concerning the stars with luminosity class III labeled with the Q3 flag (green triangles), it is interesting to note that the analysis of most of them (7 out of the 8) results in a larger value of log~$Q$ when fitting only the \ion{He}{ii}\,$\lambda$4686 (filled symbols). A closer inspection of this subsample of giant stars indicates that all of them are located well within the area of the diagram where dwarfs and sub-giants are found (see also left panel in Fig.~\ref{SPC}). This could mean that these are actually not luminosity class III stars. One possible explanation is that the \ion{He}{ii}~$\lambda$4686 is contaminated by an extra source of circumstellar emission not directly connected with the stellar wind. This may be biasing the spectral classification in that direction. Interestingly, these stars are also characterized by having a low value of the helium abundance (as provided by {\sc iacob-gbat}, see Fig.~\ref{He_micro_Q}). This may be alternatively or complementary indicating that we are actually dealing with composite spectra. Indeed, most of these stars are flagged as visual binaries with similar brightness by \citet{Sota11, Sota14}.

The only giant star labeled as Q3 and showing the contrary effect is HD~168076~AB. We note that its luminosity class has changed from III to IV in the last GOSSS revision \citep{Maiz16}, and it has been depicted as a binary star in \cite{Sana06}. In addition, this star presents a remarkably strong H$\alpha$ line in emission while this is not expected for this combination of SpT and LC (see, e.g. the spectra of HD~93250 and HD~93843 -- O4~III~(fc) and O5~III~(fc), respectively -- in Fig.~\ref{All115}, where H$\alpha$ line is in strong absorption). This resembles the case of, e.g., the O9.7~V star HD~54879 \citep{Castro15}, or the well-known magnetic O-type stars HD~37022 \citep{Stahl93, Wade06} and HD~191612 \citep{Walborn03, Sundqvist12}, whose spectra also show a strong H$\alpha$ emission which has been associated with a strong magnetic field.

Concluding, we have provided strong evidence that wind clumping effects are a likely explanation for the behavior of most of the stars labeled as Q3 with luminosity class I and II. These are targets of potential interest for further investigations of wind clumping. In addition, Q3 giants would require a different explanation, possibly associated with the presence of extra sources of circumstellar emission in \ion{He}{ii}~$\lambda$4686 (as, e.g., the case of those stars flagged as Q2, see Sect.~\ref{section41}) or hidden components. Finally, while the Q3 stars that can be explained by the inclusion of clumping in our models can still be regarded as standards for spectral classification, the status of the targets that exhibit possible signatures of contamination in the \ion{He}{ii}~$\lambda$4686 line as classification standards should be carefully reviewed.

\section{Concluding remarks and future prospects}\label{section6}

This is the first step towards the achievement of a much more ambitious objective, the full empirical characterization of the largest sample of Galactic O-type stars spectroscopically analyzed ever, including more than 300 stars. In this paper, we have presented the methodology we will follow in a series of forthcoming papers regarding the {\sc iacob-gbat/fastwind} analysis of high resolution optical spectra of O stars targeted  by the IACOB and OWN surveys. We will not only concentrate on single-snapshot observations to obtain the spectroscopic parameters but, as we have already started in our present study, we will use all the available multi-epoch observations, also including spectra from other related surveys as CAF\'E-BEANS and the ESO archives, to investigate the spectroscopic variability.

We have concentrated at this point on the results from the quantitative spectroscopic analysis, but our plan is to incorporate information about distances as soon as the \textit{Gaia} collaboration starts delivering more accurate and reliable estimations about parallaxes for our stars. This will allow us to also determine spectroscopic masses, radii and luminosities. We will also determine abundances for the whole sample of likely single and SB1 stars.

Our ultimate objective is to produce an homogeneous and statistically significant empirical overview of physical parameters and abundances of Galactic O-type stars and, eventually, use these results to assess theoretical predictions for the early phases of massive stars evolution, as well as provide empirical constraints
to some of the free parameters considered in these models.

We also plan to perform a more detailed investigation about spectroscopic variability in the O star domain using the observations compiled in the framework of the IACOB project during the last 10 years (following the lines of, e.g., \citeauthor{Fullerton96}\citeyear{Fullerton96}, \citeauthor{Martins15}\citeyear{Martins15}; see also \citeauthor{Simon15}\citeyear{Simon15} for further notes on our plans). Last, these observations will be incorporated to join efforts by the OWN, CAF\'E-BEANS and IACOB projects to characterize the orbital and physical characteristics of those stars identified as spectroscopic binaries.

\begin{acknowledgements}
We are extremely grateful to the referee for the time devoted to read the first version of the paper and the very useful and constructive comments. This research is partially supported by the Spanish Government Ministerio de Econom\'{i}a y Competitivad (MINECO/FEDER) under grants AYA2015-68012-C1-1-P, AYA2015-68012-C2-2-P and SEV2015-0548. Based on observations made with the Nordic Optical Telescope, operated by NOTSA, and the Mercator Telescope, operated by the Flemish Community, both at the Observatorio del Roque de los Muchachos (La Palma, Spain) of the Instituto de Astrof\'isica de Canarias. Based on observations at the European Southern Observatory in programmes 073.D-0609(A), 077.B-0348(A), 079.D-0564(A), 079.D-0564(C), 081.D-2008(A), 081.D-2008(B), 083.D-0589(A), 083.D-0589(B), 086.D-0997(A), 086.D-0997(B), 087.D-0946(A), 089.D-0975(A). Financial support by the Spanish Ministerio de Ciencia e Innovaci\'on under the project AYA2008-06166-C03-01.  J.M.A. acknowledges support from the Spanish Government Ministerio de Econom{\'\i}a y Competitividad (MINECO) through grants AYA2013-40\,611-P and AYA2016-75\,931-C2-2-P. M. Garcia acknowledges funding by the Spanish MINECO via grants FIS2012-39162-C06-01, ESP2013-47809-C3-1-R and ESP2015-65597-C4-1-R. CS-S acknowledges  support  from CONICYT-Chile  through  the  FONDECYT  Postdoctoral Project  No.  3170778. This work would have never been possible without the high competence of the staff working at the NOT and Mercator telescopes in the island of La Palma. 

\end{acknowledgements}



\newpage

\appendix

\section{Some notes on how {\sc iacob-gbat} computes the global $\chi^2$ distribution}\label{ErrorAppend}

An extensive description of the general structure and philosophy of the {\sc iacob} grid-based tool {\sc idl} package ({\sc iacob-gbat}) was presented in \cite{Simon11a}. Further notes on some of the considerations that must be taken into account when applying the tool to real spectra can be found in \cite{SabinSan14}. In this appendix, we explain in detail one update incorporated to {\sc iacob-gbat} after publication of \cite{Simon11a} that was not described in \cite{SabinSan14}. 

For better understanding of the information presented below, we remind the reader that the usage of \textsc{iacob-gbat} is divided in four main steps\footnote{Although the basic structure is the same, we note that there are small changes in what is considered as steps 3 and 4 with respect to what was described in \cite{Simon11a}.}:
\begin{enumerate}
        \item {\em Providing the input information to {\sc iacob-gbat}:} In this first step, the user provides the observed spectrum to be analyzed, as well as all the information required for the spectroscopic analysis. This refers to the resolving power of the spectrum, the line-broadening parameters of the star, the metallicity of the grid of {\sc fastwind} models to be used, the range of values for the various free parameters and, last, the set of H/He lines to be considered. 
	\item {\em Pre-processing of the spectra:} This step allows the user to select the spectral window around the line to be considered in the $\chi^2$ computation, and also includes the possibility to (a) correct the spectrum from radial velocity, (b) locally renormalize the diagnostic lines whenever necessary, and (d) clip part of the line to eliminate nebular lines, blends, and cosmic rays.
	\item {\em Line-by-line $\chi^2$ computation:} In this third step, \textsc{iacob-gbat} computes and stores the individual $\chi^2$ per considered diagnostic line for each combination of the 6 free parameters within the established range per parameter. This step lasts between a few minutes and less than 1 hour depending on the number of lines and the considered range in the free parameters.
	\item {\em Global $\chi^2$ computation and final results:} In this final step, \textsc{iacob-gbat} uses the information stored in the previous step to iteratively compute the global $\chi^2$ distribution and provide estimations for the central values (or upper/lower limits) and associated uncertainties for each of the free parameters of the grid. In addition, if empirical values for the absolute visual magnitude ($M_{\rm v}$) and terminal velocity of the wind (\vinf) are provided, {\sc iacob-gbat} will also compute and provide results for the stellar radius ($R$), luminosity ($L$), spectroscopic mass ($M_{\rm sp}$) and mass loss rate ($\dot{M}$). No more than 1 minute is needed to run this step.
\end{enumerate}

The last three steps can be launched independently, with the only condition that a given step requires previous steps to be performed before. This provides a great versatility to the tool. 

The present version of \textsc{iacob-gbat} (v4.0) includes a total of 37 hydrogen (\ion{H}{i}) and helium (\ion{He}{i} and \ion{He}{ii}) lines located in the optical range (3\,900\,--\,7\,000~\AA) and the J, H, K infrared bands that can be used for the stellar parameter determination. The final set of lines considered for the spectroscopic analysis can be decided by the user depending on the observed wavelength range, as well as other (more subjective) reasons based on, e.g., previous knowledge about the reliability of the available diagnostic lines or cosmic rays affecting the part of the observed spectrum. The tool also includes the possibility to establish an initial set of lines for which the line-by-line $\chi^2$ computation is performed ({\em Step 3}) and then exclude some of these lines for the computation of the global $\chi^2$ distribution and the estimation of the final parameters and uncertainties ({\em Step 4}). This possibility allows the user to evaluate in a few seconds the effect of including/excluding some of the initially considered diagnostic lines on the resulting stellar parameters. Similarly, one can quickly evaluate the effect of fixing some of the investigated parameters to a specific value by just launching {\em Step 4} in {\sc iacob-gbat}.\\

The following complements and supersedes what is described in \cite{Simon11a} regarding steps 3 and 4. The main changes incorporated in {\sc iacob-gbat} since v3.0 refer to (1) an iterative strategy implemented in the last steps of the tool to provide -- in an automatic and objective way -- weights to the diagnostic lines during the computation of the final $\chi^2$ distribution, and (2) how the central values and uncertainties for the 6 free parameters are computed from this final $\chi^2$ distribution.

As indicated above, the tool starts (in {\em Step 3}) computing the quantity $X^2_{L,M}$ for each considered line (L) and every model ($M$) in the subgrid
\begin{equation}
 X^2_{L,M} = \frac{1}{N_{\lambda}}\sum_{\lambda=1}^{N_{\lambda}}\frac{(F_{M,\lambda}-F_{O,\lambda})^2}{\sigma^2_L}
\end{equation}
where $F_{M, \lambda}$ and $F_{O, \lambda}$ are the normalized fluxes corresponding to the synthetic and observed spectrum, respectively; $\sigma_L$\,=\,(S/N)$^{-1}$ account for the S/N of the line; and $N_{\lambda}$ is the number of frequency points in the line. 

All the quantities computed in {\em Step 3} (a total of $N_M$ $\times$ $N_L$ values of $X^2_{L,M}$, where $N_M$ and $N_L$ are the number of models and lines, respectively) are then stored before proceeding with the computation of the global $\chi^2$ distribution. As indicated above, this saves a considerable amount of time in case the effect of using different strategies to combine the individual $X^2_{L,m}$ quantities to obtain this final $\chi^2$ distribution wants to be evaluated. 

Under ideal conditions (e.g., for a perfect model), $X^2_{L,M}$ corresponds to a reduced $\chi^2$. However, as indicated in \cite{Simon11a}, this situation is unlikely reached in the analysis of real spectra. Therefore, the $X^2_{L,M}$ values need to be corrected to account for possible deficiencies in the fitting process.
In its present version, {\sc iacob-gbat} proceed as follows. First, it identifies the model $M(0,L_i)$ which results in the minimum value of $X^2_{L,M}$ for line $L_i$ ($i$\,=\,1, $N_L$). At this point, it is important to stress that the model $M(0,L_i)$ may be different for each of the considered lines. Subsequently, as an initial guess and for each model, {\sc iacob-gbat} computes the quantity

\begin{equation}
 X^2_M (0) = \frac{1}{N_L}\sum_{i=1}^{N_{L}} X^2_{Li,M}\frac{\sigma^2_{Li}}{\sigma^2_{0,Li}}
\end{equation}
where $\sigma_{0,Li}$ is defined as the standard deviation of the residuals of line $L_i$ using the corresponding synthetic line from model $M(0,L_i)$. In this way, the code obtains a first global $\chi^2$ distribution using more realistic estimations of the quantities $\sigma^2_L$ per line (i.e., taking into account other sources of uncertainties in the profile fitting procedure apart from the noise in the observed spectrum).

Then, the code starts an iterative process in which, once the model in the grid resulting in the minimum value of $X^2_M$ is identified -- $M(j)$ --, a new quantity, again for all models,
\begin{equation}
 X^2_M (j) = \frac{1}{N_L}\sum_{i=1}^{N_{L}} X^2_{Li,M}\frac{\sigma^2_{Li}}{\sigma^2_{j,Li}}
\end{equation}
is computed, where now $\sigma_{j,Li}$ is defined as the standard deviation of the residuals of line $L_i$ using the corresponding synthetic line from model $M(j)$. We note that, in this case, the same model in the grid is considered to compute the quantities $\sigma_{j,Li}$. 

The process is repeated until convergence, i.e., when the best fitting model in one iteration is the same as the previous one. This is normally achieved in less than 10 iterations. Following this strategy, a lower weight is automatically and objectively given (in each iteration) to those lines which result in a worse fitting compared to the rest of lines.

Last, once the final $X^2_M$ distribution is obtained, central values and uncertainties (or upper/lower limits) are computed as described in \cite{SabinSan14}.

\section{On the effect of $v_{\infty}$ in the determination of the wind-strength $Q$-parameter}\label{vinfAppend}

\begin{table}[!t]
	\caption{Grid of reference models used for the determination of the effect of \vinf\ in the analysis. In all models $Y_{\rm He}$ and $\beta$ were fixed to 0.10 and 1.0, respectively. } 
	\label{vinf_change_table}
        \centering
        \begin{threeparttable}
            \begin{tabular}{ccccc}
		\hline \hline
        \noalign{\smallskip}
		Model & \Teff & \grav  & log~$Q$ & \vinf  \\
                      & [K]    & [dex] & [dex]   & [\kms] \\             
		\hline
\noalign{\smallskip}
D11 & 35000& 4.0 & 13.0 &  3014  \\
D12 & 35000& 4.0 & 13.5 &  3014  \\
D21 & 40000& 4.0 & 13.0 &  3054  \\
D22 & 40000& 4.0 & 13.5 &  3054  \\
D31 & 45000& 4.0 & 13.0 &  3002  \\
D32 & 45000& 4.0 & 13.5 &  3002  \\	
		\hline
S11 & 30000 & 3.5 & 12.1 &  2428  \\
S12 & 30000 & 3.5 & 12.5 &  2428  \\
S21 & 35000 & 3.5 & 12.1 &  2265  \\
S22 & 35000 & 3.5 & 12.5 &  2265  \\
S31 & 40000 & 3.5 & 12.1 &  1915  \\
S32 & 40000 & 3.5 & 12.5 &  1915  \\
\hline
	\end{tabular}
     \end{threeparttable}
\end{table}

\begin{figure*}[!t]
\centering
\includegraphics[width=0.47\textwidth]{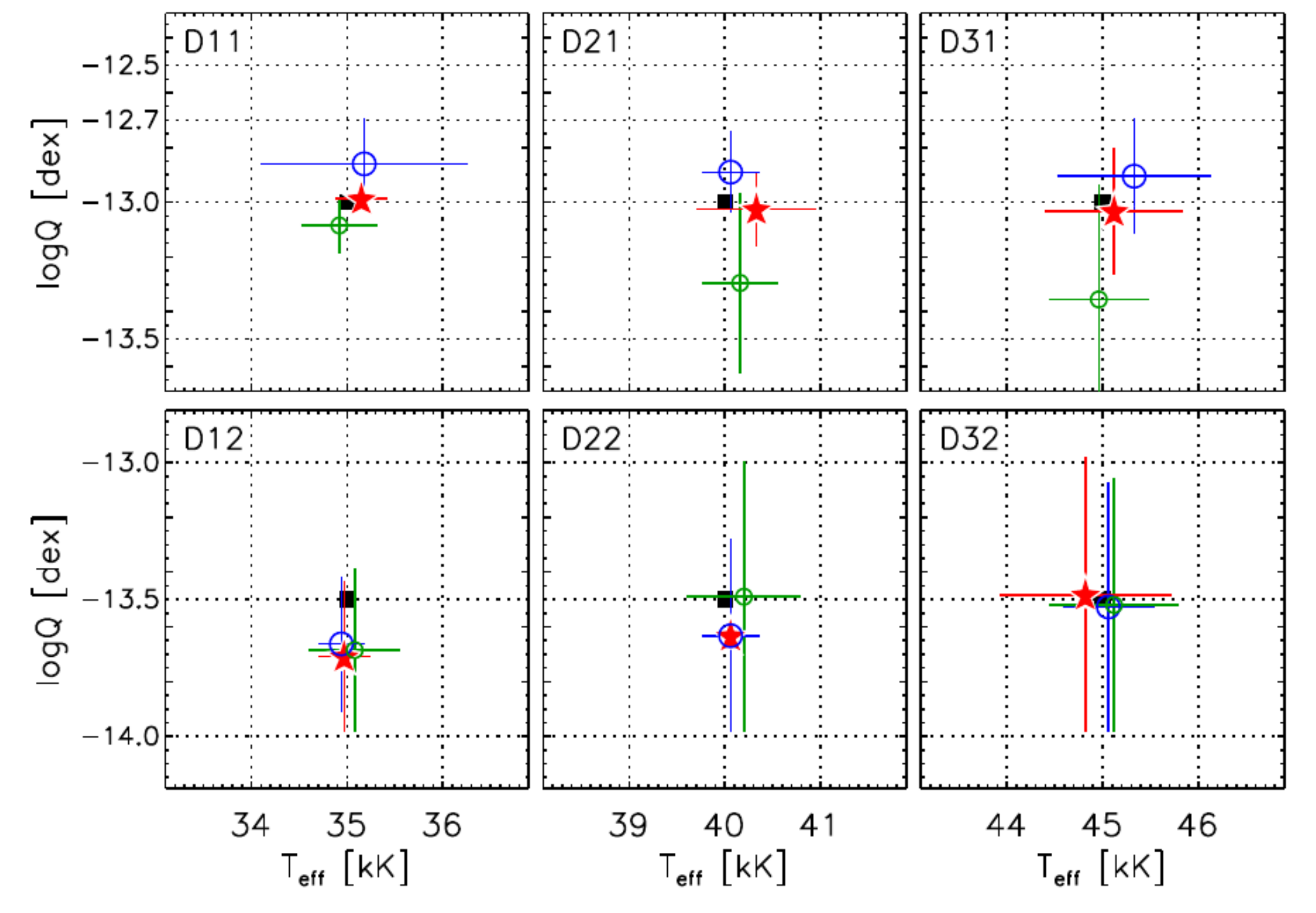}
\includegraphics[width=0.47\textwidth]{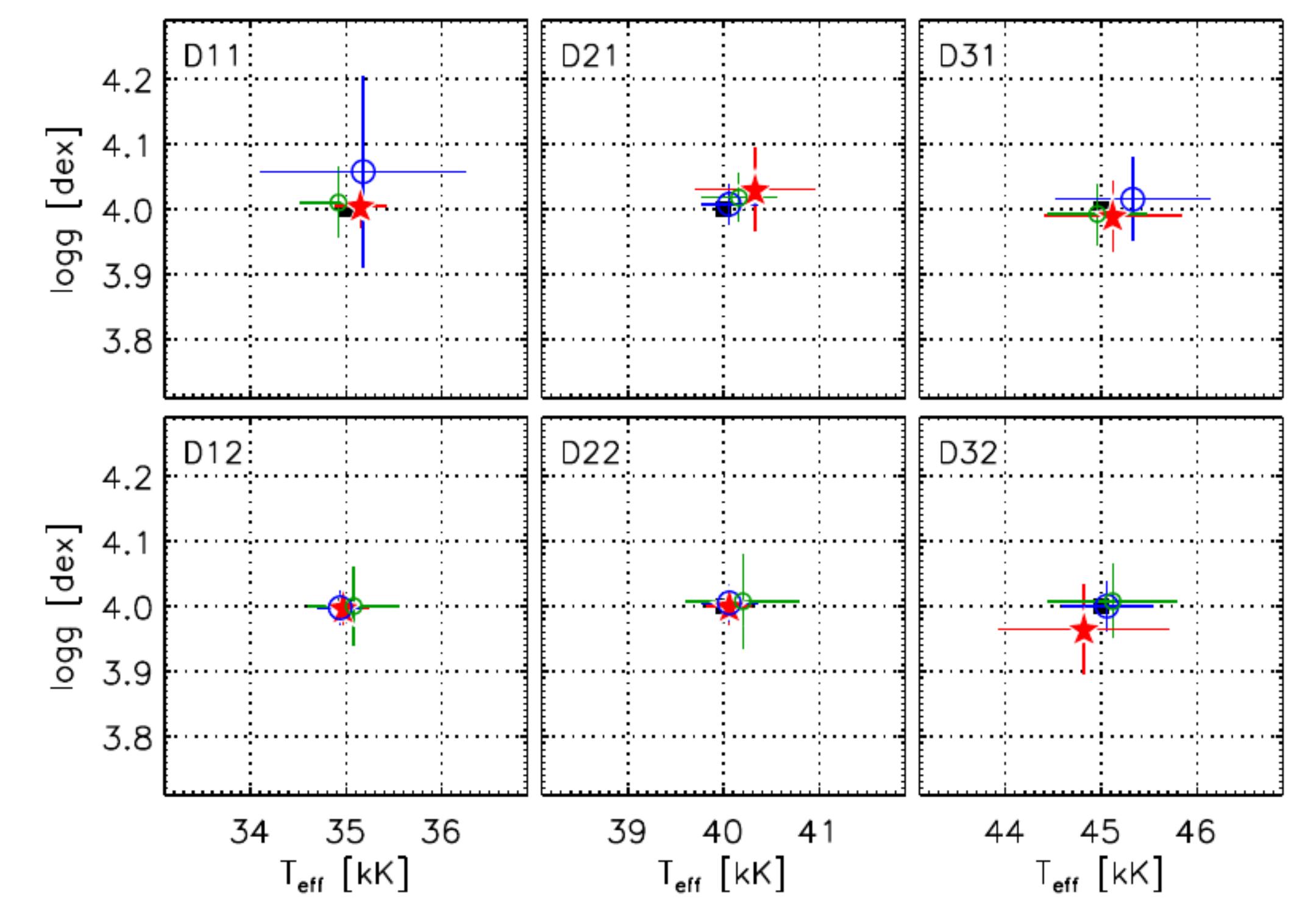}
\includegraphics[width=0.47\textwidth]{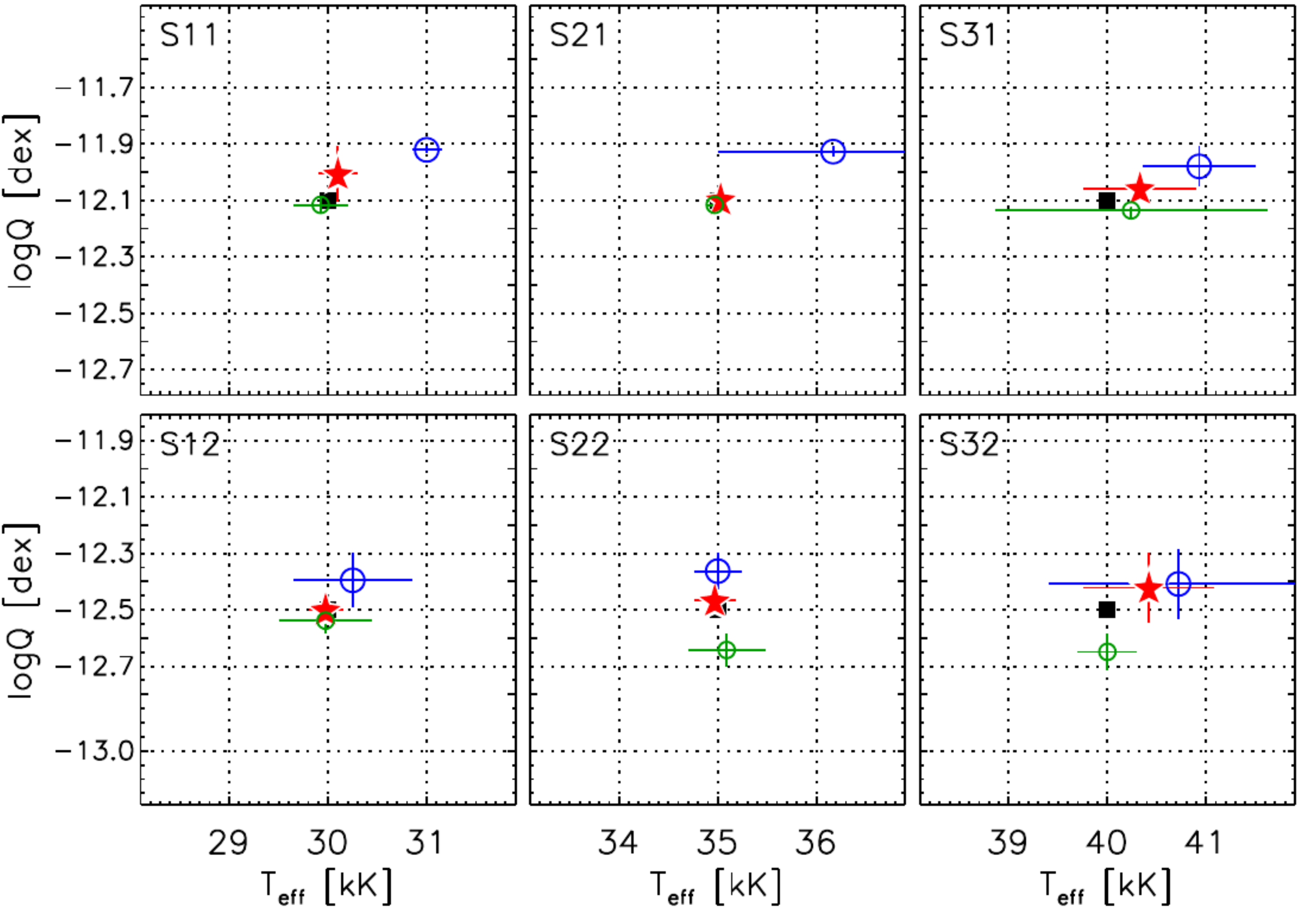}
\includegraphics[width=0.47\textwidth]{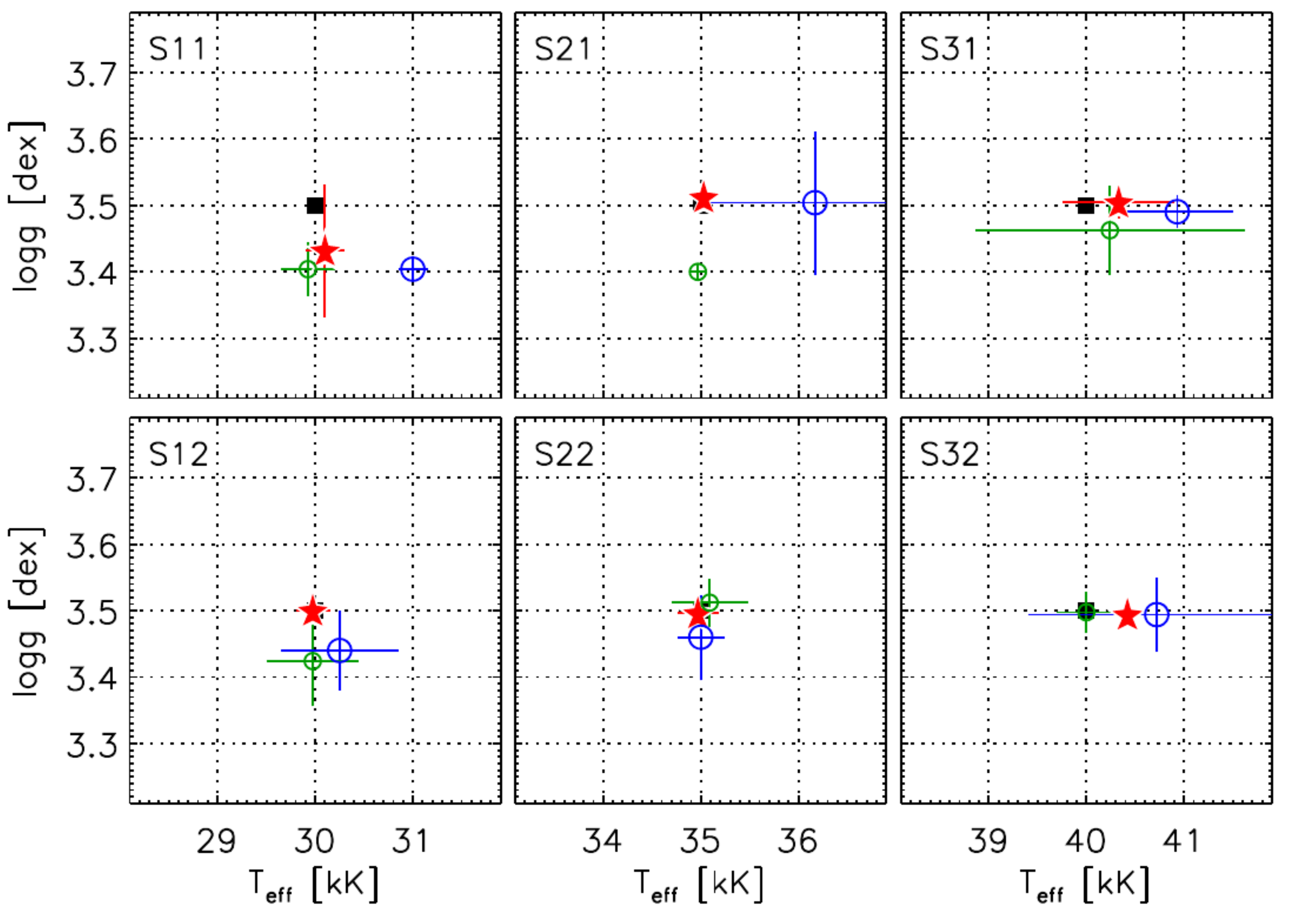}
\caption{ Results of the {\sc iacob-gbat} analysis of mock models presented in Table~\ref{vinf_change_table}. Horizontal and vertical dotted lines indicate the values considered in the grid of \fastwind\ models. Black squares indicate the input parameters of the models. The other three symbols indicate the results of the {\sc iacob-gbat} analysis (central values + associated 1-$\sigma$ uncertainties) for each set of 3 models with the same (\Teff, \grav, log$Q$, Y(He), \micro\, and $\beta$), but different values of \vinf\ (see text for explanation). Namely, red star: same \vinf\ as the one in the grid of \fastwind\ models coupled with {\sc iacob-gbat} (\vinf$_{\rm ,grid}$); large blue circle: \vinf\,=\,1.3~\vinf$_{\rm ,grid}$; small green circle: \vinf\,=\,0.7~\vinf$_{\rm ,grid}$.
}
\label{vinf_test}
\end{figure*}

The {\sc iacob-gbat} tool follows a grid-based analysis strategy in which all the information about the stellar wind properties is condensed in the wind-strength $Q$ parameter, defined as $Q$\,=\,$\dot{M}$/($v_{\infty}R$)$^{1.5}$, which is the optical-depth invariant for $\rho^2$-dependent opacities such as H$\alpha$. As indicated in Table~\ref{gridFAST}, this is one of the 6 free parameters that were considered to build the grid of {\sc fastwind} models coupled with the automatized analysis tool. Details about the specific values of $v_{\infty}$, $R$, and $\dot{M}$ that were considered in each model within the grid are provided in \cite{Simon11a}. Basically, for each pair (\Teff, \grav) in the grid a radius was assumed following the \cite{Martins05} calibrations. This radius and the scaling relation $v_{\infty}\approx$2.65$v_{esc}$\citep{Kudritzki00} were then considered to compute the terminal velocity. Last, the value of $\dot{M}$ was obtained from $Q$, $v_{\infty}$, and $R$, for all the different values of $Q$ indicated in Table~\ref{gridFAST}.

This analysis strategy assumes that the effect on the line profiles of different combinations of $v_{\infty}$, $R$, and $\dot{M}$ producing the same value of $Q$ is small whenever the assumed values of $R$ and $v_{\infty}$ do not depart very much from the actual values. While this has been already tested in the case of $R$ \citep{Puls96}, this statement has not been formally evaluated for the case of the terminal velocity.

By comparing the values of $v_{\infty}$ that would be inferred for the stars in our sample following the strategy mentioned above with those empirically obtained by \cite{Howarth97}, we have found that the mean difference in the ratio between the assumed and measured terminal velocities is 0.95 with an associated standard deviation of 0.25.

To evaluate the effect of a difference of 30\% in terminal velocity on the derived $Q$, we computed a set of \fastwind\ models with fixed stellar parameters (comprising 12 different combinations of the parameters \Teff, \grav\, and log~$Q$, see Table~\ref{vinf_change_table}) and three different combinations of $\dot{M}$ and $v_{\infty}$ providing the same value of $Q$. Basically, apart from the 12 reference models (which adopt the same values of \vinf\ as those in the grid of {\sc fastwind} models coupled with {\sc iacob-gbat}), we computed another 24 models in which the terminal velocity was changed by $\pm$30\% and the mass-loss rates where modified accordingly while keeping the same values of $R$ and log~$Q$ as in the reference models.

All the synthetic spectra associated with these 36 models (convolved with a \vsini\,=\,60~\kms and degraded to a S/N=200) were then analyzed with {\sc iacob-gbat}. The main results regarding \Teff, \grav, and log~$Q$ are summarized in Fig.~\ref{vinf_test}. We basically found that a 30\% difference in $v_{\infty}$ leads to a maximum variation in log~$Q$ of 0.15 dex, where larger values of this parameter are obtained for the models with increased terminal velocity. This uncertainty is on the order of (or a bit larger than) the uncertainty in log~$Q$ associated with the fitting process for values of the wind-strength $Q$-parameter typical of supergiants (log$Q$ > -12.7), but negligible when compared with the accuracy reached in the case of dwarfs, which is $\gtrsim$0.25 dex. The rest of the parameters are barely affected (differences up to 500 K in \Teff\, and 0.05 dex in \grav) except for very specific cases where differences up to 1000 K and 0.1 dex, respectively, have been found. In particular, we remark the very good agreement between the output of the {\sc iacob-gbat} and the input values for those cases in which $v_{\infty}$=$v_{\infty,grid}$. This demonstrate the strength of our automatized analysis strategy.

\section{On the linear dependency between log~$Q$ and log~$\mathcal{L}$}\label{appendixLinear}
 
We start from the wind-momentum luminosity relationship:
\begin{equation}\label{eqsubs1}
\log{D_{\rm mom}}=x\log{L}+D_{0} 
\end{equation}
\newline
By considering the following definitions:\\ \newline
$\mathcal{L}\,:=\,T^4_{eff}/g\,=\,L/4\pi\sigma_{\rm B} G M$ \\  \newline
$D_{\rm mom}\,:=\,\dot{M}v_{\infty}R^{1/2}$ \\  \newline
$Q\,:=\,\dot{M}/(v_{\infty}R)^{3/2}$ \\  \newline
Equation \ref{eqsubs1} can be also expressed as:
\begin{equation}
\log(\dot{M}v_{\infty}R^{1/2})=x\log{\mathcal{L}}+x\log(4\pi\sigma_{\rm B} G)+x\log{M}+D_{0}
\end{equation}
\newline
or, equivalently:
\begin{equation}
\log(Qv^{5/2}_{\infty}R^2)=x\log{\mathcal{L}}+x\log(4\pi\sigma_{\rm B} G)+x\log{M}+D_{0}
\label{eqsubs2}
\end{equation}
\newline
If now we assume the scaling relationship between $v_{\infty}$ and $v_{\rm esc}$:\\ \newline
$v_{\infty}\,=\,\eta v_{\rm esc}$ \\ \newline
with $\eta\sim2.65$ \citep{Kudritzki00} and \\ \newline
$v_{\rm esc}\,=\,(2GM(1-\Gamma)/R)^{1/2}$ \\ \newline
we end up with the following:
\begin{multline}
\log(Q\eta^{5/2}(2G(1-\Gamma))^{5/4} (M/R)^{5/4} R^2)=\\ x\log{\mathcal{L}}+x\log(4\pi\sigma_{\rm B} G)+x\log{M}+D_{0}
\end{multline}
\newline
Regrouping and isolating log~$Q$:
\begin{multline}
\log{Q}=\\
x\log{\mathcal{L}}+\frac{3}{4}\log{\frac{M}{R}}+(x-2)\log{M}-\frac{5}{4}\log(2\eta^{2}G(1-\Gamma))+\\x\log(4\pi\sigma_{\rm B} G)+D_{0}
\label{finaleq}
\end{multline}

Since $x \approx 2$, the term involving $\log M$ is negligible, and when we perform a linear regression to the data as presented in Fig.~\ref{logQ_Lum}, the slope provides us with a rough estimation of the parameter $x$. As explained in the main text, with this approach we are able to obtain the slope of the WLR without the need of information about distances. Finally, we note that, compared to the classic WLR, a larger scatter is expected in this relationship due to the rest of the terms that are not exactly constant but depend on $M$ or $M/R$.

			\section{Tables and figures}		
        \begin{landscape}
        \pagestyle{empty}
		\fontsize{8}{8}\selectfont
		\onecolumn
		{
\setlength\LTleft{-6.5cm}  
		\noindent
\begin{longtable}{llr@{\hskip 0.05in}lcccccccccccc}
	\caption{Complete list of stars considered in this work.}\\
	\multicolumn{16}{l}{\parbox{23cm}{They are separated in two groups.The first part of the table includes stars sorted by HD number when available. The second group comprises those stars that do not have an HD number, sorted alphabetically. Apart from the name(s) of each star (columns 1 and 2), an abridged spectral classification -- following Table 1 in \cite{Maiz15} (columns 3 and 4)--, and the number of available spectra per instrument (N:FIES , M:HERMES, F:FEROS, C:CAF\'E) (columns 5 to 8), we also indicate some specific details about the spectra used for the {\sc iacob-gbat} analysis (columns 9 to 11), and the diagnostic line used for the {\sc iacob-broad} analysis (columns 12 to 14). Last we include information about the radial velocity of the star. In our work this was obtained using the spectra indicated in column 9.}}\\
			\label{tableStandards}\\
			\hline \hline
			\noalign{\smallskip}
			Name          &  Alt. name        & SpT      & LC      &   N             &   M             &   F             &   C             & Spectra Used            & HJD                    & S/N                               & \multicolumn{3}{c}{iacob\_broad}              &  \multicolumn{2}{c}{$v_{\rm rad}$ [\kms]}    \\
            \cmidrule(r){12-14}
            \cmidrule{15-16}
			              &                   &          &         &                 &                 &                 &                 &                         &  -2450000              &  4500 $\AA$                   & Line             &   S/N           &  EW [m$\AA$] &   This work        &  Literat.  \\
			\hline
			\noalign{\smallskip}
			\endfirsthead
			\caption{continued.}\\
			\hline
			\hline
			\noalign{\smallskip}
			Name          &  Alt. name        & SpT      & LC      &   N             &   M             &   F             &   C             & Spectra Used            & HJD                    & S/N                               & \multicolumn{3}{c}{iacob\_broad}              &  \multicolumn{2}{c}{$v_{\rm rad}$ [\kms]}    \\
            \cmidrule(r){12-14}
            \cmidrule{15-16}
			              &                   &          &         &                 &                 &                 &                 &                         &  -2450000              &  4500 $\AA$                   & Line             &   S/N           &  EW [m$\AA$] &   This work        &  Literat.  \\
			\hline
			\noalign{\smallskip}
			\endhead
			\hline
			\noalign{\smallskip}
					\noalign{\smallskip}
		\hline\hline
		\noalign{\smallskip}
		\multicolumn{16}{l}{ Notes: *:Values from \cite{Gontcharov06}, **:Values from \cite{Kharchenko07}}\\
			\endfoot
            HD\,10125   &             ...   & O9.7   &    II   &    3   &   0   &  0   &  0   &     HD10125\_20110115\_232529\_N\_V25000   & 5577.485   & 123                               &    \ion{O}{iii}$\lambda$5592   & 264   & 146      &  -48$\pm$6   & -38.0$\pm$4.5* \\
            HD\,12323   &             ...   & O9.2   &     V   &    2   &   0   &  0   &  0   &     HD12323\_20121225\_225051\_N\_V25000   &  6287.46   & 109                               &    \ion{O}{iii}$\lambda$5592   & 167   &  92      &  -17$\pm$7   & ... \\
            HD\,12993   &             ...   & O6.5   &     V   &    3   &   0   &  0   &  0   &     HD12993\_20110912\_054359\_N\_V25000   & 5816.748   & 125                               &    \ion{O}{iii}$\lambda$5592   & 197   & 175      &  -87$\pm$3   & -101.0$\pm$6.8** \\
      HD\,14633\,AaAb   &             ...   & O8.5   &     V   &    3   &   1   &  0   &  2   &     HD14633\_20091113\_024720\_N\_V46000   & 5148.632   & 187                               &    \ion{O}{iii}$\lambda$5592   & 272   & 131      &  -34$\pm$5   & -36.8$\pm$0.9* \\
            HD\,14947   &             ...   & O4.5   & Iab/I   &    3   &   0   &  0   &  0   &     HD14947\_20081108\_231044\_N\_V46000   &  4779.48   & 133                               &    \ion{O}{iii}$\lambda$5592   & 281   &  76      &  -44$\pm$1   & -56.3$\pm$0.9* \\
            HD\,15570   &             ...   &   O4   & Iab/I   &    3   &   0   &  0   &  0   &     HD15570\_20081107\_004909\_N\_V46000   & 4777.549   &  95                               &  \ion{N}{v}$\lambda$4603   & 158   &  95          &  -41$\pm$9                              & -24.0$\pm$0.9* \\
            HD\,15629   &             ...   & O4.5   &     V   &    3   &   0   &  8   &  0   &     HD15629\_20110913\_020206\_N\_V25000   & 5817.593   & 123                               &    \ion{O}{iii}$\lambda$5592   & 254   & 153      &  -68$\pm$1   & -55.0$\pm$9.4** \\
\end{longtable}
}
\newpage

{
	\fontsize{8}{7}\selectfont
	\setlength\LTleft{-5.5cm}
	\setlength\LTright{-2cm}
	\begin{longtable}{lr@{\hskip 0.05in}l@{\hskip 0.05in}cccccccclll}
		\caption{Spectroscopic parameters for the complete sample of stars. The last columns of the table also include information about the quality of the best fitting model resulting from the {\sc iacob-gbat/fastwind} analysis and about detected spectroscopic binarity/variability.}\\
		\multicolumn{14}{l}{\parbox{21.3cm}{The stars are separated in two groups. The first part of the table includes those stars for which the analysis has been successful and the fit can be considered reliable. The second part includes stars with either a bad fit, quality flags Q4 (See Sect.~\ref{subsectionGbat}), stars with SB2 signs, or stars without spectra available. Columns include: Name, spectral type and luminosity classification, projected rotational velocity and macroturbulence used in the analysis, effective temperature, gravity, rotational corrected gravity, helium abundance, microturbulence and wind strength parameter, derived from spectra. Errors are the maximum between half of the step in the grid and the formal error from the \textsc{iacob-gbat} results. As noted in Appendix~\ref{vinfAppend} the log$Q$ is affected by an additional uncertainty of up to 0.15 dex, due to the adopted \vinf scaling, that it is not included here. Finally, this table provides the quality flag (Sect.~\ref{subsectionGbat}), and multi-epoch annotation from this work (Sect.~\ref{subsectionVar}) and the OWN project (priv comm.).}}\\
		\label{ParamStandards}\\
		\hline \hline
		\noalign{\smallskip}
		Name &  SpT & LC   & \vsini & \vmacro & \Teff & \grav & \gravt  & $Y(He)$ & $\xi_{t}$&  log\,$Q$ & qual. & Notes  & OWN Notes\\
		&      &      & [\kms] & [\kms]  &  [kK] & [dex] & [dex]   & [dex]   & [\kms]   &  [dex]    &  flag   &        &          \\
		\hline
		\noalign{\smallskip}
		\endfirsthead
		\caption{continued.}\\
		\hline \hline
		\noalign{\smallskip}
		Name &  SpT & LC   & \vsini & \vmacro & \Teff & \grav & \gravt  & $Y(He)$ & $\xi_{t}$&  log\,$Q$ & qual. & Notes  & OWN Notes\\
		&      &      & [\kms] & [\kms]  &  [kK] & [dex] & [dex]   & [dex]   & [\kms]   &  [dex]    &  flag   &        &          \\
		\hline
		\noalign{\smallskip}
		\endhead
		\noalign{\smallskip}
		\hline\hline
		\noalign{\smallskip}
		\multicolumn{14}{l}{ Notes: C:Constant, LPV:Line profile variability, WV:Wind variable.  $g_{true}=g+g_{cent}$, $g_{cent}\approx\frac{(V_{rot}\sin i)^2}{R_{*}} $}\\
		\endfoot
		HD\,190429\,A &    O4 &          I\,f &  90 & 113 & 36.6$\pm$0.8 & 3.52$\pm$0.10 & 3.53$\pm$0.09 & 0.17$\pm$0.04 &      $<$10 &  -12.1$\pm$0.1 & Q1 & WVe,MD2 &    . \\
		HD\,14947 &  O4.5 &          I\,f & 114 &  22 & 39.1$\pm$1.1 & 3.64$\pm$0.11 & 3.65$\pm$0.10 & 0.15$\pm$0.04 & 5.0-19.9 &  -12.0$\pm$0.1 & Q3 & LPV,WVe &    . \\
		CPD\,-47\,2963\,AB &    O5 &         I\,fc &  67 & 110 & 37.1$\pm$0.5 & 3.51$\pm$0.05 & 3.51$\pm$0.04 & 0.11$\pm$0.03 &       $>$8 &  -12.3$\pm$0.1 & Q3 &     LPV &    . \\
		CYG\,OB2-11 &  O5.5 &         I\,fc &  75 &  85 & 37.3$\pm$1.5 & 3.63$\pm$0.15 & 3.63$\pm$0.14 & 0.11$\pm$0.03 & 5.0-19.9 &  -12.3$\pm$0.1 & Q3 &       C &    . \\
		HD\,169582 &    O6 &         Ia\,f &  66 &  97 & 38.9$\pm$1.3 & 3.70$\pm$0.22 & 3.70$\pm$0.21 & 0.23$\pm$0.07 &      $>$13 &  -12.3$\pm$0.2 & Q1 &     MD1 &  VAR \\
		HD\,163758 &  O6.5 &        Ia\,fp &  76 &  87 & 34.6$\pm$0.6 & 3.28$\pm$0.08 & 3.29$\pm$0.07 & 0.20$\pm$0.03 &      $>$15 &  -12.3$\pm$0.1 & Q3 & LPV,WVe &  VAR \\
		HD\,157857 &  O6.5 &       II\,(f) & 114 &  69 & 36.7$\pm$0.7 & 3.58$\pm$0.08 & 3.60$\pm$0.08 & 0.13$\pm$0.03 &      $>$15 &  -12.4$\pm$0.1 & Q1 &     WVa & VAR? \\
		HD\,69464 &    O7 &       Ib\,(f) &  73 & 105 & 35.8$\pm$1.0 & 3.37$\pm$0.09 & 3.38$\pm$0.08 & 0.16$\pm$0.03 &      $>$16 &  -12.4$\pm$0.1 & Q2 & LPV,WVa &  SB1 \\
\end{longtable}
}

\end{landscape}

	\newpage
	
	{
		\fontsize{8}{7}\selectfont
		\begin{longtable}{lr@{\hskip 0.05in}lccccccc}\\
			\caption{Spectroscopic parameters for the sample of stars labeled as Q3 (H$\alpha$ and 
				\ioni{He\,\sc{ii}$\lambda$4686}{ii} not fitting at the same time). }\\
			\multicolumn{9}{l}{\parbox{14cm}{For each star we include three lines. The first one refers to the parameters obtained from the conventional analysis, i.e., fitting all diagnostic H/He lines. The other two lines are the parameters obtained by forcing to fit either H$\alpha$ or He\,{\sc ii}$\lambda$4686, respectively, and still fitting the other diagnostic lines at the same time. Columns include: Name, spectral type and luminosity classification, multi-epoch annotation, line forced in the fitting, effective temperature, gravity, wind strength parameter (without uncertainties for the forced cases), helium abundance, and microturbulence.
}}\\
			\label{ParamStandards3}\\
			\hline \hline
			\noalign{\smallskip}
			Name         &  SpT & LC   & Notes   & mass-loss    & \Teff & \grav   &  log\,$Q$ & $Y(He)$  & $\xi_{t}$  \\
			             &      &      &  &  diagnostic              &  [kK] & [dex]   &  [dex]    &  [dex]   &   [\kms]    \\
			\hline
			\noalign{\smallskip}
			\endfirsthead
			\caption{continued.}\\
			\hline \hline
			\noalign{\smallskip}
			Name         &  SpT & LC   & Notes   & mass-loss    & \Teff & \grav   &  log\,$Q$ & $Y(He)$  & $\xi_{t}$  \\
			             &      &      &  &  diagnostic              &  [kK] & [dex]   &  [dex]    &  [dex]   &   [\kms]    \\
			\hline
			\noalign{\smallskip}
			\endhead
			\noalign{\smallskip}
			\hline\hline
			\noalign{\smallskip}
			\endfoot
          HD\,10125 & O9.7 &         II &     SB1 & both lines & 30.9$\pm$0.5 & 3.38$\pm$0.06 & -12.7$\pm$0.1 & 0.11$\pm$0.03 &      $>$16 \\
                    &      &            &         &        He\,\sc{ii}$\lambda$4686 &        $<$30.4 &         $<$3.23 &         -13.5 &         $>$0.10 & 11$\pm$4 \\
                    &      &            &         &        H$\alpha$ &        $<$30.4 &         $<$3.26 &         -13.0 &         $>$0.09 & 13$\pm$3 \\
                    \noalign{\smallskip}
                    HD\,14947 & O4.5 &   Iab/I\,f & LPV,WVe & both lines & 39.1$\pm$1.1 & 3.64$\pm$0.11 & -12.0$\pm$0.1 & 0.15$\pm$0.04 & 5.0-19.9 \\
                    &      &            &         &        He\,\sc{ii}$\lambda$4686 & 39.0$\pm$1.0 & 3.64$\pm$0.11 &         -12.1 & 0.15$\pm$0.04 & 5.0-19.9 \\
                    &      &            &         &        H$\alpha$ & 39.5$\pm$0.9 & 3.65$\pm$0.09 &         -11.9 & 0.15$\pm$0.03 &      $<$16 \\
          \noalign{\smallskip}
          HD\,47432 & O9.7 &         Ib & LPV,WVe & both lines & 29.1$\pm$0.5 & 3.01$\pm$0.05 & -12.5$\pm$0.1 & 0.10$\pm$0.03 &      $>$19 \\
                    &      &            &         &        He\,\sc{ii}$\lambda$4686 & 28.1$\pm$0.5 &         $<$2.84 &         -13.5 &         $>$0.13 & 10$\pm$3 \\
\end{longtable}
	}

\newpage


	\begin{figure}
		\centering
		\includegraphics[width=0.49\textwidth]{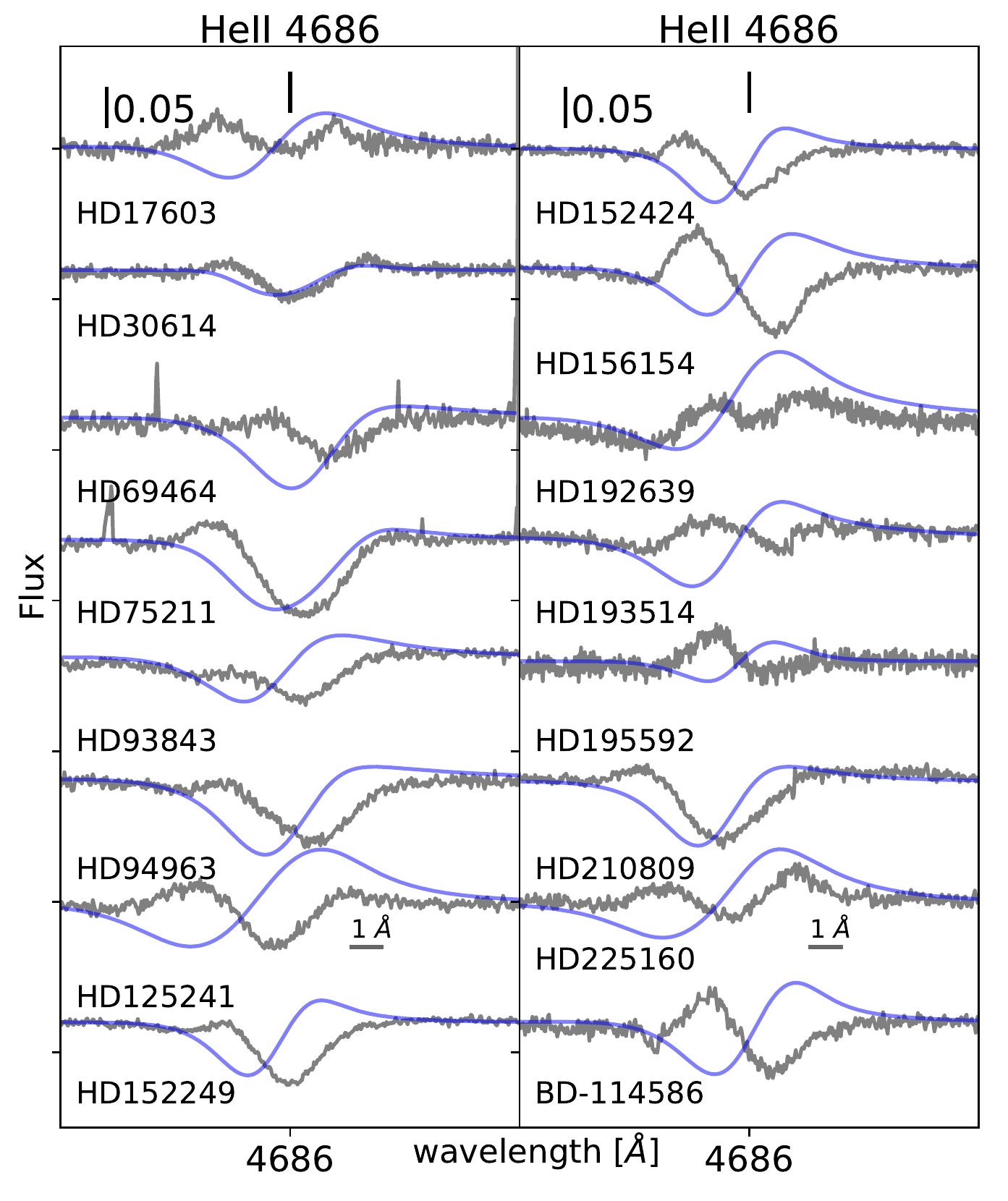}
		\caption{Comparison between observed spectra (gray) and the best fitting \fastwind\ model (blue) for the 16 stars cataloged as Q2 (\ioni{He}{ii}\,$\lambda$4686 in P-Cygni or double peak). No model was able to reproduce these features.}
		\label{ExamplesQ2}
	\end{figure}

	\begin{figure}
		\centering
		\includegraphics[width=0.49\textwidth]{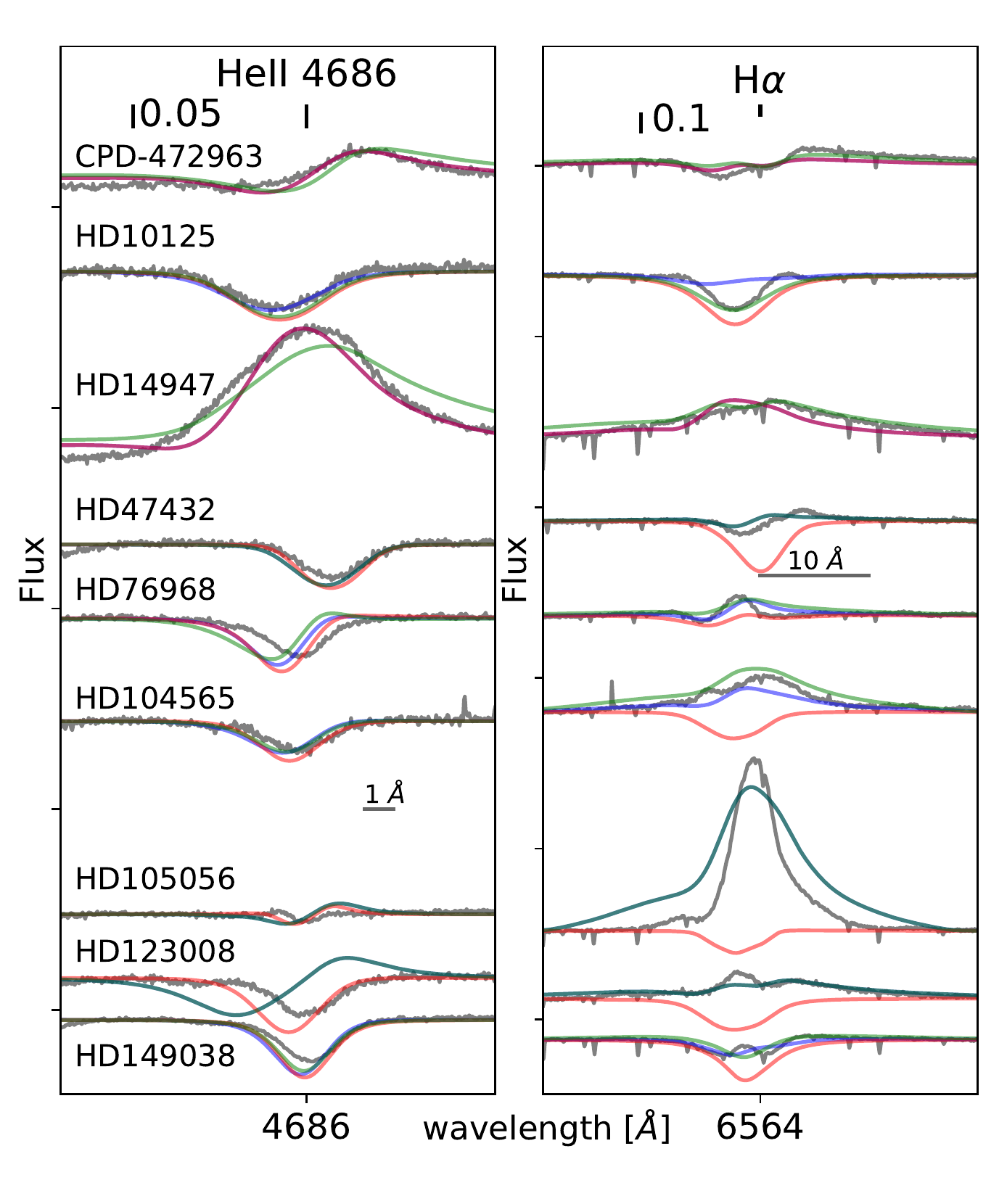}
		\caption{Comparison between observed (gray) and synthetic (blue,green,red) spectra associated with the 3 sets of parameters indicated in table~\ref{ParamStandards3}. In this figure, we concentrate on stars labeled as Q3, with luminosity class I or II, and where the fitting of H$\alpha$ implies a higher value of log~$Q$ compared to \ion{He}{ii}\,$\lambda$4686. Blue lines correspond to the {\sc iacob-gbat} best fitting models obtained by considering the full set of H/He lines. Green/red lines correspond to the higher/lower log~$Q$ values indicated in table~\ref{ParamStandards3} for each star}.
		\label{ExamplesQ3_1}
	\end{figure}
	
	\begin{figure}
		\centering
		\includegraphics[width=0.49\textwidth]{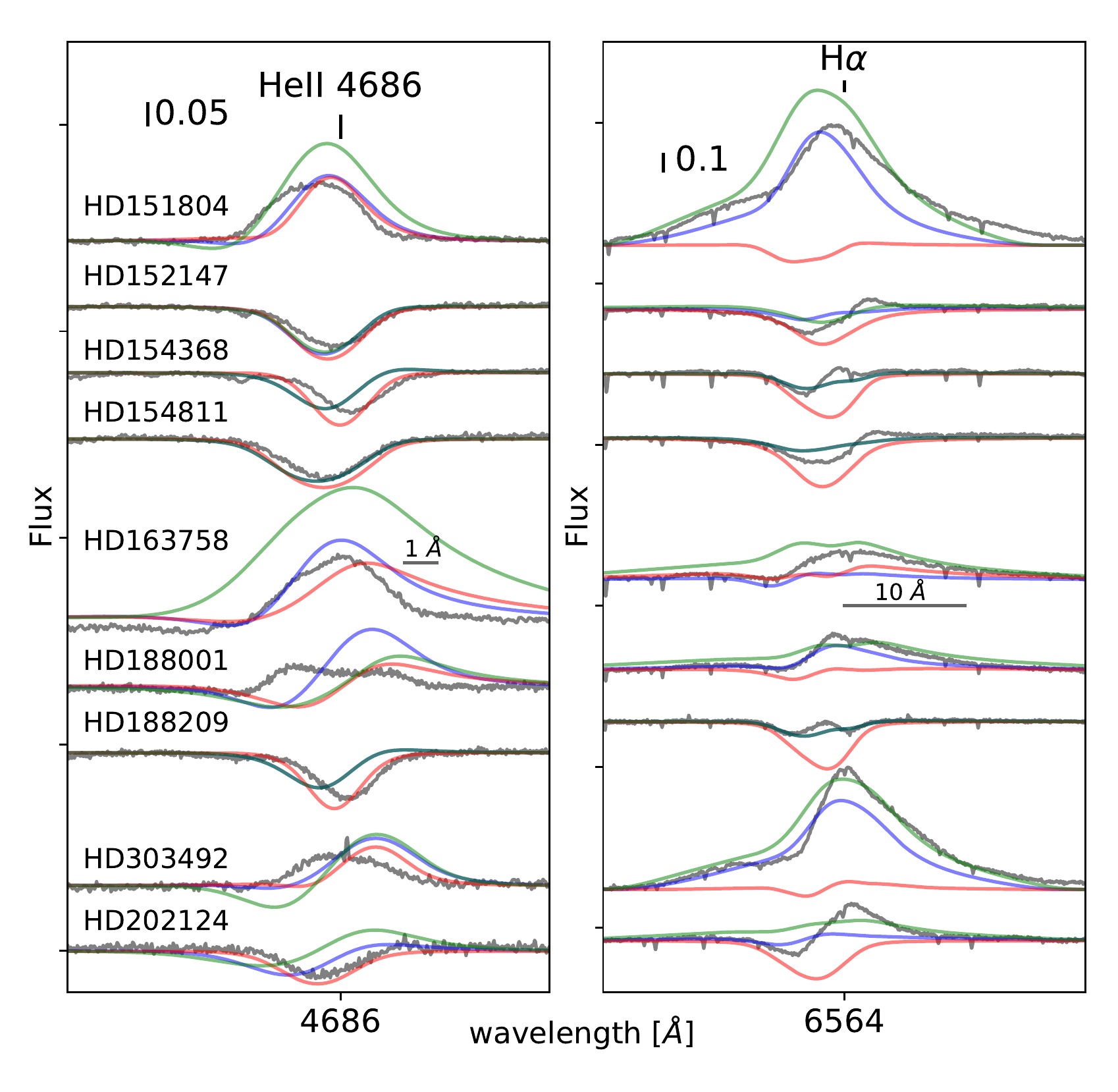}
		\caption{Continuation of Fig.~\ref{ExamplesQ3_1}.}
		\label{ExamplesQ3_2}
	\end{figure}

	\begin{figure}
		\centering
		\includegraphics[width=0.49\textwidth]{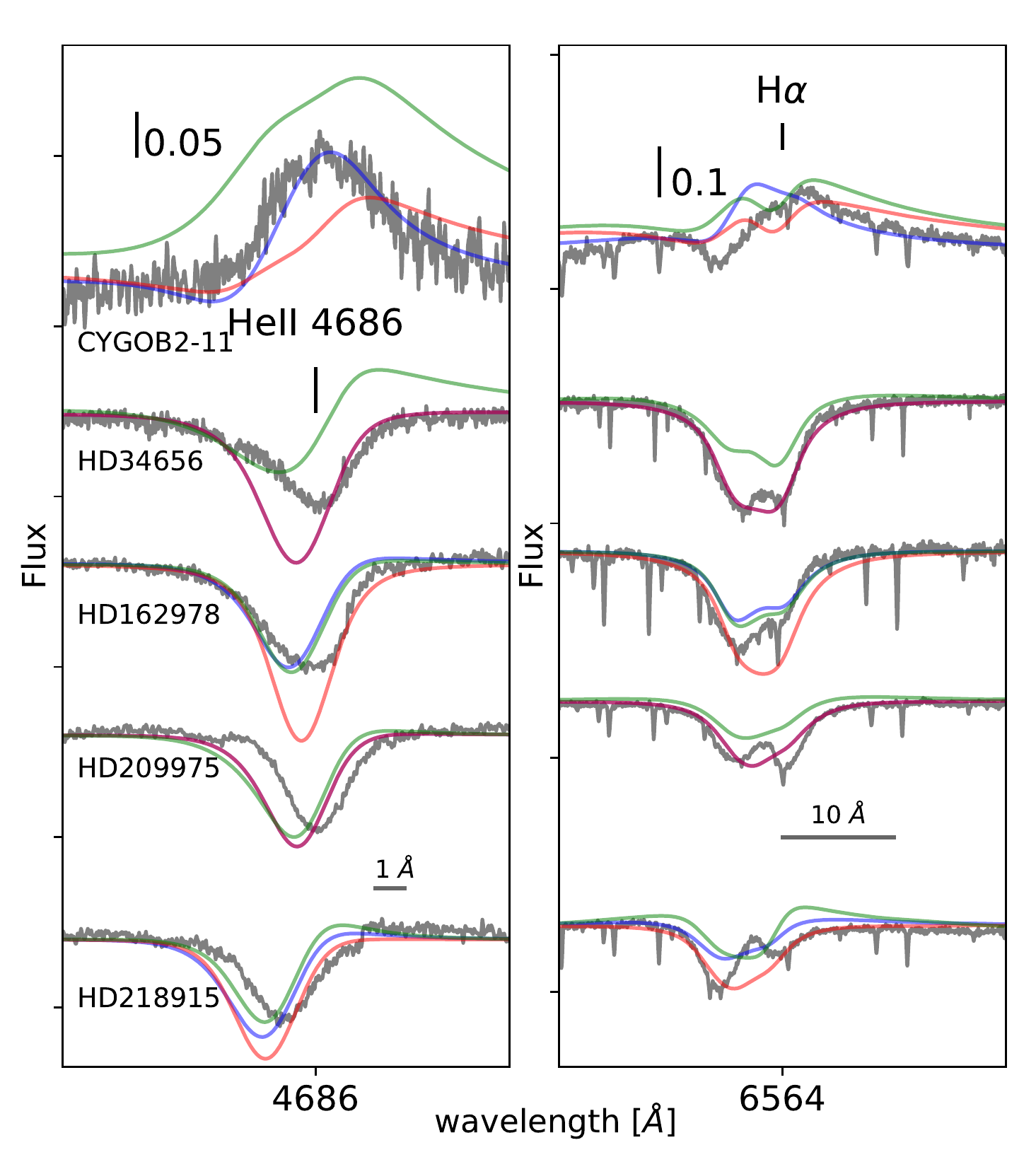}
		\caption{As Fig.~\ref{ExamplesQ3_1}, but for stars labeled as Q3, with luminosity class I or II, and where the fitting of H$\alpha$ implies a lower value of log~$Q$ compared to \ion{He}{ii}\,$\lambda$4686}.
		\label{ExamplesQ3_3}
	\end{figure}

	\begin{figure}
		\centering
		\includegraphics[width=0.49\textwidth]{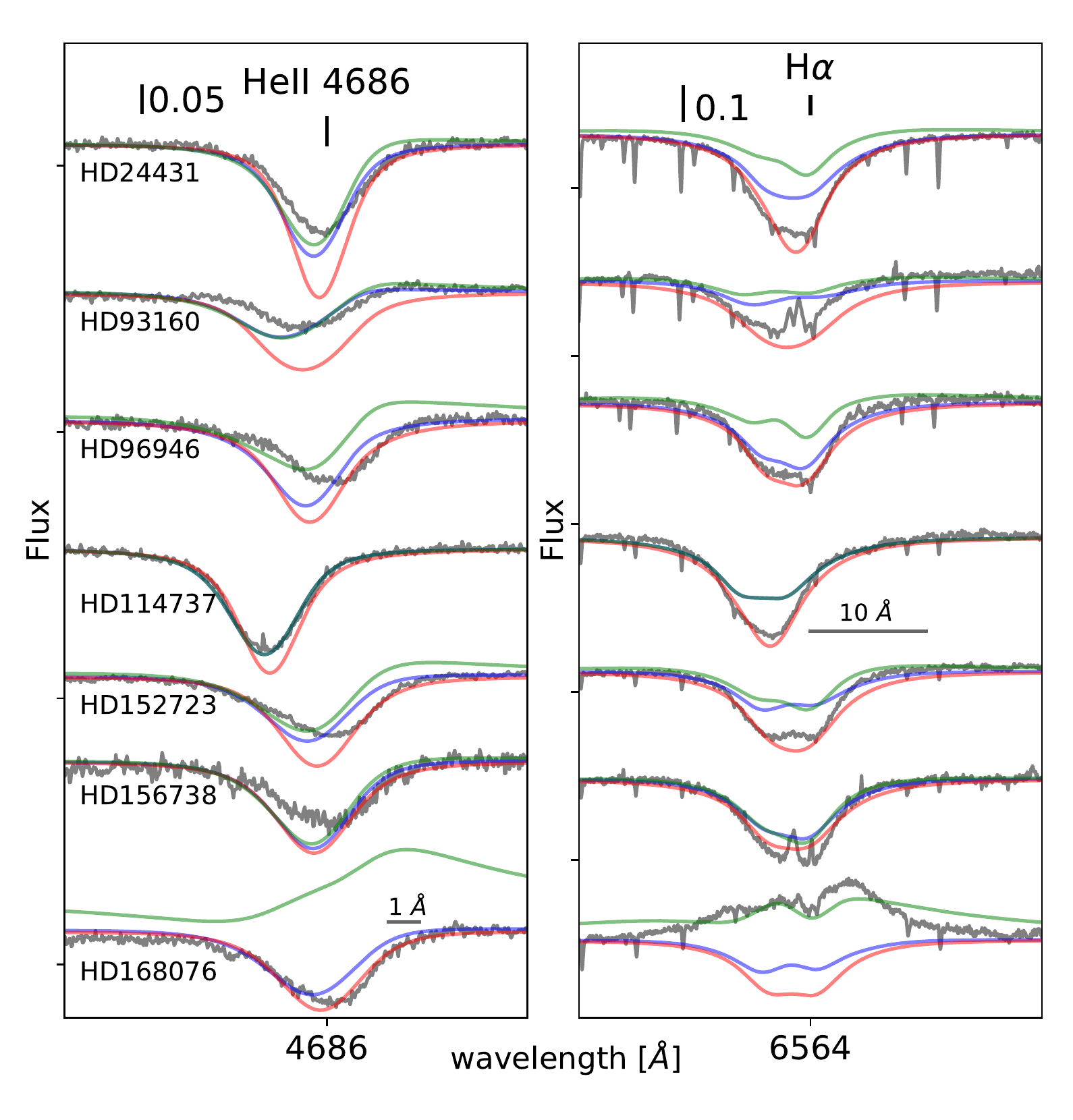}
		\caption{As Fig.~\ref{ExamplesQ3_1} but for luminosity class III stars. All but one (HD~168076~AB) objects present a lower log~$Q$ value when fitting only H$\alpha$.
}
		\label{ExamplesQ3_4}
	\end{figure}

	\begin{figure}
		\centering
		\includegraphics[width=0.49\textwidth]{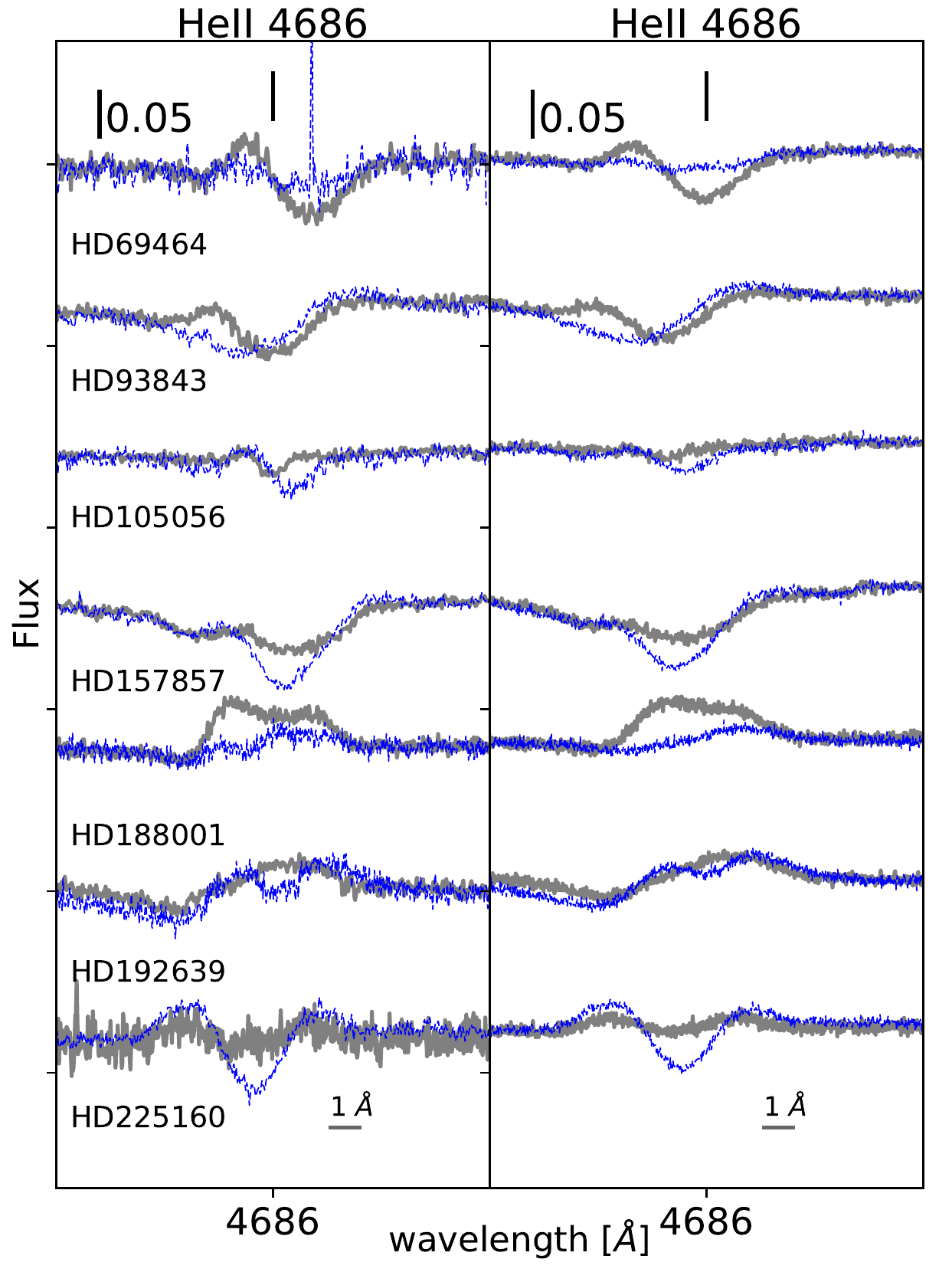}
		\caption{Impact of resolving power for stars with very strong variability in \ioni{He}{ii}\,$\lambda$4686. The comparison is between two extreme cases. [Left] Spectra with the original, high resolution. There is one star (HDE\,229196, O6~II, \vsini\,$\sim$\,250\kms) which does not appear in the histogram. [Right] Spectra downgraded to R=2500, the standard resolution used for spectral classification.}
		\label{HeII4686+R2500}
	\end{figure}

				\begin{landscape}	
			\begin{figure}
				\centering
				\includegraphics[scale=1.5]{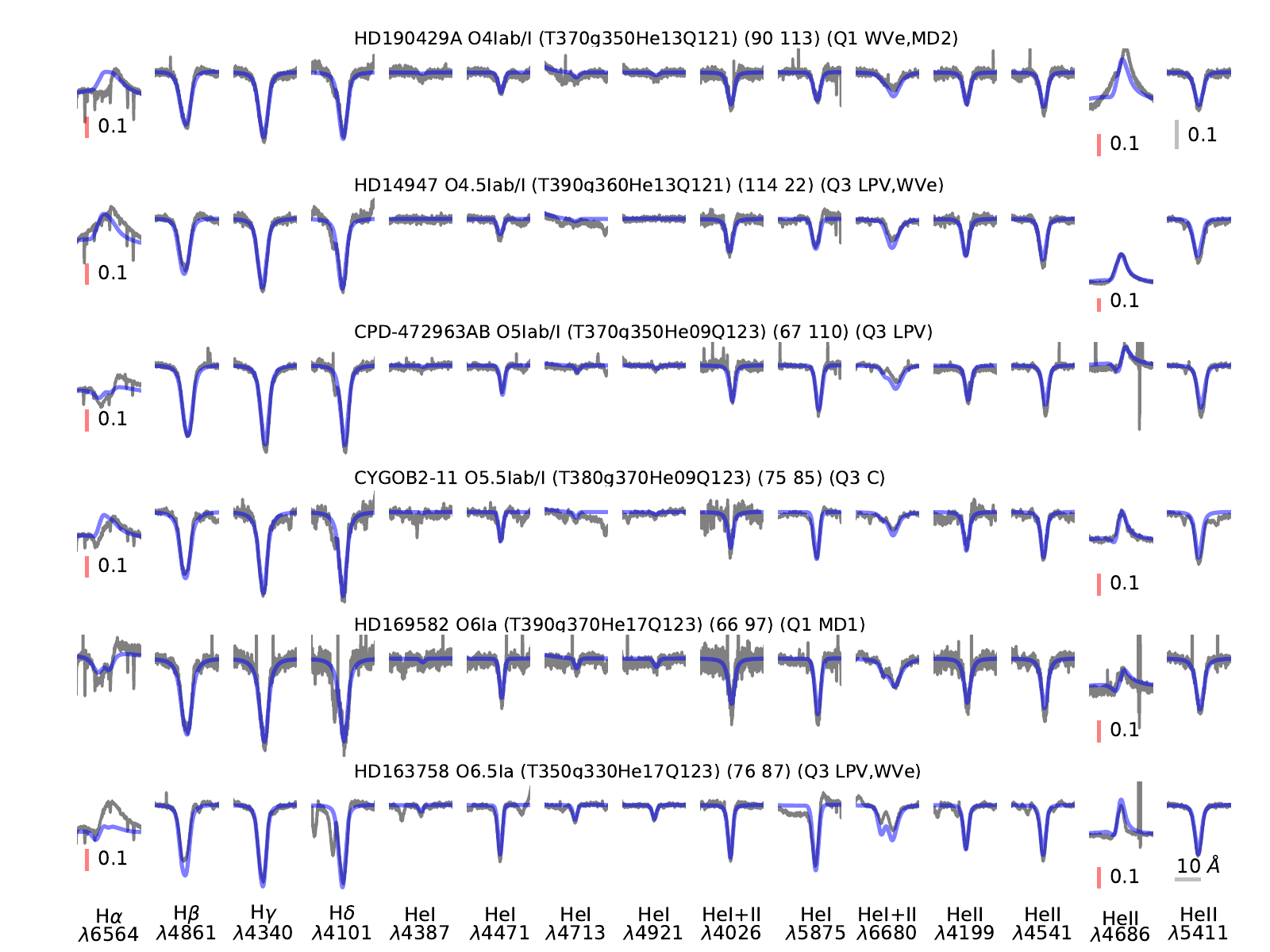}
				\caption{Observed spectra (grey) and best fitting model (blue) comparison for the main diagnostic lines. The sample is sorted following table~\ref{ParamStandards}. Each star is labeled with its name and spectral class, the model used for comparison (properly convolved with resolution and \vrad), \vsini\ and \vmacro\ values, the quality flag and notes on binarity. The scale is the same for all lines except those marked with a red scale.}
				\label{All0}
			\end{figure}
						\begin{figure}
							\centering
							\includegraphics[scale=1.5]{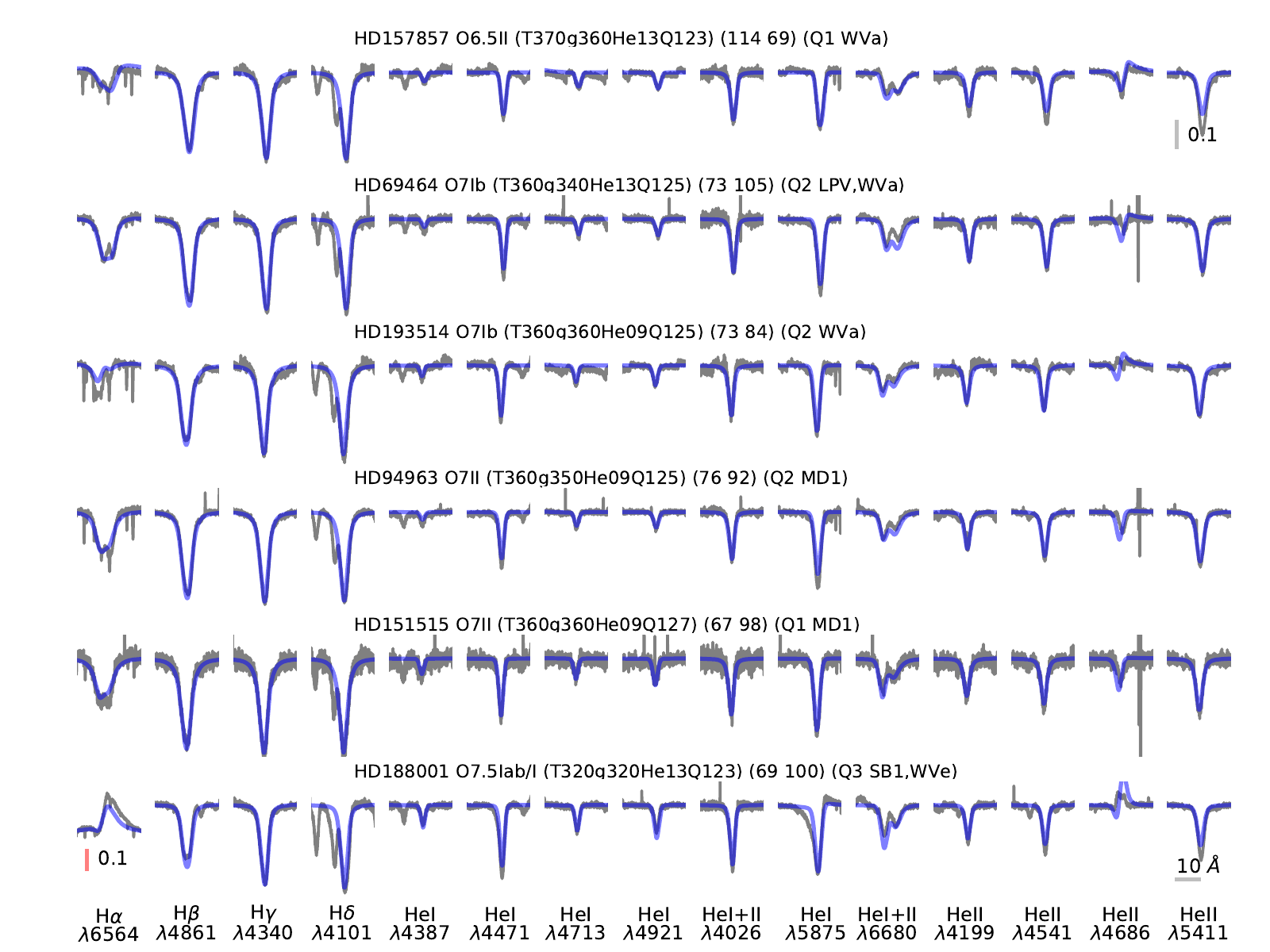}
							\caption{Same as Fig.~\ref{All0}.}
							\label{All5}
						\end{figure}
							\begin{figure}
								\centering
								\includegraphics[scale=1.5]{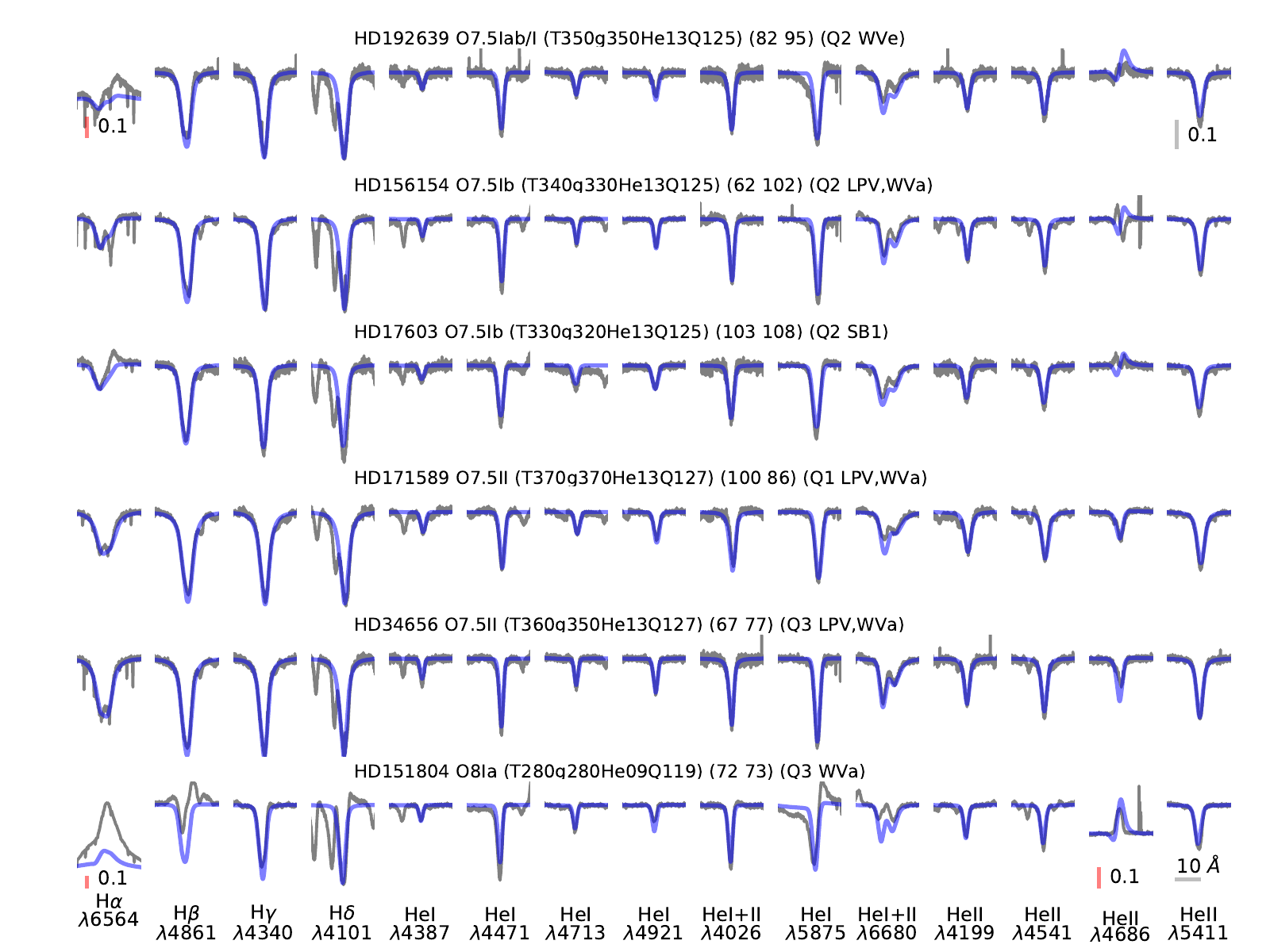}
								\caption{Same as Fig.~\ref{All0}.}
								\label{All10}
							\end{figure}
								\begin{figure}
									\centering
									\includegraphics[scale=1.5]{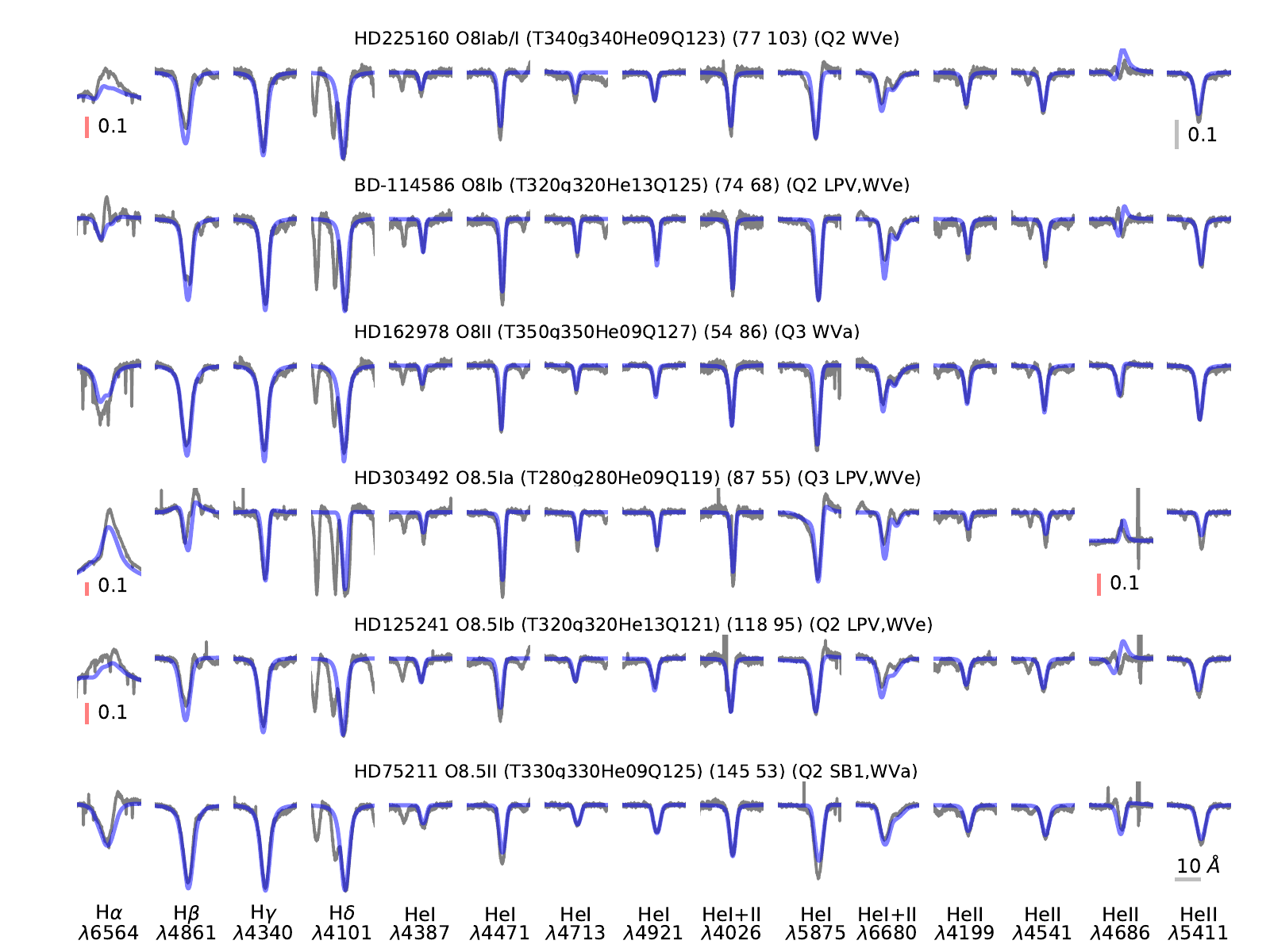}
									\caption{Same as Fig.~\ref{All0}.}
									\label{All15}
								\end{figure}
									\begin{figure}
										\centering
										\includegraphics[scale=1.5]{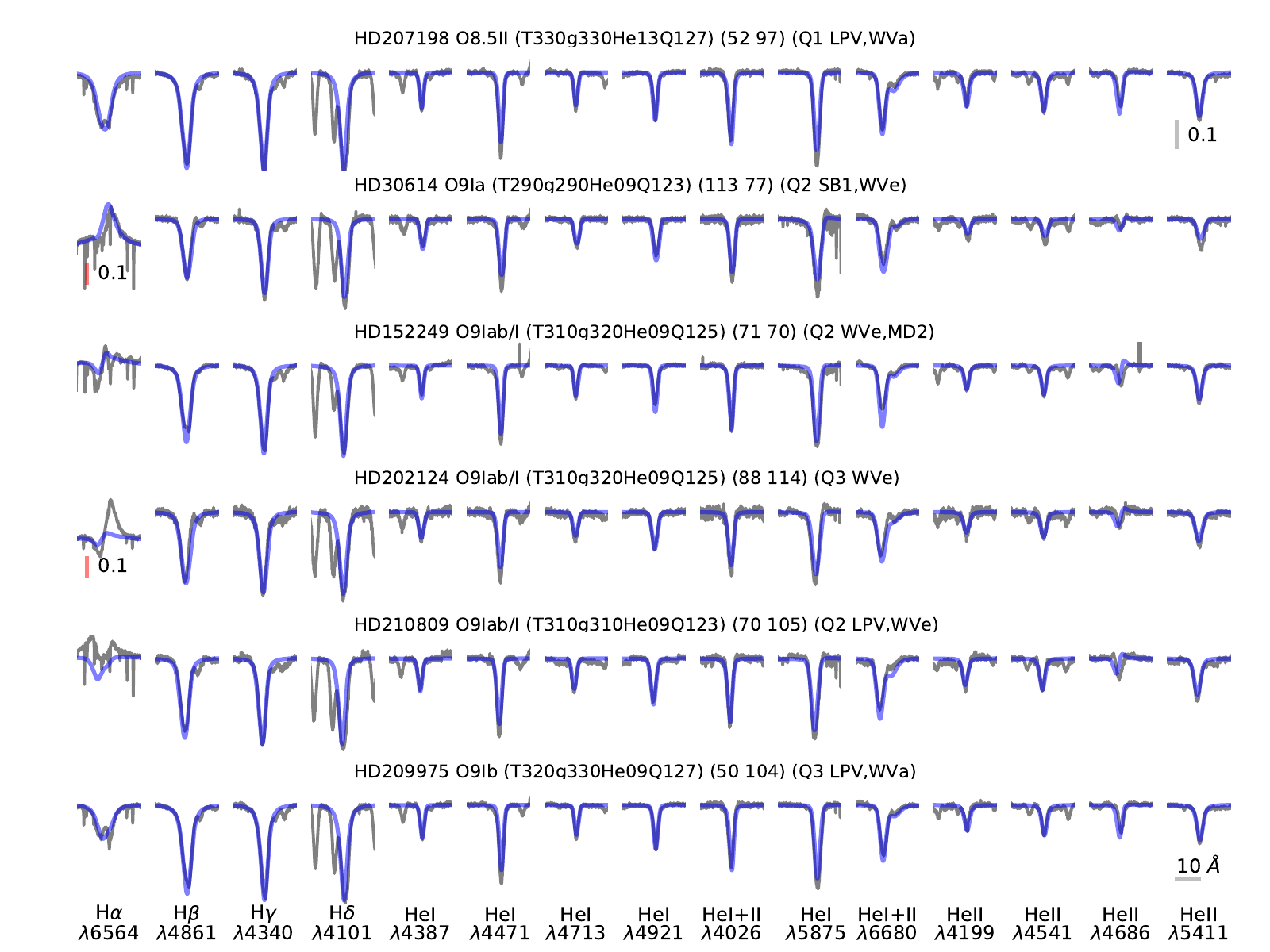}
										\caption{Same as Fig.~\ref{All0}.}
										\label{All20}
									\end{figure}
										\begin{figure}
											\centering
											\includegraphics[scale=1.5]{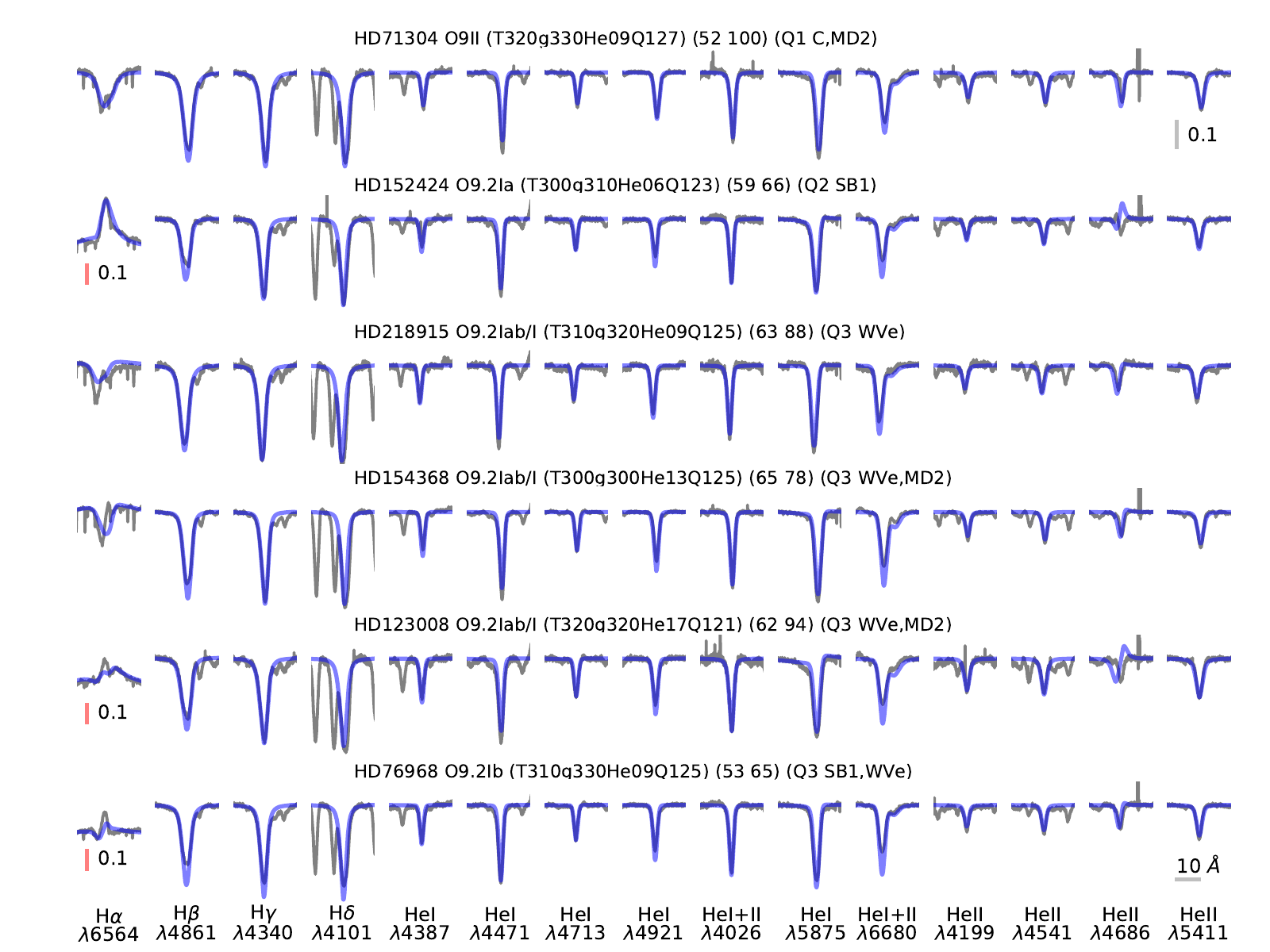}
											\caption{Same as Fig.~\ref{All0}.}
											\label{All25}
										\end{figure}
											\begin{figure}
												\centering
												\includegraphics[scale=1.5]{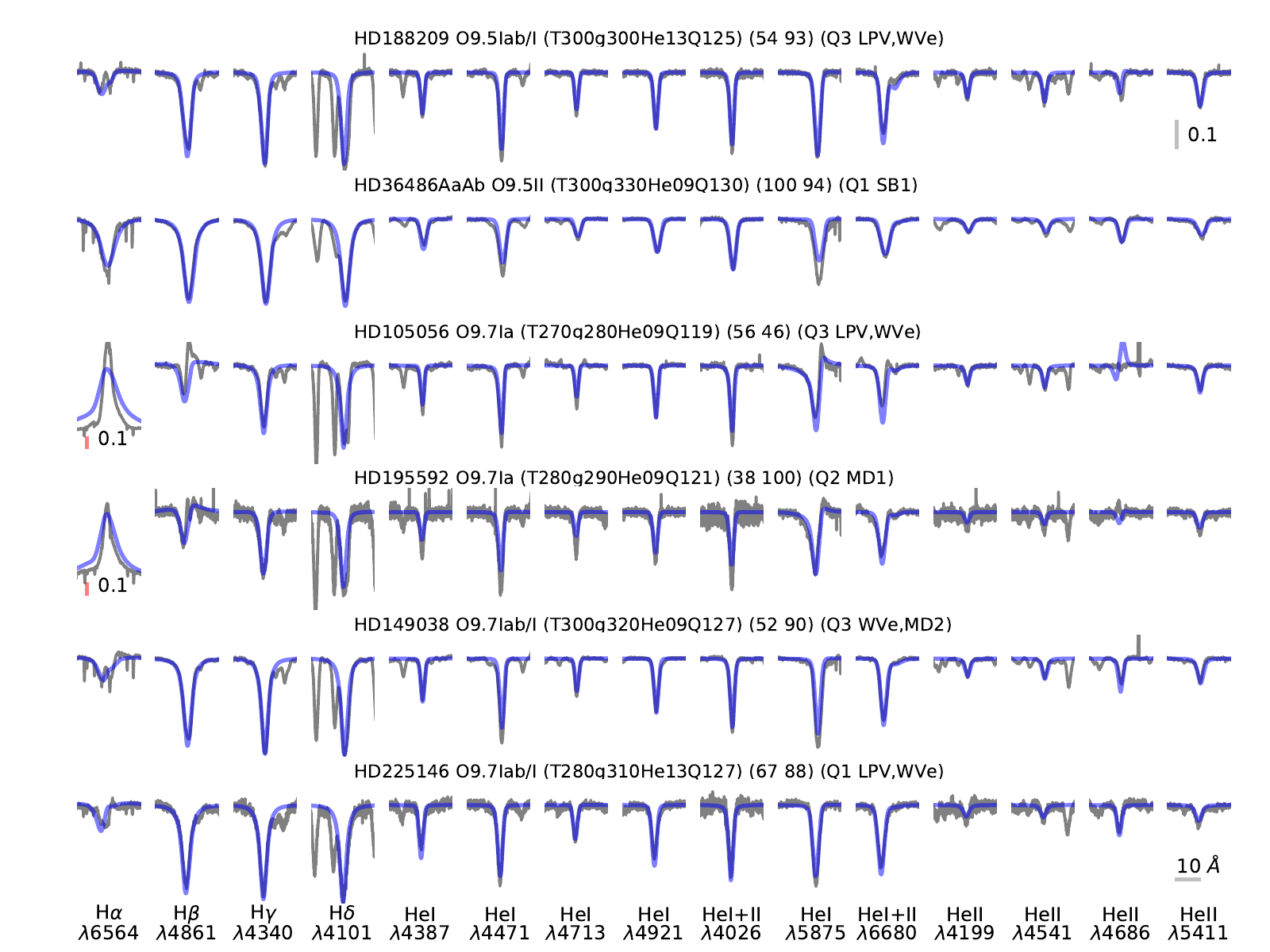}
												\caption{Same as Fig.~\ref{All0}.}
												\label{All30}
											\end{figure}
												\begin{figure}
													\centering
													\includegraphics[scale=1.5]{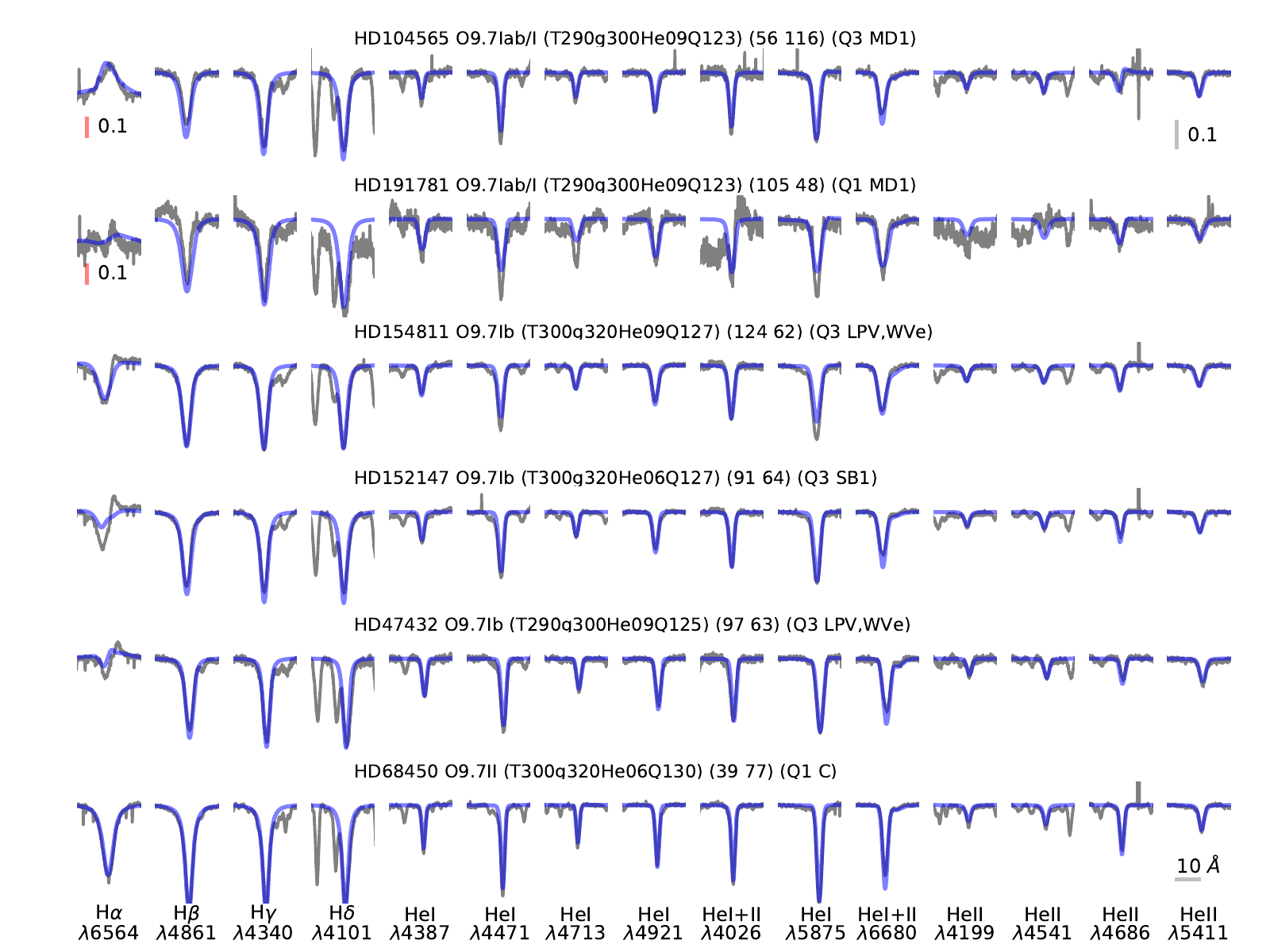}
													\caption{Same as Fig.~\ref{All0}.}
													\label{All35}
												\end{figure}
													\begin{figure}
														\centering
														\includegraphics[scale=1.5]{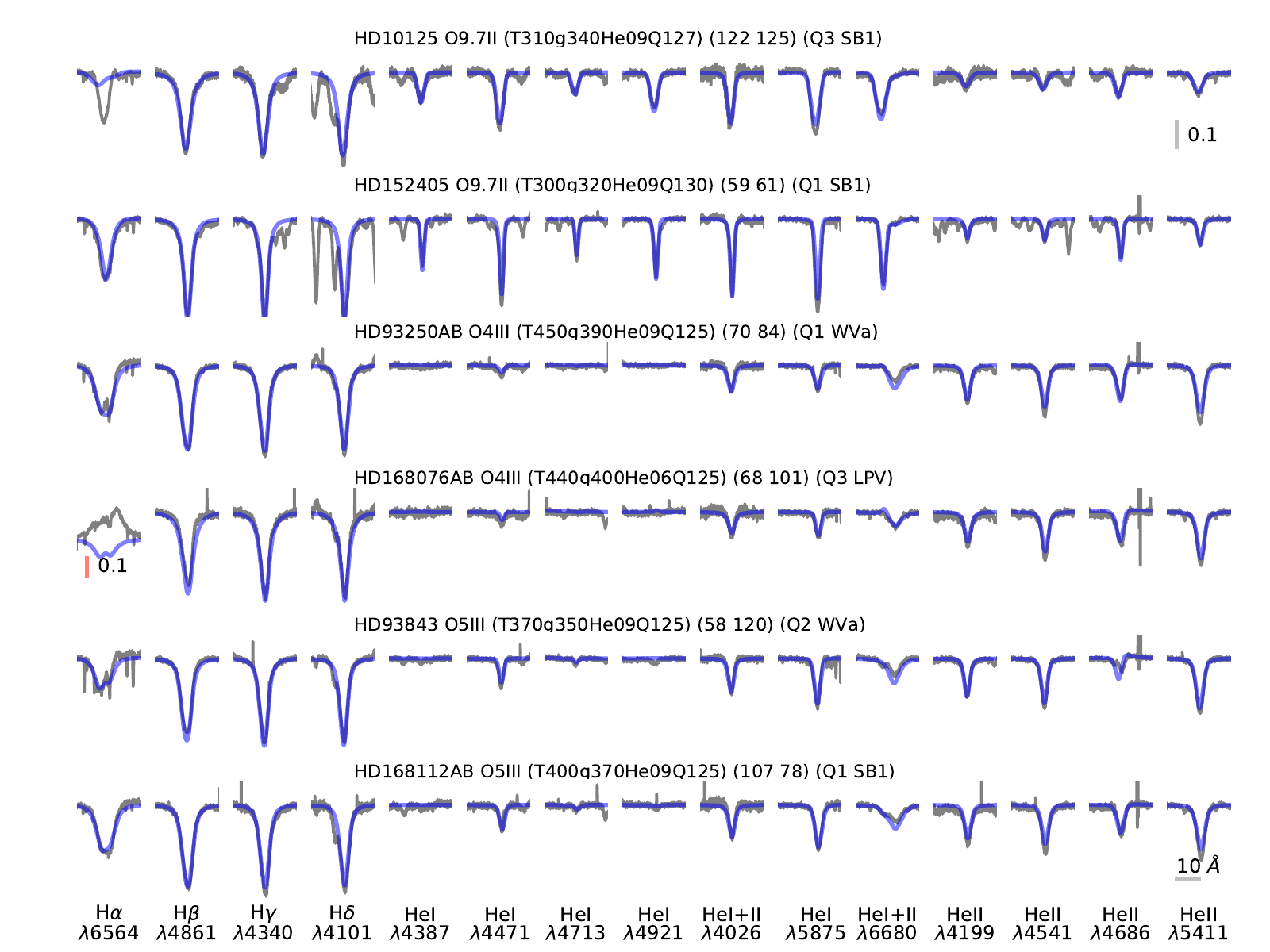}
														\caption{Same as Fig.~\ref{All0}.}
														\label{All40}
													\end{figure}
														\begin{figure}
															\centering
															\includegraphics[scale=1.5]{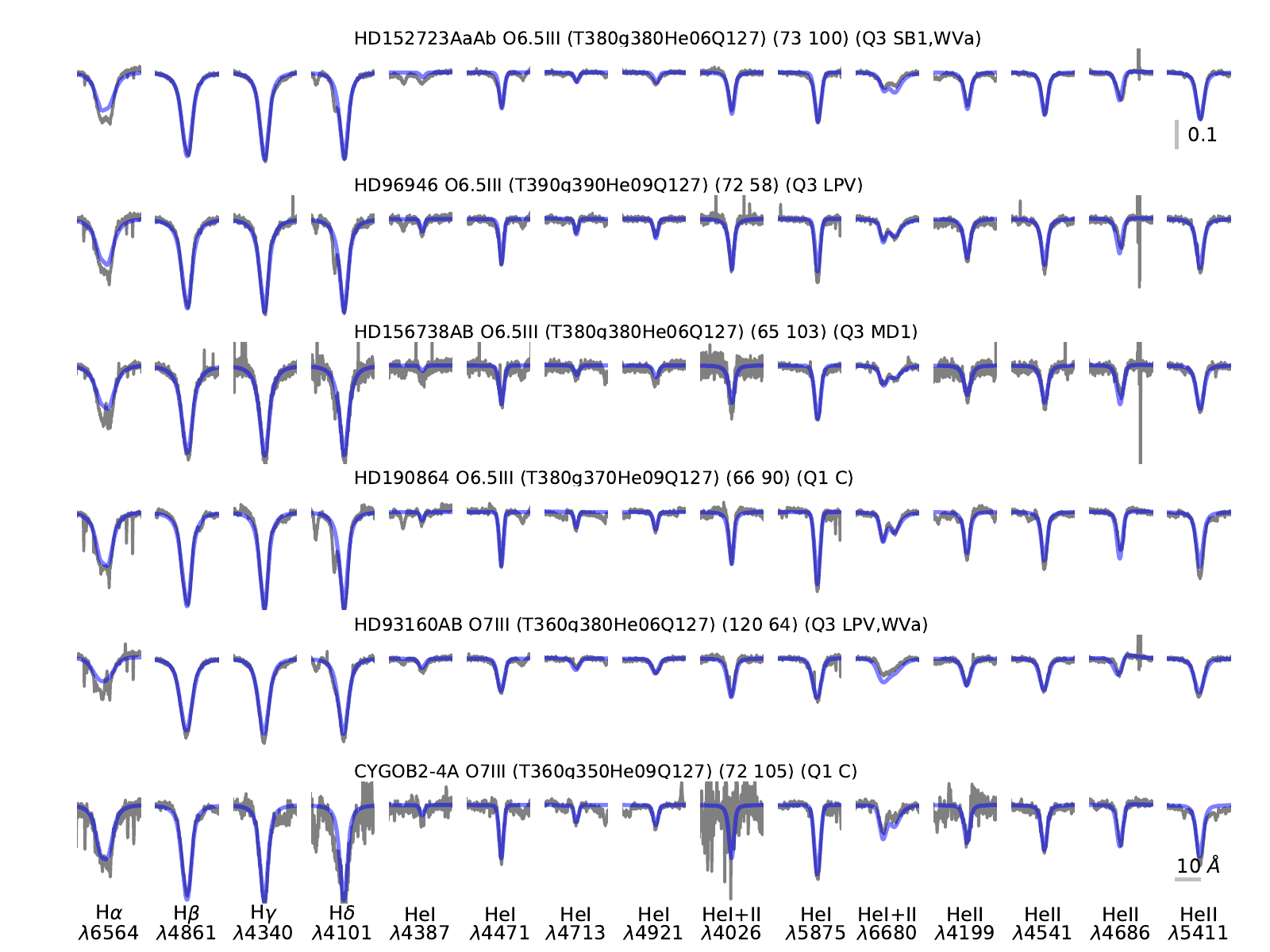}
															\caption{Same as Fig.~\ref{All0}.}
															\label{All45}
														\end{figure}
															\begin{figure}
																\centering
																\includegraphics[scale=1.5]{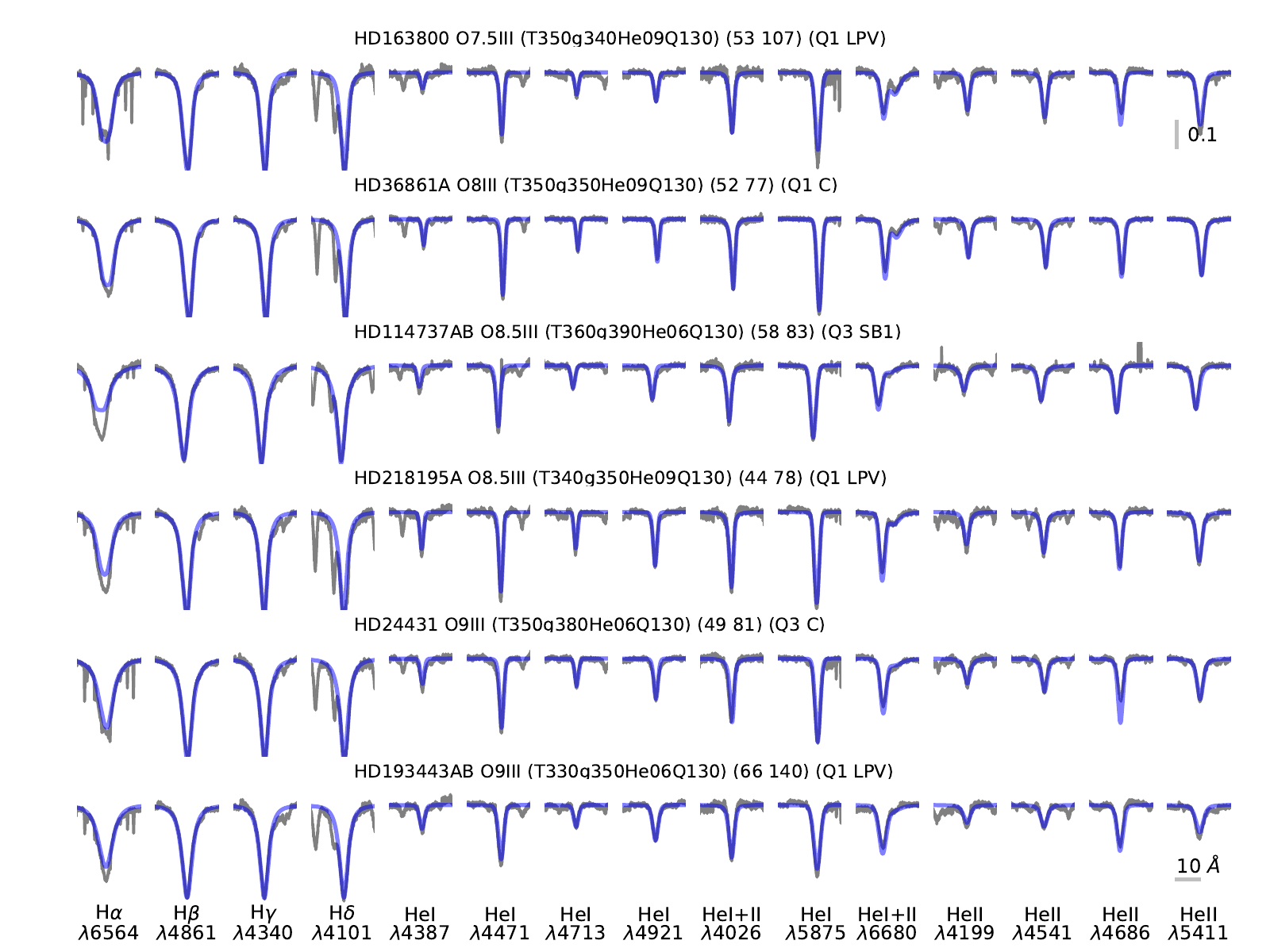}
																\caption{Same as Fig.~\ref{All0}.}
																\label{All50}
															\end{figure}
																\begin{figure}
																	\centering
																	\includegraphics[scale=1.5]{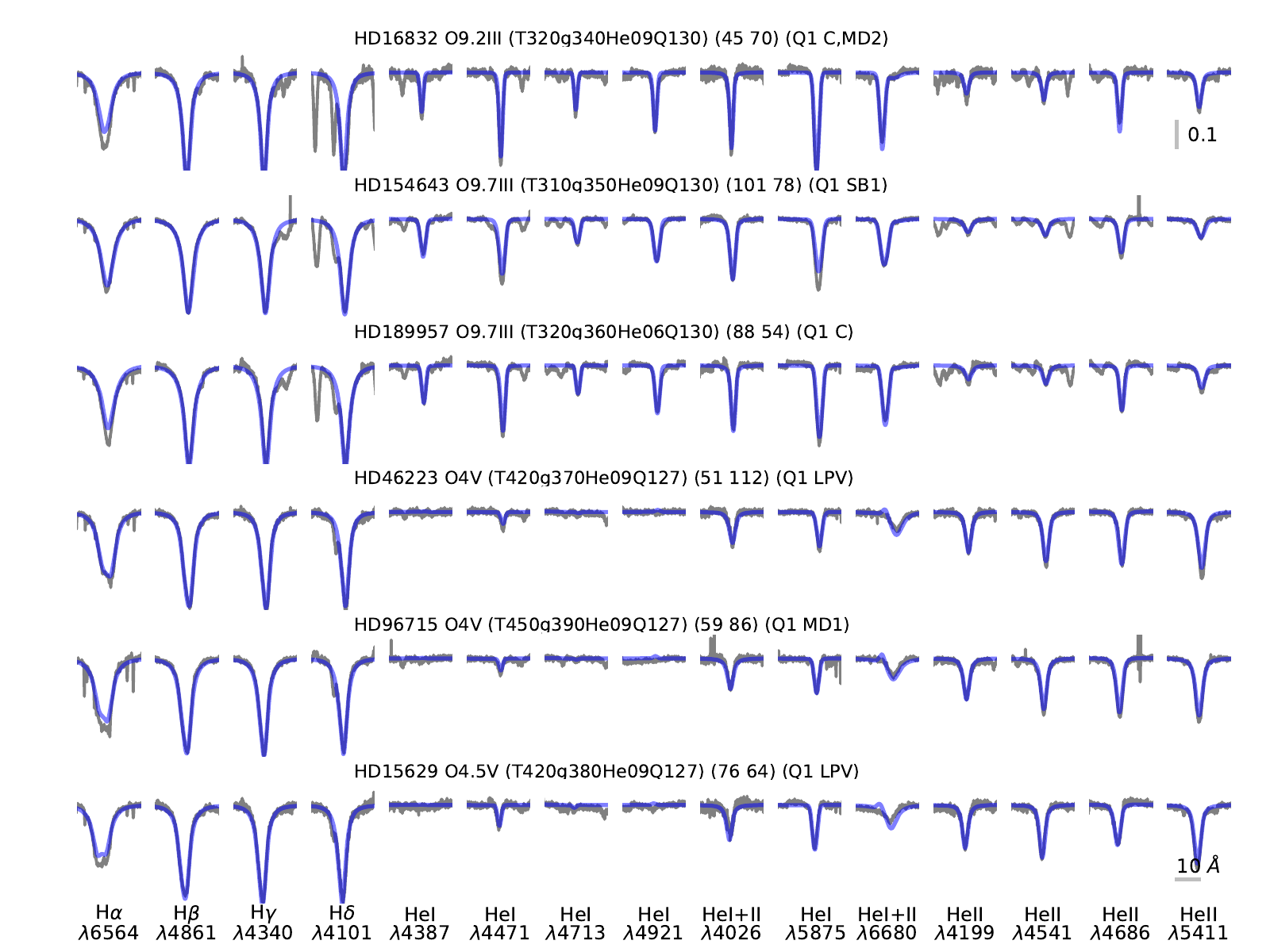}
																	\caption{Same as Fig.~\ref{All0}.}
																	\label{All55}
																\end{figure}
																	\begin{figure}
																		\centering
																		\includegraphics[scale=1.5]{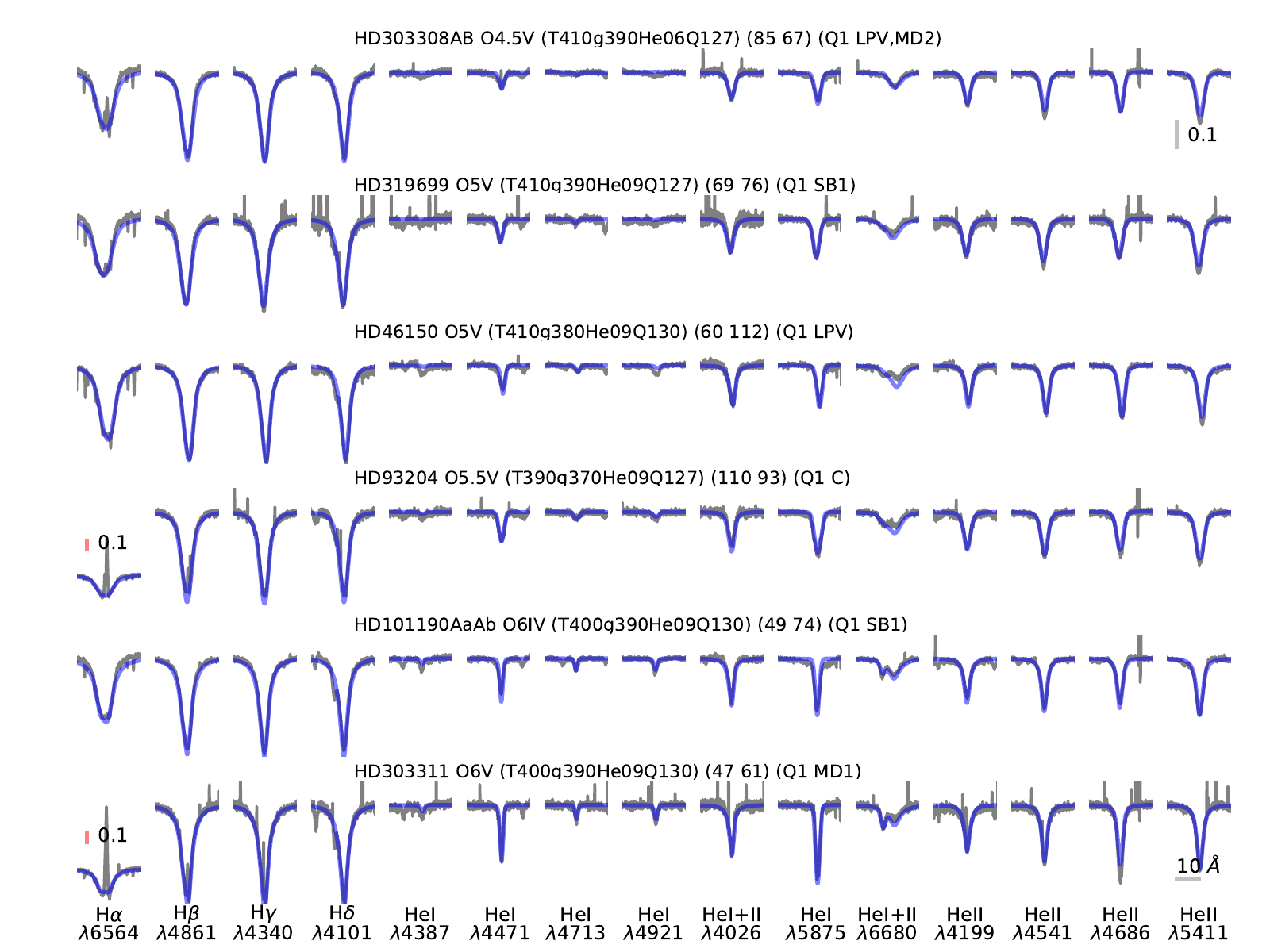}
																		\caption{Same as Fig.~\ref{All0}.}
																		\label{All60}
																	\end{figure}
																		\begin{figure}
																			\centering
																			\includegraphics[scale=1.5]{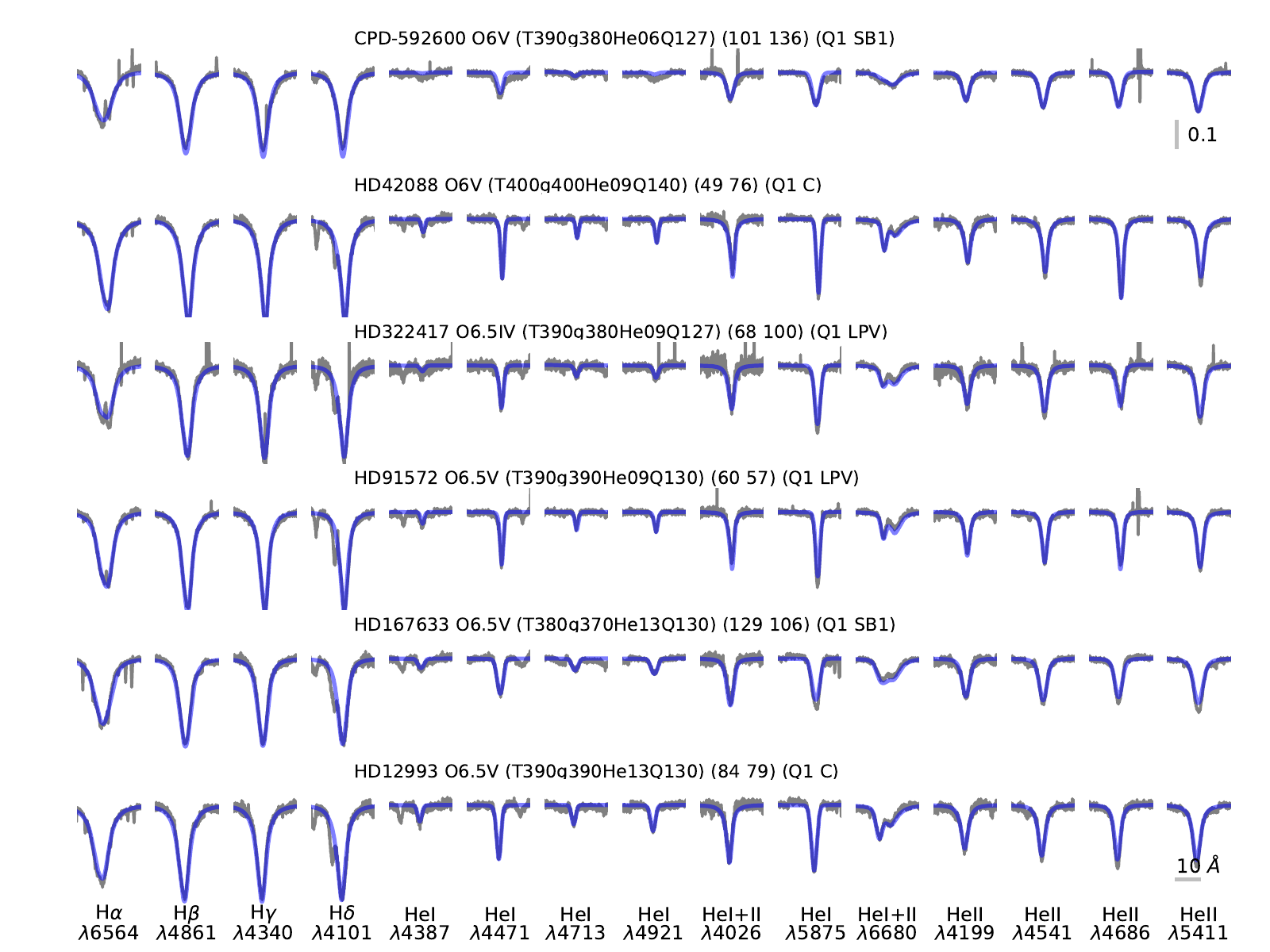}
																			\caption{Same as Fig.~\ref{All0}.}
																			\label{All65}
																		\end{figure}
																			\begin{figure}
																				\centering
																				\includegraphics[scale=1.5]{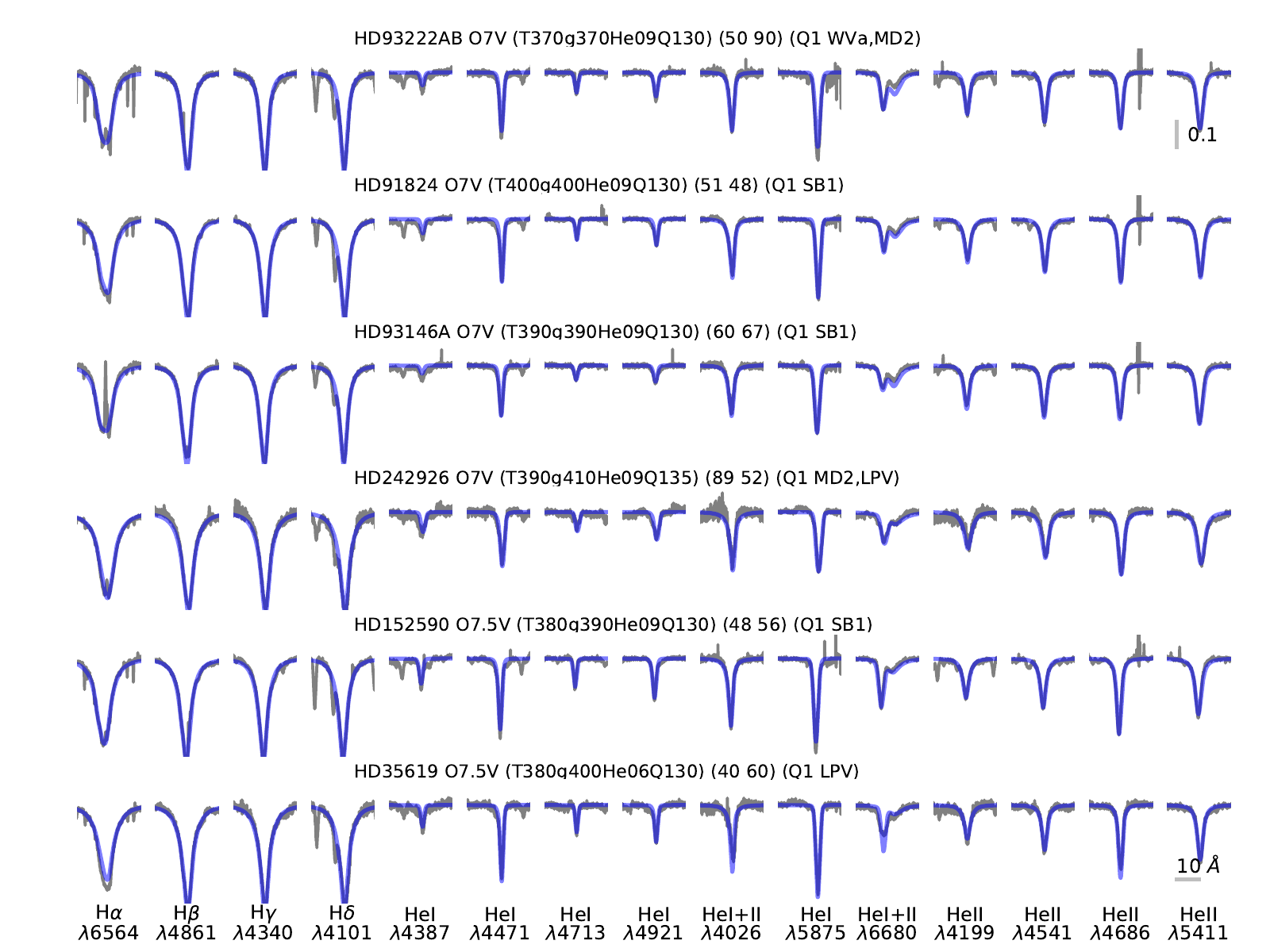}
																				\caption{Same as Fig.~\ref{All0}.}
																				\label{All70}
																			\end{figure}
																			\clearpage
																\begin{figure}
																	\centering
																	\includegraphics[scale=1.5]{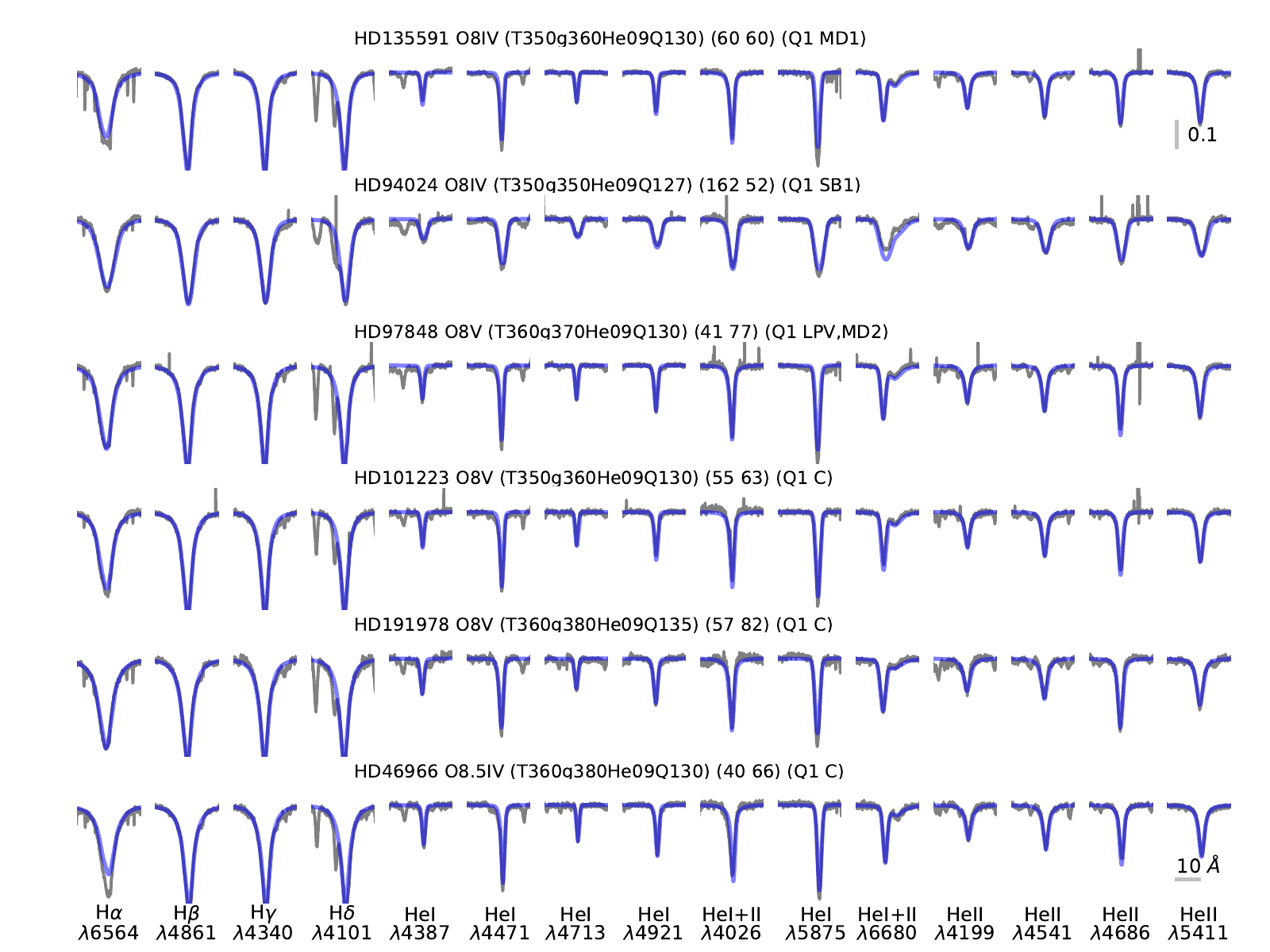}
																	\caption{Same as Fig.~\ref{All0}.}
																	\label{All75}
																\end{figure}
														\begin{figure}
															\centering
															\includegraphics[scale=1.5]{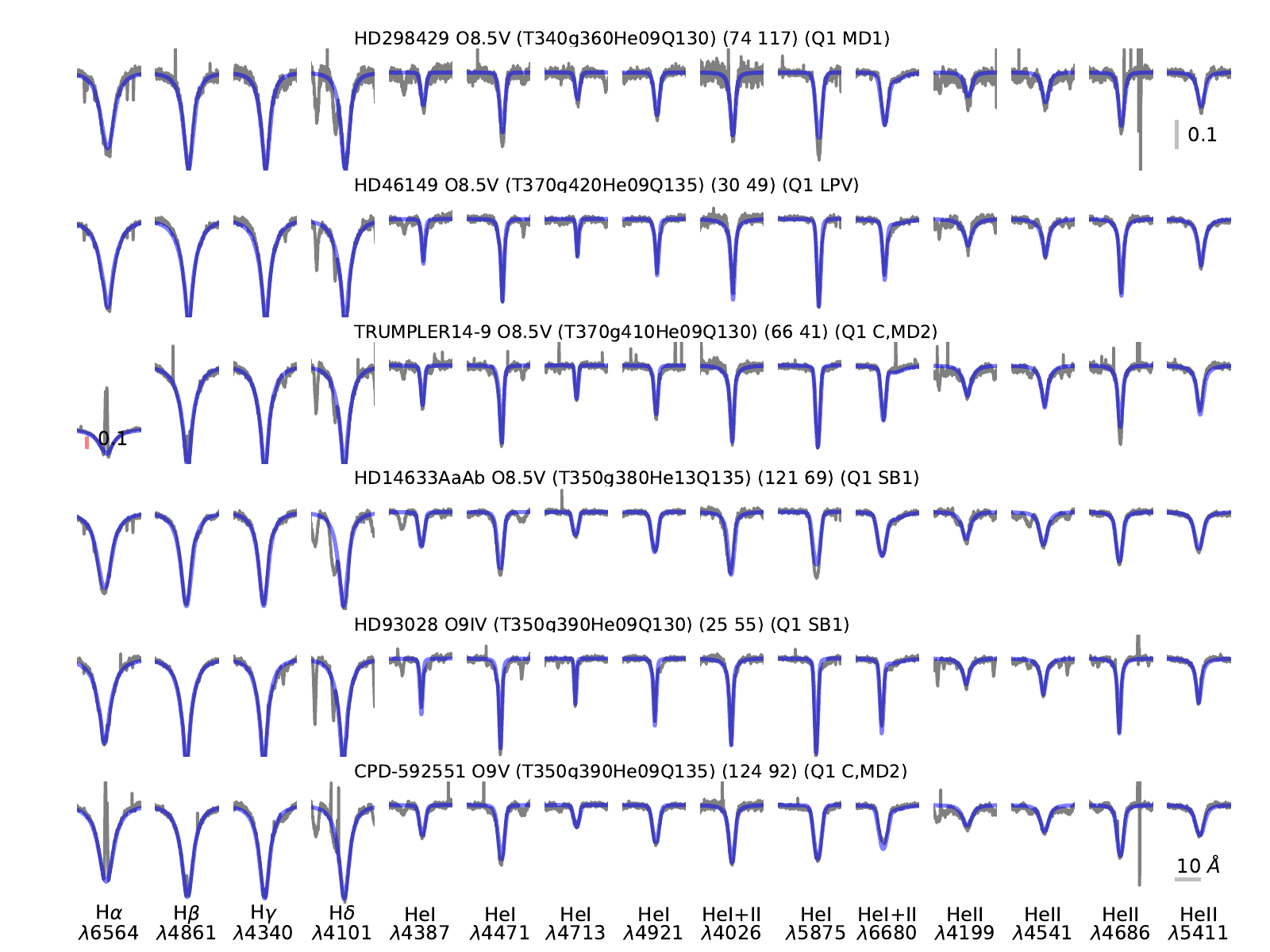}
															\caption{Same as Fig.~\ref{All0}.}
															\label{All80}
														\end{figure}
													\begin{figure}
														\centering
														\includegraphics[scale=1.5]{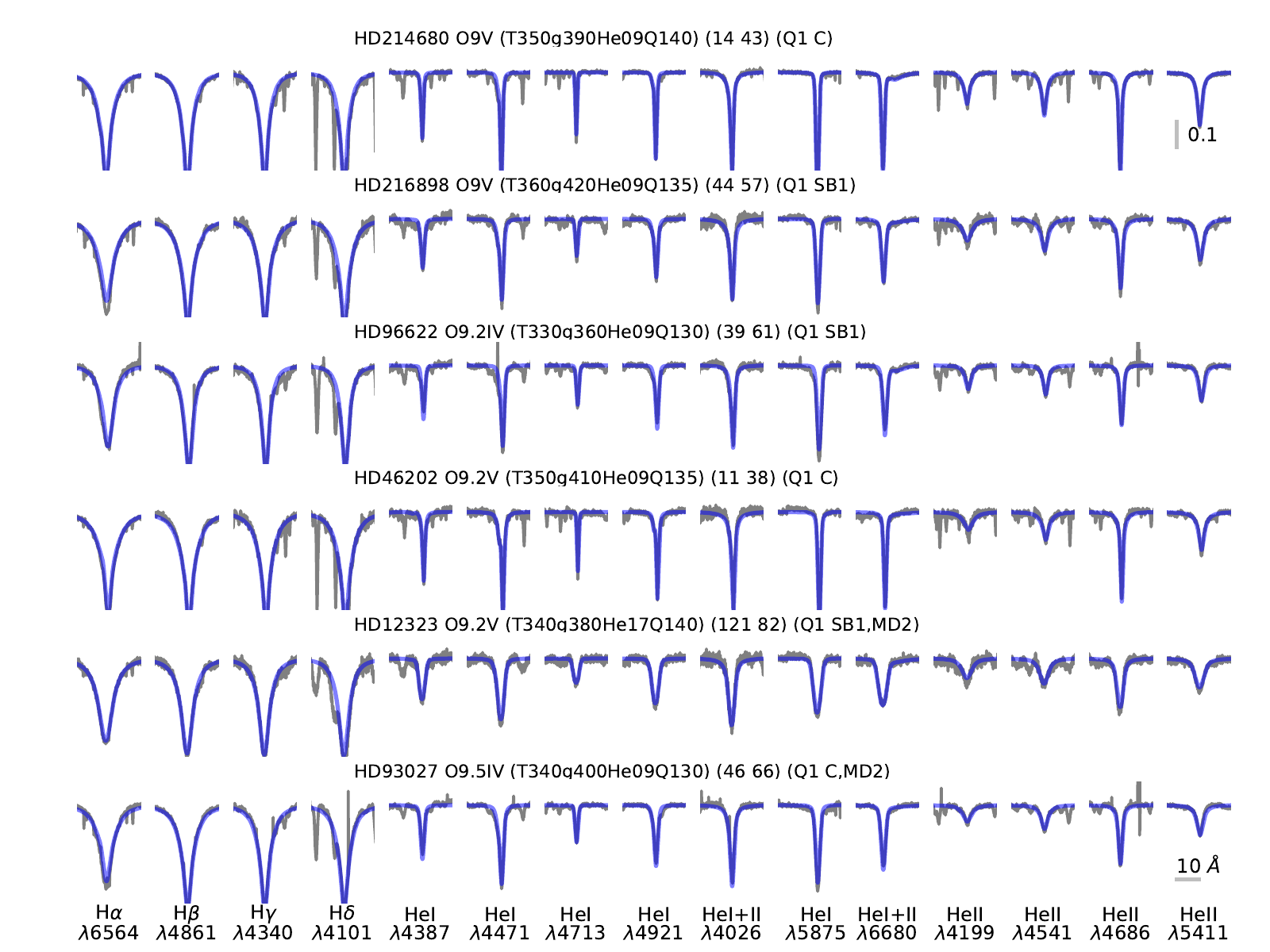}
														\caption{Same as Fig.~\ref{All0}.}
														\label{All85}
													\end{figure}
												\begin{figure}
													\centering
													\includegraphics[scale=1.5]{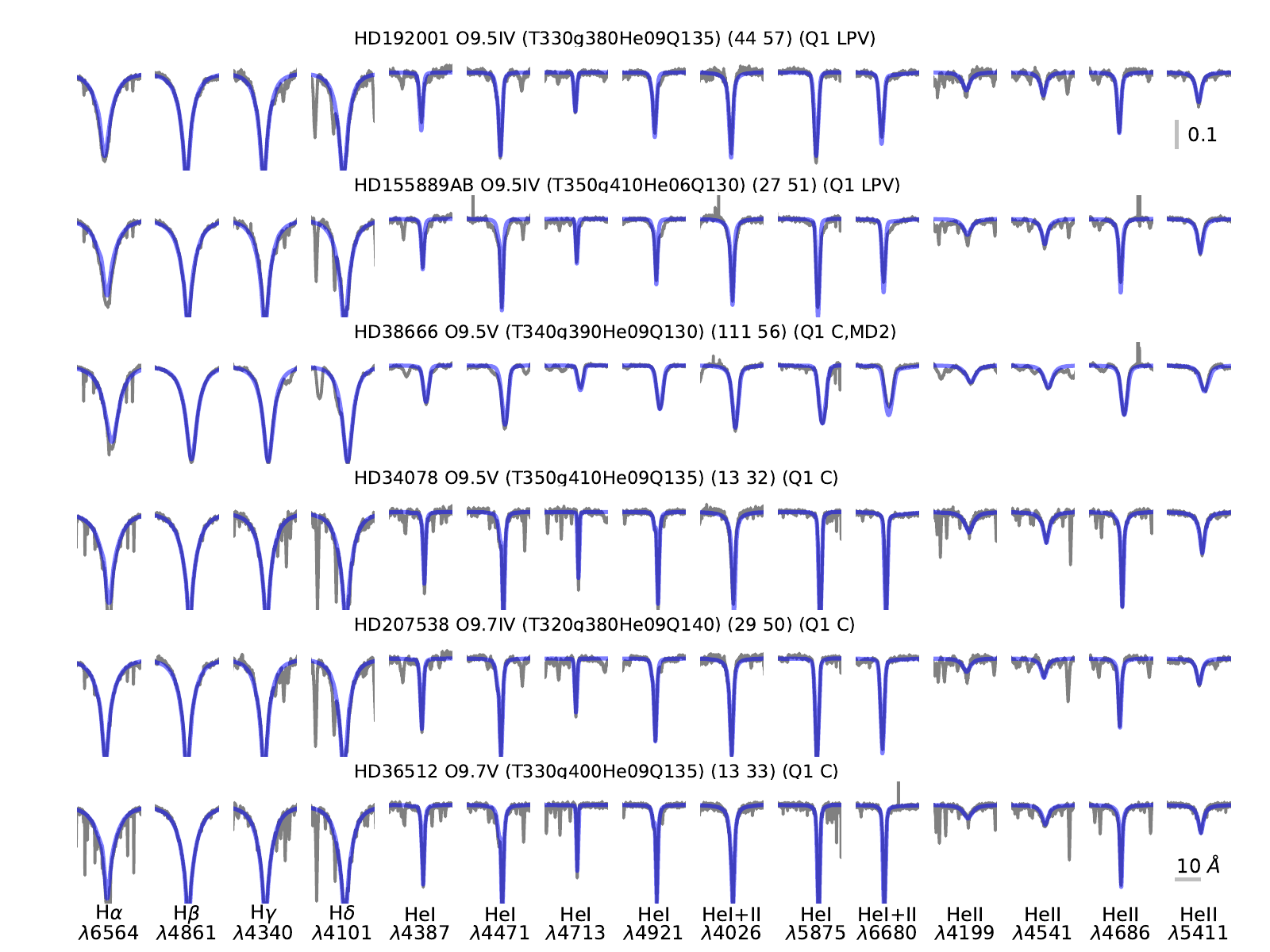}
													\caption{Same as Fig.~\ref{All0}.}
													\label{All90}
												\end{figure}
											\begin{figure}
												\centering
												\includegraphics[scale=1.5]{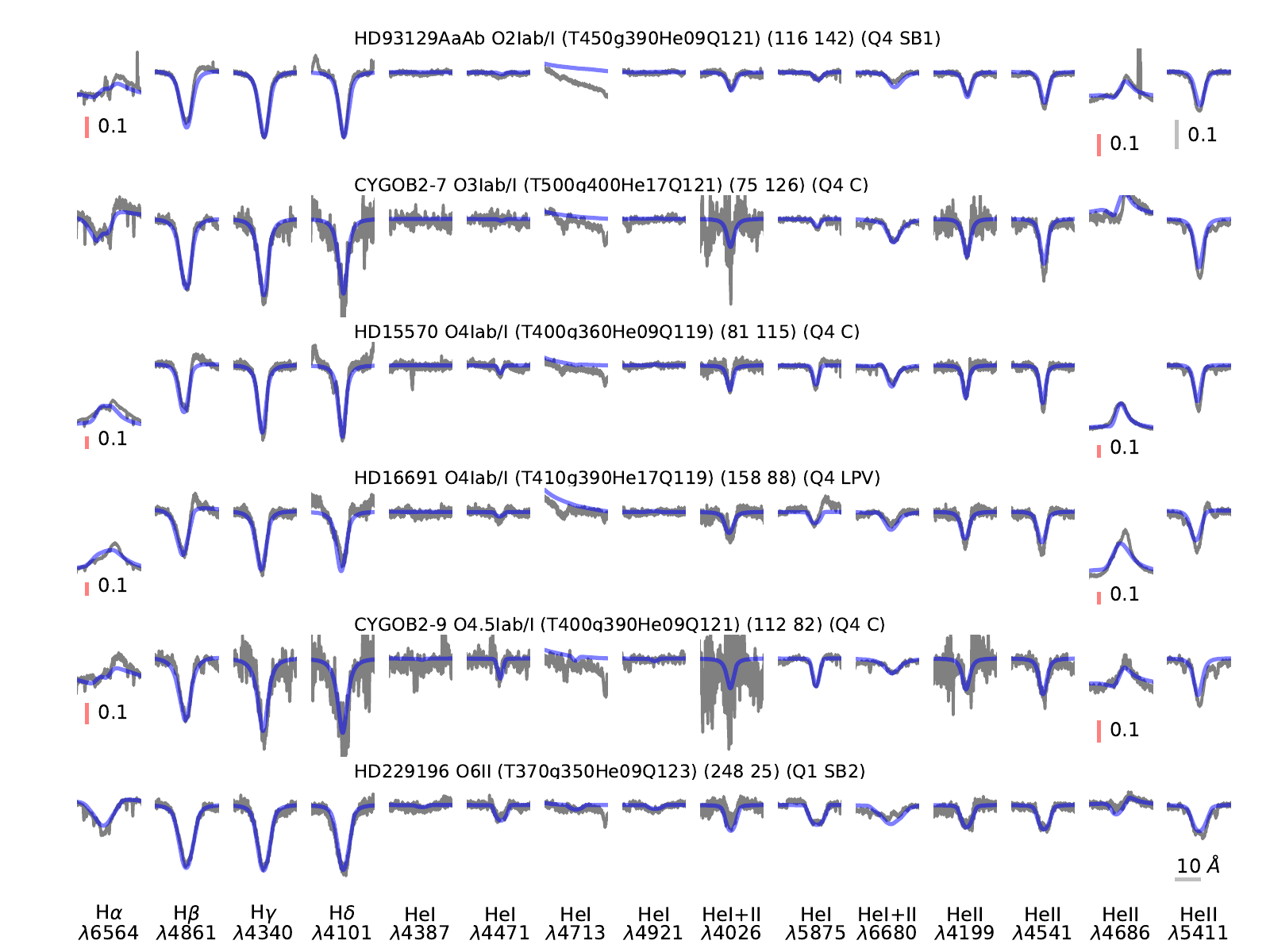}
												\caption{Same as Fig.~\ref{All0}.}
												\label{All115}
											\end{figure}
											\begin{figure}
												\centering
												\includegraphics[scale=1.5]{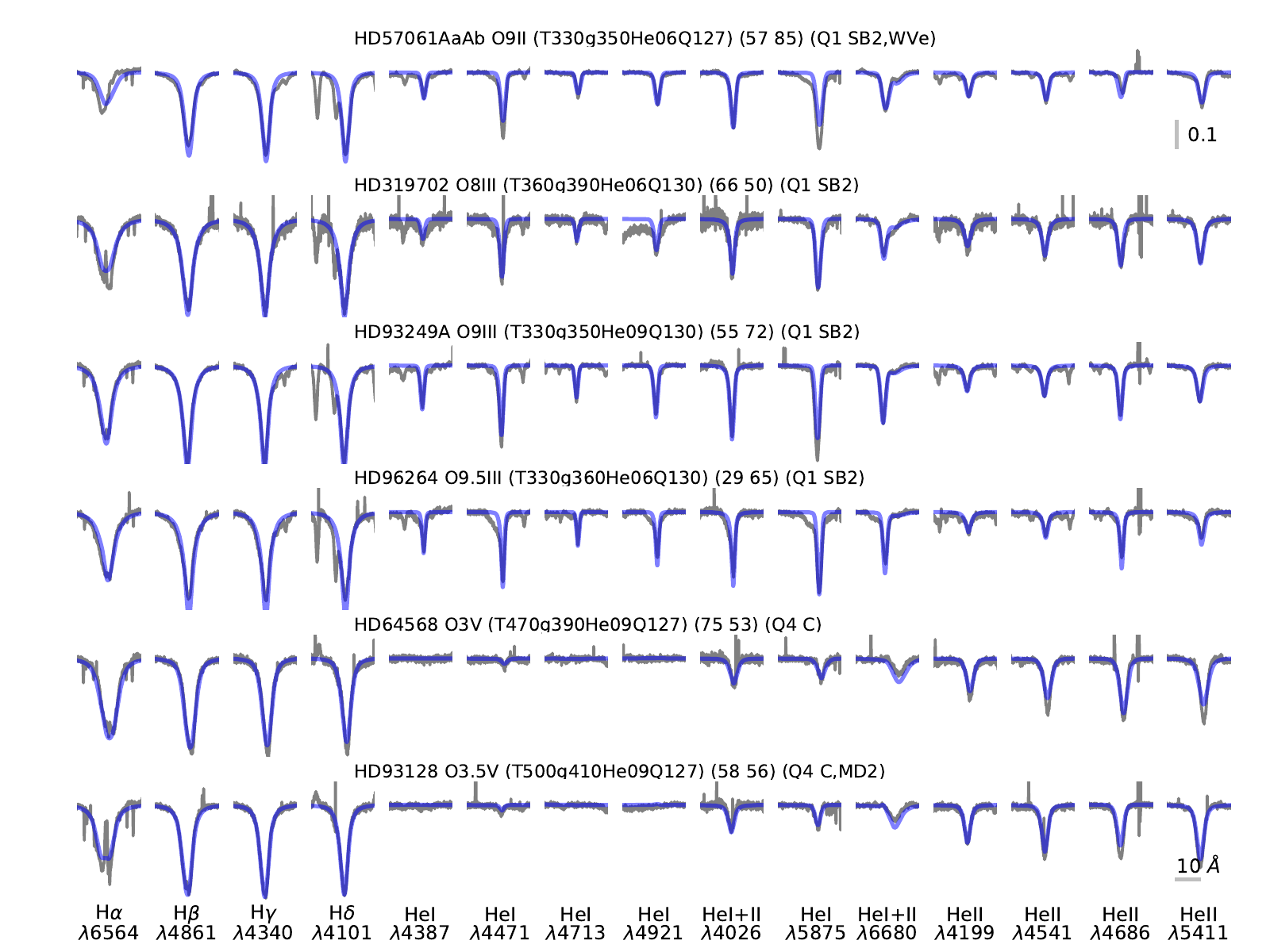}
												\caption{Same as Fig.~\ref{All0}.}
												\label{All121}
											\end{figure}
											\begin{figure}
												\centering
												\includegraphics[scale=1.5]{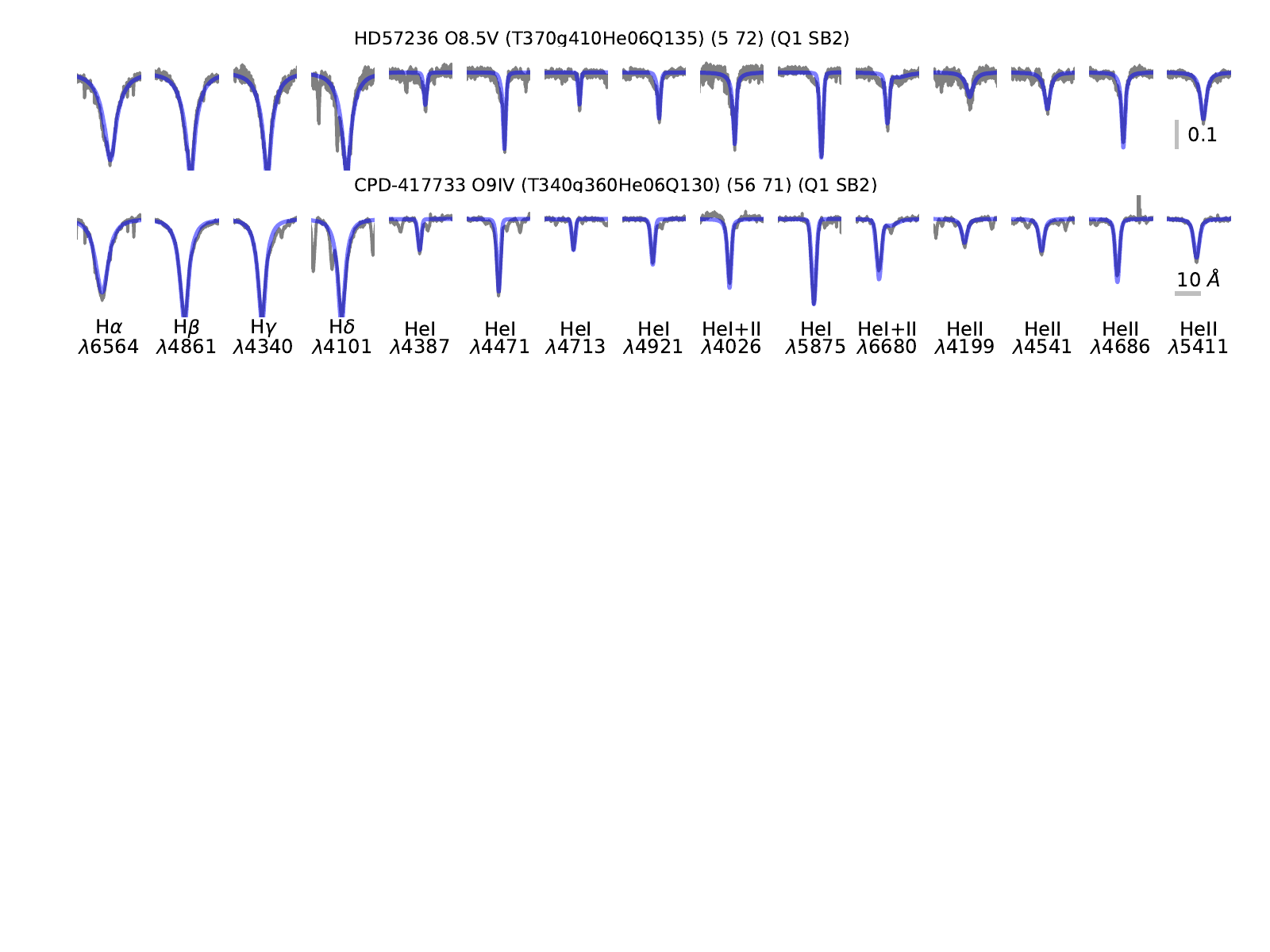}
												\caption{Same as Fig.~\ref{All0}.}
												\label{All127}
											\end{figure}
		\end{landscape}	

%
	\end{document}